\documentclass[twocolumn]{aastex63}
\usepackage[colorinlistoftodos,textsize=small]{todonotes}

\usepackage[modulo, pagewise]{lineno}
\usepackage{multirow}
\usepackage{ulem}
\usepackage{xspace}
\usepackage{ifthen}
\usepackage{comment}
\usepackage{soul}
\usepackage[T1]{fontenc}

\hypersetup{linkcolor=red,citecolor=gray,filecolor=cyan,urlcolor=magenta}

\newcommand{\h}{\mbox{HESS J0632+057}\xspace}
\newcommand{\tpsr}{\mbox{PSR J2032+4127/MT91 213}\xspace}
\newcommand{\hbin}{\mbox{PSR B1259-63/LS 2883}\xspace}
\newcommand{\lsi}{\mbox{LS I +61$^\circ$ 303}\xspace}
\newcommand{\lsf}{\mbox{LS 5039}\xspace}
\newcommand{\hjei}{\mbox{HESS J1832-093}\xspace}
\newcommand{\fgl}{\mbox{1FGL J1018.6-5856}\xspace}

\def \hess {H.E.S.S.\xspace}

\def\fordiscussion{2}
\def\fordiscussion{1}

\newcommand{\RN}[1]{\uppercase\expandafter{\romannumeral#1}}

\accepted{September 22, 2020}

\submitjournal{ApJ}

\shorttitle{Observations of \h}
\shortauthors{The \hess, MAGIC, and VERITAS Collaborations}

\begin{document}

\title{Observation of the gamma-ray binary \h with the \hess, MAGIC, and VERITAS telescopes}

\author{C.~B.~Adams}
\affiliation{Physics Department, Columbia University, New York, NY 10027, USA}

\author[0000-0003-2098-170X]{W.~Benbow}
\affiliation{Center for Astrophysics $|$ Harvard \& Smithsonian, Cambridge, MA 02138, USA}

\author{A.~Brill}
\affiliation{Physics Department, Columbia University, New York, NY 10027, USA}

\author{J.~H.~Buckley}
\affiliation{Department of Physics, Washington University, St. Louis, MO 63130, USA}

\author[0000-0002-8136-9461]{M.~Capasso}
\affiliation{Department of Physics and Astronomy, Barnard College, Columbia University, NY 10027, USA}

\author{A.~J.~Chromey}
\affiliation{Department of Physics and Astronomy, Iowa State University, Ames, IA 50011, USA}

\author[0000-0002-1853-863X]{M.~Errando}
\affiliation{Department of Physics, Washington University, St. Louis, MO 63130, USA}

\author{A.~Falcone}
\affiliation{Department of Astronomy and Astrophysics, 525 Davey Lab, Pennsylvania State University, University Park, PA 16802, USA}

\author{K.~A~Farrell}
\affiliation{School of Physics, University College Dublin, Belfield, Dublin 4, Ireland}

\author{Q.~Feng}
\affiliation{Department of Physics and Astronomy, Barnard College, Columbia University, NY 10027, USA}

\author{J.~P.~Finley}
\affiliation{Department of Physics and Astronomy, Purdue University, West Lafayette, IN 47907, USA}

\author{G.~M~Foote}
\affiliation{Department of Physics and Astronomy and the Bartol Research Institute, University of Delaware, Newark, DE 19716, USA}

\author[0000-0002-1067-8558]{L.~Fortson}
\affiliation{School of Physics and Astronomy, University of Minnesota, Minneapolis, MN 55455, USA}

\author[0000-0003-1614-1273]{A.~Furniss}
\affiliation{Department of Physics, California State University - East Bay, Hayward, CA 94542, USA}

\author{A.~Gent}
\affiliation{School of Physics and Center for Relativistic Astrophysics, Georgia Institute of Technology, 837 State Street NW, Atlanta, GA 30332-0430}

\author{G.~H.~Gillanders}
\affiliation{School of Physics, National University of Ireland Galway, University Road, Galway, Ireland}

\author{C.~Giuri}
\affiliation{DESY, Platanenallee 6, 15738 Zeuthen, Germany}

\author[0000-0002-9440-2398]{O.~Gueta}
\affiliation{DESY, Platanenallee 6, 15738 Zeuthen, Germany}

\author{D.~Hanna}
\affiliation{Physics Department, McGill University, Montreal, QC H3A 2T8, Canada}

\author[0000-0002-4758-9196]{T.~Hassan}
\affiliation{DESY, Platanenallee 6, 15738 Zeuthen, Germany}

\author{O.~Hervet}
\affiliation{Santa Cruz Institute for Particle Physics and Department of Physics, University of California, Santa Cruz, CA 95064, USA}

\author{J.~Holder}
\affiliation{Department of Physics and Astronomy and the Bartol Research Institute, University of Delaware, Newark, DE 19716, USA}

\author[0000-0002-7609-343X]{B.~Hona}
\affiliation{Department of Physics and Astronomy, University of Utah, Salt Lake City, UT 84112, USA}

\author{T.~B.~Humensky}
\affiliation{Physics Department, Columbia University, New York, NY 10027, USA}

\author[0000-0002-1089-1754]{W.~Jin}
\affiliation{Department of Physics and Astronomy, University of Alabama, Tuscaloosa, AL 35487, USA}

\author[0000-0002-3638-0637]{P.~Kaaret}
\affiliation{Department of Physics and Astronomy, University of Iowa, Van Allen Hall, Iowa City, IA 52242, USA}

\author{M.~Kertzman}
\affiliation{Department of Physics and Astronomy, DePauw University, Greencastle, IN 46135-0037, USA}

\author[0000-0003-4785-0101]{D.~Kieda}
\affiliation{Department of Physics and Astronomy, University of Utah, Salt Lake City, UT 84112, USA}

\author[0000-0002-4260-9186]{T.~K~Kleiner}
\affiliation{DESY, Platanenallee 6, 15738 Zeuthen, Germany}

\author{F.~Krennrich}
\affiliation{Department of Physics and Astronomy, Iowa State University, Ames, IA 50011, USA}

\author{S.~Kumar}
\affiliation{Physics Department, McGill University, Montreal, QC H3A 2T8, Canada}

\author[0000-0003-4641-4201]{M.~J.~Lang}
\affiliation{School of Physics, National University of Ireland Galway, University Road, Galway, Ireland}

\author{M.~Lundy}
\affiliation{Physics Department, McGill University, Montreal, QC H3A 2T8, Canada}

\author[0000-0001-9868-4700]{G.~Maier}
\affiliation{DESY, Platanenallee 6, 15738 Zeuthen, Germany}

\author{C.~E~McGrath}
\affiliation{School of Physics, University College Dublin, Belfield, Dublin 4, Ireland}

\author[0000-0002-1499-2667]{P.~Moriarty}
\affiliation{School of Physics, National University of Ireland Galway, University Road, Galway, Ireland}

\author[0000-0002-3223-0754]{R.~Mukherjee}
\affiliation{Department of Physics and Astronomy, Barnard College, Columbia University, NY 10027, USA}

\author[0000-0003-3343-0755]{D.~Nieto}
\affiliation{Institute of Particle and Cosmos Physics, Universidad Complutense de Madrid, 28040 Madrid, Spain}

\author[0000-0002-8321-9168]{M.~Nievas-Rosillo}
\affiliation{DESY, Platanenallee 6, 15738 Zeuthen, Germany}

\author[0000-0002-9296-2981]{S.~O'Brien}
\affiliation{Physics Department, McGill University, Montreal, QC H3A 2T8, Canada}

\author[0000-0002-4837-5253]{R.~A.~Ong}
\affiliation{Department of Physics and Astronomy, University of California, Los Angeles, CA 90095, USA}

\author[0000-0002-5955-6383]{A.~N.~Otte}
\affiliation{School of Physics and Center for Relativistic Astrophysics, Georgia Institute of Technology, 837 State Street NW, Atlanta, GA 30332-0430}

\author[0000-0002-4282-736X]{N.~Park}
\affiliation{Department of Physics, Queen's University, Kingston, ON K7L 3N6, Canada}

\author[0000-0003-3049-0808]{S.~Patel}
\affiliation{Department of Physics and Astronomy, University of Iowa, Van Allen Hall, Iowa City, IA 52242, USA}

\author[0000-0002-7990-7179]{K.~Pfrang}
\affiliation{DESY, Platanenallee 6, 15738 Zeuthen, Germany}

\author[0000-0002-9955-901X]{A.~Pichel}
\affiliation{Instituto de Astronomía y Física del Espacio (IAFE, CONICET-UBA), CC 67 - Suc. 28, (C1428ZAA) Ciudad Autónoma de Buenos Aires, Argentina}

\author[0000-0002-5680-0766]{M.~Pohl}
\affiliation{Institute of Physics and Astronomy, University of Potsdam, 14476 Potsdam-Golm, Germany and DESY, Platanenallee 6, 15738 Zeuthen, Germany}

\author{R.~R.~Prado}
\affiliation{DESY, Platanenallee 6, 15738 Zeuthen, Germany}

\author[0000-0002-4855-2694]{J.~Quinn}
\affiliation{School of Physics, University College Dublin, Belfield, Dublin 4, Ireland}

\author[0000-0002-5351-3323]{K.~Ragan}
\affiliation{Physics Department, McGill University, Montreal, QC H3A 2T8, Canada}

\author{P.~T.~Reynolds}
\affiliation{Department of Physical Sciences, Munster Technological University, Bishopstown, Cork, T12 P928, Ireland}

\author{D.~Ribeiro}
\affiliation{Physics Department, Columbia University, New York, NY 10027, USA}

\author{E.~Roache}
\affiliation{Center for Astrophysics $|$ Harvard \& Smithsonian, Cambridge, MA 02138, USA}

\author{A.~C.~Rovero}
\affiliation{Instituto de Astronomía y Física del Espacio (IAFE, CONICET-UBA), CC 67 - Suc. 28, (C1428ZAA) Ciudad Autónoma de Buenos Aires, Argentina}

\author{J.~L.~Ryan}
\affiliation{Department of Physics and Astronomy, University of California, Los Angeles, CA 90095, USA}

\author[0000-0001-7297-8217]{M.~Santander}
\affiliation{Department of Physics and Astronomy, University of Alabama, Tuscaloosa, AL 35487, USA}

\author{S.~Schlenstedt}
\affiliation{CTAO, Saupfercheckweg 1, 69117 Heidelberg, Germany}

\author{G.~H.~Sembroski}
\affiliation{Department of Physics and Astronomy, Purdue University, West Lafayette, IN 47907, USA}

\author{R.~Shang}
\affiliation{Department of Physics and Astronomy, University of California, Los Angeles, CA 90095, USA}

\author{D.~Tak}
\affiliation{DESY, Platanenallee 6, 15738 Zeuthen, Germany}

\author{V.~V.~Vassiliev}
\affiliation{Department of Physics and Astronomy, University of California, Los Angeles, CA 90095, USA}

\author{A.~Weinstein}
\affiliation{Department of Physics and Astronomy, Iowa State University, Ames, IA 50011, USA}

\author[0000-0003-2740-9714]{D.~A.~Williams}
\affiliation{Santa Cruz Institute for Particle Physics and Department of Physics, University of California, Santa Cruz, CA 95064, USA}

\author{T.~J~Williamson}
\affiliation{Department of Physics and Astronomy and the Bartol Research Institute, University of Delaware, Newark, DE 19716, USA}
\collaboration{65}{(The VERITAS Collaboration)}

%
\author[0000-0001-8307-2007]{V.~A.~Acciari}
\affiliation{Instituto de Astrof\'isica de Canarias and Dpto. de  Astrof\'isica, Universidad de La Laguna, E-38200, La Laguna, Tenerife, Spain}

\author[0000-0002-5613-7693]{S.~Ansoldi}
\affiliation{Universit\`a di Udine and INFN Trieste, I-33100 Udine, Italy}
\affiliation{also at International Center for Relativistic Astrophysics (ICRA), Rome, Italy}

\author[0000-0002-5037-9034]{L.~A.~Antonelli}
\affiliation{National Institute for Astrophysics (INAF), I-00136 Rome, Italy}

\author[0000-0001-9076-9582]{A.~Arbet Engels}
\affiliation{ETH Z\"urich, CH-8093 Z\"urich, Switzerland}

\author[0000-0002-4899-8127]{M.~Artero}
\affiliation{Institut de F\'isica d'Altes Energies (IFAE), The Barcelona Institute of Science and Technology (BIST), E-08193 Bellaterra (Barcelona), Spain}

\author[0000-0001-9064-160X]{K.~Asano}
\affiliation{Japanese MAGIC Group: Institute for Cosmic Ray Research (ICRR), The University of Tokyo, Kashiwa, 277-8582 Chiba, Japan}

\author[0000-0002-2311-4460]{D.~Baack}
\affiliation{Technische Universit\"at Dortmund, D-44221 Dortmund, Germany}

\author[0000-0002-1444-5604]{A.~Babi\'c}
\affiliation{Croatian MAGIC Group: University of Zagreb, Faculty of Electrical Engineering and Computing (FER), 10000 Zagreb, Croatia}

\author[0000-0002-1757-5826]{A.~Baquero}
\affiliation{IPARCOS Institute and EMFTEL Department, Universidad Complutense de Madrid, E-28040 Madrid, Spain}

\author[0000-0001-7909-588X]{U.~Barres de Almeida}
\affiliation{Centro Brasileiro de Pesquisas F\'isicas (CBPF), 22290-180 URCA, Rio de Janeiro (RJ), Brazil}

% ...
\collaboration{196}{(The MAGIC collaboration)}






\author{H.~Abdalla}
\affiliation{University of Namibia, Department of Physics, Private Bag 13301, Windhoek 10005, Namibia}

\author{F.~Aharonian}
\affiliation{Dublin Institute for Advanced Studies, 31 Fitzwilliam Place, Dublin 2, Ireland}
\affiliation{Max-Planck-Institut f\"ur Kernphysik, P.O. Box 103980, D 69029 Heidelberg, Germany}
\affiliation{High Energy Astrophysics Laboratory, RAU,  123 Hovsep Emin St  Yerevan 0051, Armenia}

\author{F.~Ait~Benkhali}
\affiliation{Max-Planck-Institut f\"ur Kernphysik, P.O. Box 103980, D 69029 Heidelberg, Germany}

\author{E.O.~Ang\"uner}
\affiliation{Aix Marseille Universit\'e, CNRS/IN2P3, CPPM, Marseille, France}

\author{C.~Arcaro}
\affiliation{Centre for Space Research, North-West University, Potchefstroom 2520, South Africa}

\author[0000-0002-2153-1818]{H.~Ashkar}
\affiliation{IRFU, CEA, Universit\'e Paris-Saclay, F-91191 Gif-sur-Yvette, France}

\author[0000-0002-9326-6400]{M.~Backes}
\affiliation{University of Namibia, Department of Physics, Private Bag 13301, Windhoek 10005, Namibia}
\affiliation{Centre for Space Research, North-West University, Potchefstroom 2520, South Africa}

\author[0000-0002-5085-8828]{V.~Barbosa~Martins}
\affiliation{DESY, D-15738 Zeuthen, Germany}

\author{M.~Barnard}
\affiliation{Centre for Space Research, North-West University, Potchefstroom 2520, South Africa}

\author{R.~Batzofin}
\affiliation{School of Physics, University of the Witwatersrand, 1 Jan Smuts Avenue, Braamfontein, Johannesburg, 2050 South Africa}

% ...

\collaboration{152}{(The \hess Collaboration)
}

\author{Y. Moritani}
\affiliation{Kavli Institute for the Physics and Mathematics of the Universe (WPI), The University of Tokyo Institutes for Advanced Study, The University of Tokyo, 5-1-5 Kashiwanoha, Kashiwa, Chiba 277-8583, Japan}

\author{D.F. Torres}
\affiliation{Institute of Space Sciences (ICE, CSIC), Campus UAB, Carrer de Can Magrans s/n, 08193 Barcelona, Spain; \\
 Institut d'Estudis Espacials de Catalunya (IEEC), Gran Capit\`a 2-4, 08034 Barcelona, Spain; \\
 Instituci\'o Catalana de Recerca i Estudis Avan\c cats (ICREA), 08010 Barcelona, Spain \\}

\correspondingauthor{G.~Maier}
\email{gernot.maier@desy.de}
\correspondingauthor{D.~Hadasch, A.~L\'opez-Oramas}
\email{contact.magic@mpp.mpg.de}
\correspondingauthor{D.~Malyshev, G.~Puehlhofer}
\email{contact.hess@hess-experiment.eu}



\begin{abstract}
The results of gamma-ray observations of the binary system \h collected during 450\,hours over 15 years, between 2004 and 2019, are presented. Data taken with the atmospheric Cherenkov telescopes H.E.S.S., MAGIC, and VERITAS at energies above 350\,GeV were used together with observations at X-ray energies obtained with \emph{Swift}-XRT, Chandra, \textit{XMM}-Newton, NuSTAR, and Suzaku. Some of these observations were accompanied by measurements of the H$\alpha$ emission line. 
A significant detection of the modulation of the VHE gamma-ray fluxes with a period of $316.7\pm4.4$\,days is reported, consistent with the period of $317.3\pm0.7$\,days obtained with a refined analysis of X-ray data. 
The analysis of data of four orbital cycles with dense observational coverage reveals short timescale variability, with flux-decay timescales of less than 20\,days at very high energies. 
Flux variations observed over the time scale of several years indicate orbit-to-orbit variability.
The analysis confirms the previously reported correlation of X-ray and gamma-ray emission from the system at very high significance, but can not find any correlation of optical H$\alpha$ parameters with X-ray or gamma-ray energy fluxes in  simultaneous observations. 
The key finding is that the emission of \h in the X-ray and gamma-ray energy bands is highly variable on different time scales.
The ratio of gamma-ray to X-ray flux shows the equality or even dominance of the gamma-ray energy range.
This wealth of new data is interpreted taking into account the insufficient knowledge of the ephemeris of the system, and discussed in the context of results reported on other gamma-ray binary systems.
\end{abstract}



\section{Introduction}
\label{sec:intro}

The most common definition of gamma-ray binaries is based on their composition and their energy output: a gamma-ray binary consists of a compact object orbiting a star, 
with periodic releases of large amounts of non-thermal emission at energies $>$1\,MeV \citep{2013A&ARv..21...64D}.
The gamma-ray binary source class consists of fewer than ten members, and every member shows different characteristics.
For all of them, except \hjei \citep{2020A&A...637A..23M}, it is known  that the massive star is either a Be-type star surrounded by an equatorial circumstellar disk, or an O-type star. 

In contrast, the nature of the compact object is, for many of the systems, unknown.
%
Exceptions are the gamma-ray pulsar binaries \hbin and \tpsr\replaced{, both}{, each of which consists} of a neutron star and a Be-type star. Pulsations were first discovered in the radio regime for the former one \citep{Johnston1992} and in the gamma-ray regime for the latter \citep{Abdo2009}.
Recently, a transient periodic radio signal in the direction of \lsi was claimed \citep{2021ATel14297....1W}. Together with the magnetar-like flares and the super-orbital modulation detected in \lsi, these are strong hints for a strongly magnetized neutron star in the system \citep{2002ApJ...575..427G,2012ApJ...744..106T, Ackermann:2013d}.
For the binary system  \h, which also hosts a Be star (MWC 148), the observational characteristics point towards a neutron star as the compact object, interacting with the circumstellar disk of the Be star. 
Spectroscopic measurements of H$\alpha$ emission from the Be star indicate a $\leq$ 2.5\,M$_{\odot}$ neutron star \citep{Li_2011,Moritani-2018}.
However, a microquasar scenario, in which particles are accelerated in the jets of a black hole, cannot be ruled out (see \citet{2009IJMPD..18..347B} for a review).

\h was serendipitously discovered as a point-like source at energies above 400\,GeV with the \hess experiment during a scan of the Monoceros Loop and Rosette Nebula region \citep{Aharonian-2007}. This gamma-ray source is coincident with the massive emission-line star MWC 148, which is of spectral type B0pe at a distance of 1.1 -- 1.7\,kpc
\citep{1955ApJS....2...41M,2009ApJ...690L.101H, Aragona2010}. 
Associated with this Be star, an X-ray source was detected in soft X-ray energies using \textit{XMM}-Newton observations, named XMMU J063259.3+054801 \citep{2009ApJ...690L.101H}. 

The non-detection with the VERITAS experiment \citep{Acciari-2009} provided evidence for variability of \h at gamma-ray energies.
\h was later detected again in 2011 February by MAGIC and VERITAS at VHE \citep{Aleksic-2012,Aliu-2014}. This detection took place during a high-state in X-rays measured with the Neil Gehrels \textit{Swift} Observatory X-Ray Telescope (\textit{Swift}-XRT; \citet{2011ATel.3152....1F}).
The period of $321\pm 5$\,days found in X-ray data taken with \textit{Swift}-XRT established finally the binary nature of \h \citep{Bongiorno-2011}. \cite{Aliu-2014} later confirmed the periodicity with a larger \textit{Swift}-XRT data set to be $315^{+6}_{-4}$\,days.
\h was detected 
after 9\,years of accumulating \textit{Fermi}-LAT data in the orbital phase range 0.0 -- 0.5 \citep{2017ApJ...846..169L} but, because of low statistical significance of the results, the orbital period in the HE band could not be determined.
The spectral energy distribution (SED) of \h shows two components at high energies, which is typical for gamma-ray binaries \replaced{\citep{2020A&A...637A..23M}}{\citep{2013A&ARv..21...64D}}.

In the radio band, observations were conducted with the Giant Metrewave Radio Telescope (GMRT; 5\,GHz) and the Very Large Array  \citep[VLA; 1280\,GHz;][]{Skilton2009}.  They revealed a point-like  radio  source  at  the  position  of  MWC 148,  at both radio wavelengths, and variability on time scales of roughly one month in the 5\,GHz emission. Extended and variable radio emission at 50 -- 100\,AU scales at 1.6\,GHz was discovered by \cite{Moldon2011} using the European  VLBI  Network  (EVN). They monitored the source over the 2011 January -- February X-ray outburst, 
detecting a shift of the peak of the emission by 21\,AU over 30\,days.
Similar behavior has been found in other gamma-ray binaries, such as \hbin \citep{2011ApJ...732L..10M}, \lsf \citep{2011ASSP...21....1M}, or \lsi \citep{2001A&A...376..217M}, which also show a radio morphology with a central core and one-sided extended emission of a few AU length.


\begin{figure}
\plotone{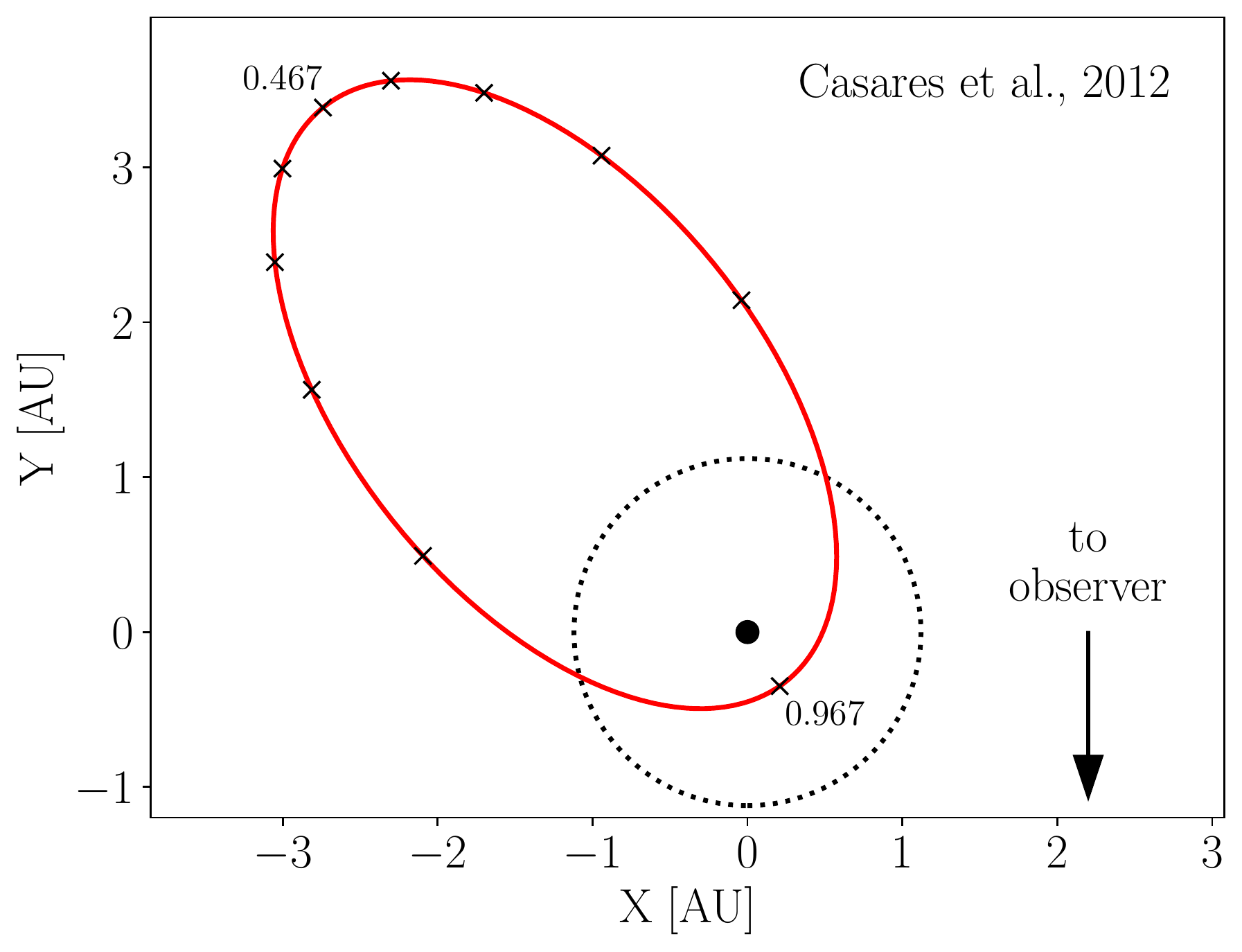}
\plotone{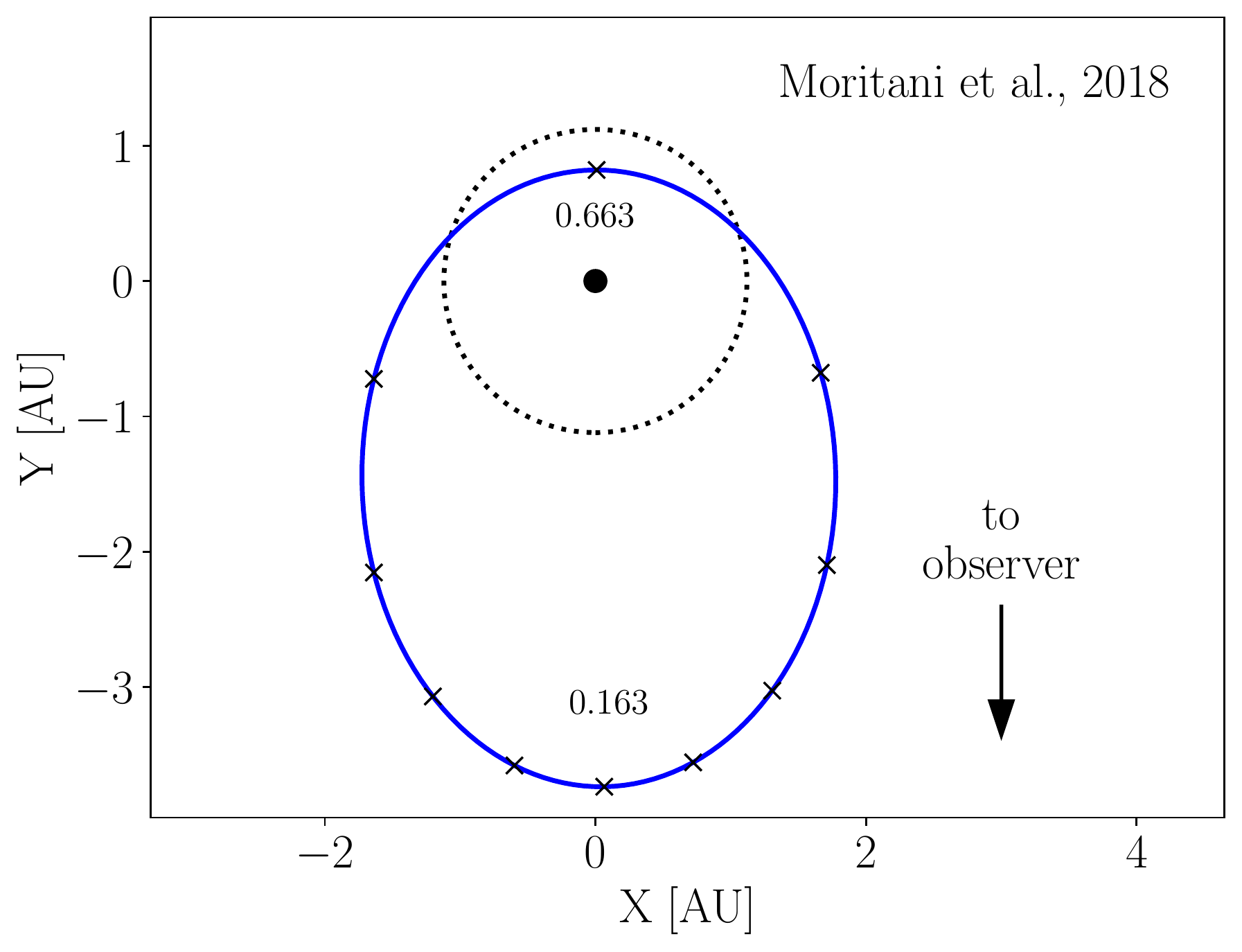}
\caption{\label{fig:OrbitalSolutions}
Illustration of the two suggested orbital solutions for \h \added{as seen from above the orbital plane}.
The very large uncertainties reported for each solution are not shown.
Top: Orbital solution derived from measurements of photospheric lines of the Be star \citep{Casares-2012}.
Bottom: Orbital solution derived from measurements of the radial velocity of the H$\alpha$ line \citep[][Table 1, for $P_{orb}=313$ days]{Moritani-2018}.
The massive star is indicated by the black marker.
Crosses mark intervals of 0.1 in orbital phase, numbers indicate values for apastron and periastron orbital phase values.
The dashed line illustrates roughly the size of the disk of the Be star as derived from  
H$\alpha$ measurements \citep{2015ApJ...804L..32M, 2016A&A...593A..97Z}.
}
\end{figure}

The actual orbital geometry including eccentricity, position of apastron and periastron, and inclination of the \h binary system is essentially unknown.
Two different orbital solutions derived from optical measurements exist with large uncertainties quoted for all relevant parameters.
They are derived through different methods and are based on different data sets (see Figure \ref{fig:OrbitalSolutions}).
They follow the same definition of phase $\phi=0$ as MJD=54857.0, 
arbitrarily set to the date of the first \textit{Swift}-XRT observations \citep{Bongiorno-2011}.
One solution comes from \cite{Casares-2012} using the photospheric absorption lines (mainly {He}~{\scshape{i}}) of data taken 2008-2011, and taking the centroid velocity of the H$\alpha$ emission line to check for the orbital period of the binary (which was fixed to 321\,days). 
The other solution is from \cite{Moritani-2018} based on the bisector velocity of the \added{H$\alpha$} wing, regarding the wavelength where the wing intensity is the same between blue and red side as the center \citep{1986ApJ...308..765S} of the H$\alpha$ emission line from data taken 2013-2017. The main differences between the two solutions lie in the eccentricity of the orbit and in the orbital phase of periastron passage: $\epsilon\simeq$0.8 vs. $\epsilon \simeq$0.6 and $\phi$=0.967 vs. $\phi$=0.663 \citep{Casares-2012, Moritani-2018}, respectively. 
The second significant difference is the prediction on the mass of the compact object as a function of inclination of the system.
The orbital solution provided by \cite{Casares-2012} sets an inclination limit of $>60^\circ$ for masses of the compact object consistent with a non-accreting pulsar. 
The differences may be due to the different methods applied (absorption vs.~emission spectra, folding period fixed vs.~Fourier analysis, epoch of observations, etc.) or due to the sparse dataset used by \cite{Casares-2012}, in which the observations are slightly clustered around periastron.

The Be star in the \h binary, MWC 148, together with the properties of its disk, has been studied through optical spectroscopic observations by \cite{2015ApJ...804L..32M} and \cite{2016A&A...593A..97Z, 2021arXiv210201971Z}. Applying the geometry derived by \cite{Casares-2012}, \cite{2016A&A...593A..97Z} suggest that the compact object passes deeply into the disk during the periastron passage and that the circumstellar disk  is  likely truncated  by  the  orbiting  compact  object. 
In contrast, \cite{2015ApJ...804L..32M} conclude from the non-detection of variations in Fe \RN{2} emission lines that the interaction between the compact object and the Be star \added{disk} occurs at a distance of two to seven stellar radii from the Be star.
Furthermore, they deduced that the detected short-term ($\leq$50 days) disk variability is caused by the interaction of the pulsar wind with the disk, rather than by tidal forces.

Optical polarization studies by \cite{2017MNRAS.464.4325Y} revealed variation in both polarization degree and polarization angle. This finding can be explained by perturbations of the circumstellar material close to periastron passage, applying the orbital solution by \cite{Casares-2012}, or before and after apastron passage, following the solution of \cite{Moritani-2018}. However, an additional \replaced{source of polarization}{source of polarized emission}, such as a jet, might also explain the changes in polarization \citep{2017MNRAS.464.4325Y}.

\cite{2011ApJ...737L..12R} performed an X-ray timing and spectral analysis of \h during its high and low states to obtain better insight into the nature of the compact object. No periodic or quasi--periodic signal in any of the emission states was found. 
Their calculated upper limit on the amplitude of a sinusoidal signal - defined as X-ray pulsed fraction of \h - is comparable to limits found for the well-studied binaries \lsi and \lsf. 
\added{Recently, however, \cite{ls5039_pulsar} claim to have detected a period of $\sim$9 s in \lsf. 
We caution that the significance after trials is not large for the different observations, 
that the period detected during different observations varies and
requires a high and positive $\dot P$, 
that the orbital corrections do not increase the pulse significance as expected, and that after
background subtraction two data sets give rather different pulse fractions. 
A recent paper by \citet{2021arXiv210304403V} 
finds that the statistical significance of this feature is too low to claim it as a real detection.}

\deleted{However}
Comparing the spectral evolution of \h with the \replaced{the latter TeV binaries reveals differences}{TeV binaries \lsi and \lsf, some differences are revealed}. Specifically, \h shows a steeper X-ray spectrum when the flux is higher, whereas the other two binaries show the opposite behaviour. A comparison of the X-ray spectra during the different orbital phases of \h will be discussed later in this paper.

The most recent study of simultaneous observations at hard X-rays with NuSTAR and in TeV gamma-rays with VERITAS was performed by \cite{2020ApJ...888..115A}. The data were taken during a period of rising flux.
The authors applied a leptonic model, assuming the compact object to be a pulsar, and \replaced{found}{claim} that, under this \added{simple} assumption,  \h follows the same trend as other binaries: the magnetization of the wind decreases with greater distance from the compact object. \replaced{This}{According to \cite{2020ApJ...888..115A} this} finding strengthens the pulsar scenario for \h.

This paper presents results from 15 years of gamma-ray observations, several years of X-ray observations, and optical observation of \h during different orbital phases. 
The paper is structured as follows. In section \ref{sec:observations} the instruments used for the studies are described. The results are then presented in section \ref{sec:results}. There the details of the overall light curve, the phase-folded light curves, and the spectra measured during different phases are shown. In section \ref{sec:correlations} the  correlation studies between measurements in the optical (H$\alpha$ observations), X-ray, and VHE are shown. Paragraph \ref{sec:outburstJan2018} presents the data of the unusual bright outburst from 2018 January. The results are discussed in section \ref{sec:discussion} and summarized with final remarks in section \ref{sec:summary}.

\section{Observations}
\label{sec:observations}


\subsection{\hess Gamma-Ray Observations}
The \hess (High Energy Stereoscopic System) experiment is an array of five imaging atmospheric Cherenkov telescopes located in the southern hemisphere in the Khomas Highland of Namibia, ($23^{\circ}16^{\prime}$~S, $16^{\circ}30^{\prime}$~E), at an elevation of 1800\,m above sea level.
Four telescopes of the array (CT~1--4) have 12-meter diameter mirror dishes. A larger, 28-meter diameter telescope (CT5) was deployed at the site in July 2012 during the \hess phase-II upgrade.

The CT~1--4 telescopes have been in operation since 2004~\citep{hinton2004} and have a $5^\circ$-diameter field of view (FoV). 
CT~5 is equipped with a camera with a $3.5^\circ$ diameter FoV.
Using only CT~1--4, as it is done in this work, \hess is sensitive  to gamma-rays with energies from $\sim0.2$~TeV  up to tens of TeV for observations at zenith.  

The \hess observations of \h cover almost 14~years (2004 March -- 2018 February), totalling 120\,h of data after quality cuts, dead-time corrected and taken under dark sky conditions (Table~\ref{table:ObservationSummary}). Analysis of $73$\,h of data obtained prior to 2012 February  (MJD~55951) was reported in~\citet{Aliu-2014} and these data are re-analyzed in this work, applying improved calibration and reconstruction methods for this paper. All available \hess data were processed with the \textit{Image Pixel-wise fit for Atmospheric Cherenkov Telescope} (ImPACT) analysis \citep[][]{Parsons2014} with \texttt{std\_ImPACT} cut configuration, which requires a minimum of 60 photo-electrons per image. 

The signal was extracted from a circular region of radius
\replaced{$0.071^\circ$)}{$0.07^\circ$} centered on the position of \h.
For the  background  estimation,  the  reflected  background  method~\citep{Fomin94,2007A&A...466.1219B},  which ensures that the regions used for signal (ON) and background (OFF) extraction have the same acceptance in the FoV of the camera, was applied. A cross-check of the main analysis was performed using the Model++  analysis \citep{2009APh....32..231D}, yielding compatible results.

\subsection{MAGIC Gamma-Ray Observations}

The MAGIC (Major Atmospheric Gamma-Ray Imaging Cherenkov Telescopes) experiment is a stereoscopic system of two 17\,m-diameter imaging atmospheric Cherenkov telescopes. It is located on the Canary Island of La Palma, Spain, at the observatory El Roque de Los Muchachos ($28^{\circ}45^{\prime}$~N, $17^{\circ}53^{\prime}$~W, 2200 m above the sea level). It is equipped with pixelized cameras covering a field of view with a diameter of $\sim3.5^{\circ}$. The MAGIC telescopes reach a trigger threshold down to 50\,GeV and a sensitivity of $\sim0.6\%$ of the Crab Nebula flux above 250\,GeV in 50\,hours of observations at zenith angles $<$30$^\circ$ \citep{Aleksic-2016}.

MAGIC observed \h from 2010 October until 2017 November for a total of 68\,hours (Table \ref{table:ObservationSummary}). 
The dataset taken during an X-ray outburst occurring from 2010 October to 2011 March was published in~\citet{Aleksic-2012}. 
The most recent data cover 2015 December to 2017 November. After quality cuts and correction for dead-time, 57.3\,hours of good data remain. These observations were taken both under dark (30.9\,hours) and under moderate-to-strong moonlight conditions (26.4 hours). They were carried out in wobble mode, pointing at two regions symmetrically $0.4^{\circ}$ away from the source position to simultaneously evaluate the background \citep{Fomin94}. The data have been analyzed using the standard MAGIC procedure \citep{Zanin2013}. The recorded images were calibrated, cleaned, and parameterized \citep{Hillas85,Albert2008a}, while applying different cleaning levels and size cuts for the moderate-to-strong moon data according to \cite{Ahnen2017Moon}. The background rejection and the estimation of the gamma-ray arrival direction was performed using the random forest method \citep{magic:RF}. 

\subsection{VERITAS Gamma-Ray Observations}

The VERITAS (Very Energetic Radiation Imaging Telescope Array System) telescope array is composed of four imaging atmospheric Cherenkov telescopes located at the Fred Lawrence Whipple Observatory in southern Arizona (1268\,m a.s.l., 31$^\circ$40'N, 110$^\circ$57'W; \cite{Weekes-2002}).
Each telescope has a 12-m diameter segmented mirror of $\approx$100\,m$^2$ area.  
Cherenkov light emitted in extensive air showers is reflected onto photomultiplier cameras  covering a field of view of 3.5$^\circ$ diameter.

The VERITAS observations of \h \ (VER J0633+057) have been obtained over  11\,years, from 2008 December to 2019 January.
For a detailed discussion of the evolution of the performance of the VERITAS instrument with time, see \cite{Park-2015}.

VERITAS observed \h \ for a total dead-time corrected exposure time of 260 hours, after the application of standard data selection cuts to account for non-optimal weather or hardware conditions (see Table \ref{table:ObservationSummary} for details).
\h was observed at an average elevation angle of 60$^\circ$ and at a fixed offset of $0.5^\circ$ from the center of the camera field of view. 
Observations were taken typically in dark sky conditions, with the exception of a small fraction ($\sim 10$\,hours) of the dataset which was obtained during 30\%--65\% moon illumination.
The latter results in a higher energy threshold and reduced sensitivity compared to observations taken under nominal sky conditions  \citep{Archambault-2017}.

Data were analysed using VERITAS standard analysis software tools utilising boosted decision trees for gamma-hadron separation based on image parameters \citep[see e.g.][]{Krause:2017, Maier-2017}.
The source region was defined as a circle with radius $0.09\degr$ around the nominal position of the X-ray source XMMU J063259.3+054801.
Background event rates are estimated using the reflected background method \citep{Fomin94}, assuming typically six background regions located symmetrically to the signal region around the camera centre.
Light-intensity calibration factors obtained from regular monitoring of the optical throughput and of the detector performance are applied to take time-dependent changes in the instrument response into account \citep{Nievas-2021}.
The observations up to March 2017 (MJD~57822) have been presented before \citep{Aliu-2014, Maier-2017b}.
These data have been re-analysed applying improved calibration and background suppression methods.
 
%
%
%
%
\begin{deluxetable*}{lccccc}[!htb]
\centering
\tablecolumns{6}
\tablecaption{
Summary of gamma-ray observations of HESS J0632+057.
\label{table:ObservationSummary}}
\tablehead{
\colhead{Observatory} &
\colhead{Observation} &
\colhead{Range of} &
\colhead{Range of } &
\colhead{Observation } &
\colhead{Observation }  \\
\colhead{} &
\colhead{Type\tablenotemark{a}} &
\colhead{Elevations} & 
\colhead{Energy Thresholds} &
\colhead{Time\tablenotemark{c}} &
\colhead{Range} \\
\colhead{} &
\colhead{} &
\colhead{(deg)} & 
\colhead{(GeV)\tablenotemark{b}} &
\colhead{(min)} &
\colhead{(years)}}
\startdata
VERITAS & nominal HV & 43 -- 64 & 160 -- 630 & 14995 & 2008 -- 2019 \\
VERITAS  & reduced HV & 53 -- 64 & 420 -- 630 &  630 & 2012 -- 2018 \\
\hline
\hess & CT1--4 & 32 -- 62 & 260 -- 680  &  7200 & 2004 -- 2018\\
\hline 
MAGIC & Stereo &  38 -- 67  & 147 -- 251  &  4080 & 2010 -- 2017 \\
\enddata
\tablenotetext{a}{VERITAS Observations under bright moonlight conditions require different PMT camera settings and are labeled with reduced high voltage (HV).
}
\tablenotetext{b}{The energy threshold is defined as the lowest energy for events to be used in the analysis.
Note that these definitions vary between the observatories, as different spectral reconstruction methods are applied.
}
\tablenotetext{c}{Observation time after quality cuts and dead-time correction.}
\end{deluxetable*}

\begin{deluxetable}{lcccc}[!htb]
\centering
\tablecolumns{3}
\tablecaption{
Summary of H$\alpha$ profile measurements of the Be star MWC 148 in \h using HIDES.
\label{table:ObservationSummary_optical}}
\tablehead{
\colhead{Observation} &
\colhead{Number of } &
\colhead{Total} \\
\colhead{Time Range} &
\colhead{Pointings} & 
\colhead{Exposure Time} \\
\colhead{(YYYYMMDD)} &
\colhead{} &
\colhead{(min)}}
\startdata
20131031 -- 20140410  & 14 & 2809 \\
20141104 -- 20150402  & 17 & 2350 \\
20160107 -- 20160330  & 6 &  1020 \\
20161206 -- 20161223  & 2 & 443 \\
20170224 -- 20170413 & 7 & 1210 \\
\enddata
\end{deluxetable}


\subsection{X-ray Observations}

This study includes data obtained with the \emph{Swift}-XRT, \textit{XMM}-Newton, Chandra and NuSTAR instruments. All data presented in \cite{2019AN....340..465M} were used, adding also \emph{Swift}-XRT and Chandra data obtained after the publication of the aforementioned paper. Here, only analysis details for these additional data are provided; for a description of the analysis of NuSTAR and Suzaku data, see \added{section 2.4 and 2.5 of} \cite{2019AN....340..465M}.

The \emph{Swift}-XRT data taken between Dec. 05 and 14, 2018, were analysed with the \texttt{heasoft v.6.22} software package and reprocessed with \texttt{xrtpipeline v.0.13.4}, as suggested by the \emph{Swift}-XRT team\footnote{See the \emph{Swift}-XRT User's Guide (\url{https://swift.gsfc.nasa.gov/analysis/xrt_swguide_v1_2.pdf})}. 
The spectra were extracted with \texttt{xselect},
using a $36''$ circle for source counts and an annulus centered at the source position with inner/outer radii of $60''$/$300''$ for the background. 
Chandra data were analyzed using \texttt{CIAO v.4.9} software and CALDB 4.7.6. The data were reprocessed with the \texttt{chandra\_repro} utility, source and background spectra with corresponding redistribution matrix (RMF) and ancillary response files (ARF) were extracted from circular regions of radii $11''$ (source) and $50''$(background) with the \texttt{specextract} tool.
Two Chandra observations (obs.~IDs 20974 and 20975; total exposure $\sim 70$~\,ksec) were performed over 2018 February 19-21, about two weeks after the last Chandra observation presented in \cite{2019AN....340..465M}. 

For each analyzed X-ray energy spectrum, neighboring energy bins were
merged until each bin contained at least one count. Obtained spectra were then fitted in the 0.3-10~\,keV energy range. 
The reported fluxes were extracted from the fit of the spectra with an absorbed power-law model (\texttt{cflux*phabs*po}) with \texttt{XSPEC v.12.9.1m}. 
The fit was performed with Cash-statistics, suitable for the analysis of low-statistics data~\citep{cash:79}.


\subsection{Optical H\texorpdfstring{$\alpha$}\ Observations} 
\label{sec:optical}

High-dispersion H$\alpha$ profiles of the Be star in \h were obtained  using the HIDES (High-Dispersion Echelle Spectrograph), a fiber-fed system \citep{Kambe-2013} deployed on a 188\,cm telescope at Okayama Astrophysical Observatory (OAO)\footnote{The current Subaru Telescope Okayama Branch Office.}. Furthermore, spectra using ESPaDOnS (Echelle SpectroPolarimetric Device for the Observation of Stars) at the Canada -- France -- Hawaii Telescope \citep{Manset-2003} from 2013 October 31 to 2017 April 13 were obtained, see Table~\ref{table:ObservationSummary_optical}.
The resolving power, $R$, of HIDES and ESPaDOnS is $\sim$50000 and $\sim$68000, respectively.


The observations and the data reduction steps are described in detail in \cite{Moritani-2018}; a short summary is provided here.
The HIDES data were reduced using IRAF. 
For the ESPaDOnS data were reduced  using the Libre-Esprit/Upena pipeline, provided by the instrument team. The continuum level around the H$\alpha$ line was re-fitted and the spectrum was normalized in order to compare with the HIDES data.

The radial velocity was measured as the bi-sector velocity of the wing of the H$\alpha$ profile \citep{Moritani-2018} and the full width half maximum (FWHM) was determined by fitting the middle and bottom of the profiles with a single Gaussian function.
The equivalent width (EW) was estimated by simply integrating the profiles. The observations are summarized in Table~\ref{table:ObservationSummary_optical}.

%
%
%

\section{Results}
\label{sec:results}

The gamma-ray observations of \h with \hess, MAGIC, and VERITAS amount to a large, $\sim 450$\,h dataset  covering 15\,years, from 2004 to 2019 (see Table \ref{table:ObservationSummary}).
The aggregated detection significance over the entire dataset, calculated at the position of the X-ray source XMMU J063259.3+054801, is 17.9\,$\sigma$ for \hess, 8.9\,$\sigma$ for MAGIC and 27.0\,$\sigma$ for VERITAS.
Statistical significances are calculated using equation 17 from \citet{Li:1983}.
Systematic uncertainties for gamma-ray measurements originate mostly from variations in the atmospheric throughput to Cherenkov photons, mirror reflectivity, camera response, and errors in the background estimations as well as in the signal-selection cut efficiencies.
The systematic uncertainty is estimated to be \replaced{11-20\%}{10-20\%} on the gamma-ray flux and 0.10-0.15 on the spectral index \citep{Aleksic-2016,hess_syst_crab}. 
Gamma-ray observations are grouped in intervals of approximately 10\% of the orbital period due to the generally weak detection significance of individual observations.
Shorter periods are used for the flux calculations during high states.

Integral gamma-ray fluxes are calculated throughout this paper above an energy of 350\,GeV adopting a gamma-ray spectral index of $\Gamma$=2.6, which is representative for the data sample (see Table \ref{table:EnergySpectrum} and Section \ref{sec:spectra}).

%
%
%
%
\begin{figure*}
\centering
\includegraphics[width=0.75\textwidth]{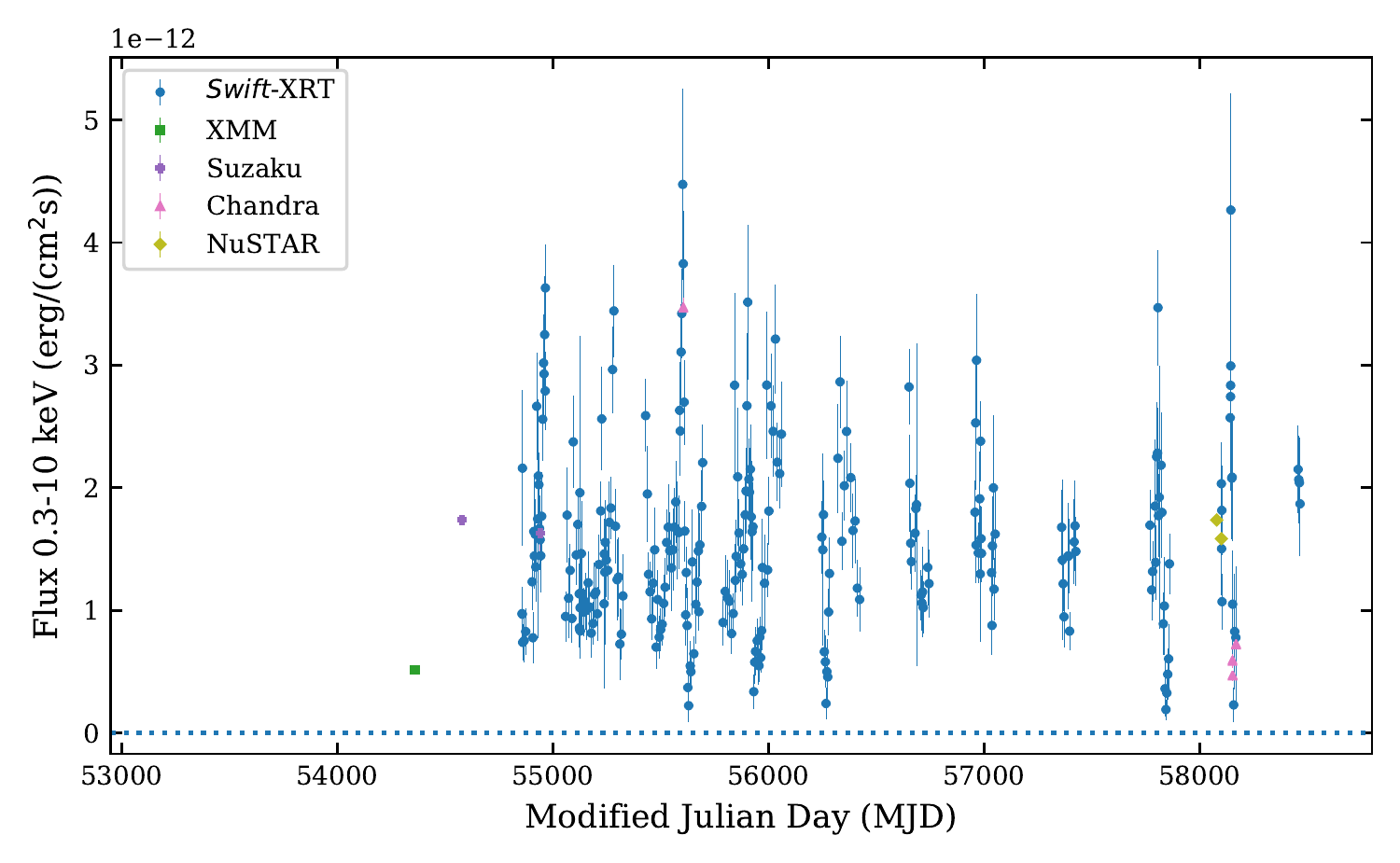}
\includegraphics[width=0.75\textwidth]{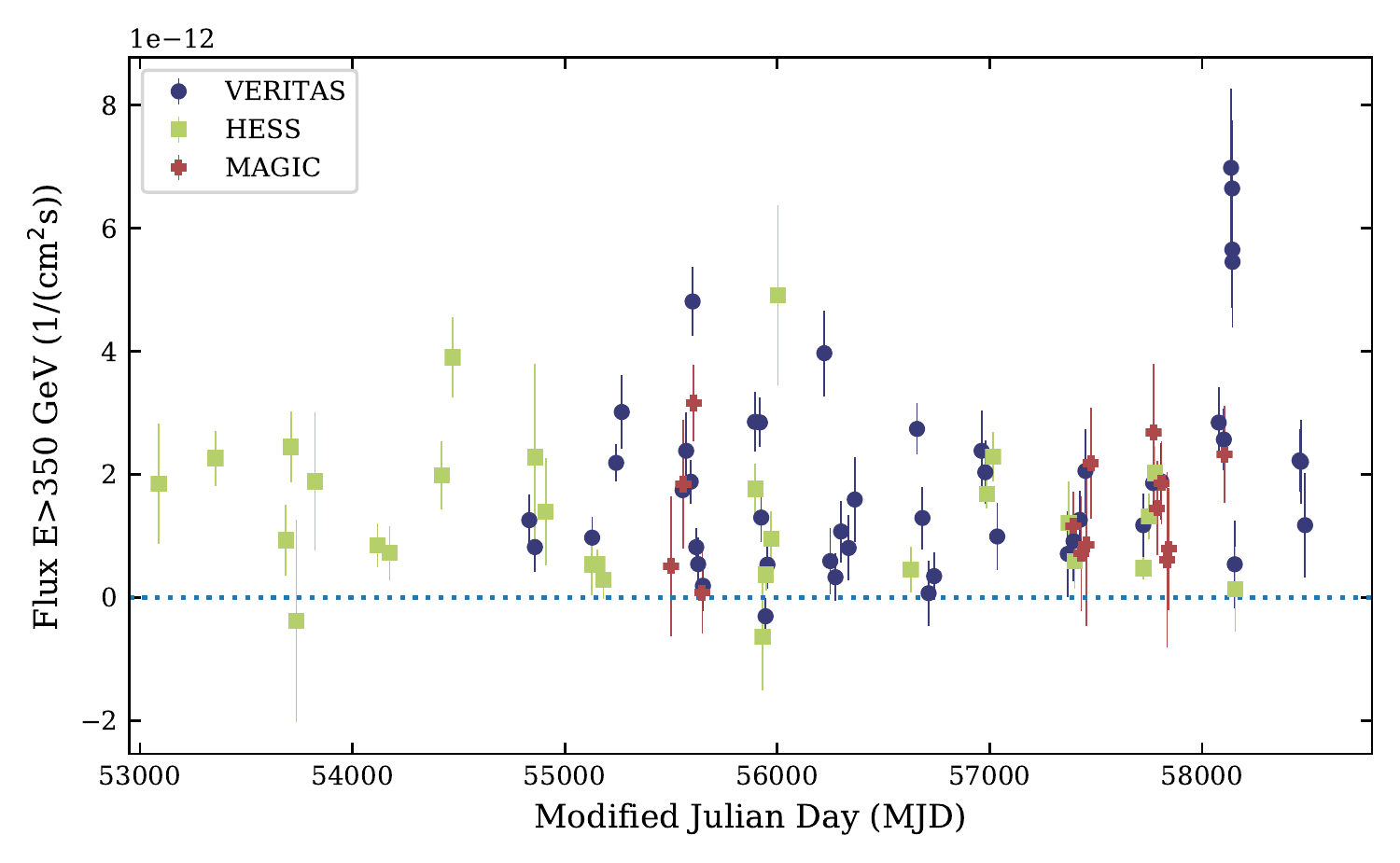}
\caption{\label{fig:light curves}
Top: Long-term observations of \h \ in X-rays  (0.3--10\,keV) with \emph{Swift}-XRT, \emph{XMM-Newton}, \emph{Suzaku}, \emph{Chandra} and {NuSTAR}.
Bottom: Long-term observations of \h \ in gamma rays ($>$350 GeV) with H.E.S.S., MAGIC, and VERITAS.
Vertical lines indicate 68\% statistical uncertainties; note that these are smaller than the marker size for all X-ray instruments except \emph{Swift}-XRT.
}
\end{figure*}

\subsection{Light-curve modulation and orbital period determination}
\label{sec:period}

The long-term light curves (Figure \ref{fig:light curves}) confirm that \h is highly variable at X-ray and gamma-ray energies (see also \cite{Acciari-2009, Bongiorno-2011, Aliu-2014}).
Above 350\,GeV, the flux varies between 5 and 10\% with respect to the flux of the Crab Nebula (above the same energy) dropping below the detection limits of the different ground-based instruments for some periods.

The large datasets allow us to search for periodic variability patterns on different timescales. The period was first determined  to be $321\pm5$\,days by \cite{Bongiorno-2011} using approximately two years of \emph{Swift}-XRT measurements (MJD 54857 to 55647). 
Subsequent updates to this analysis, based on extended datasets from \emph{Swift}-XRT, yielded comparable results of $P_{orb} = 313-321$ days \citep{Aliu-2014, Moritani-2018, 2019AN....340..465M}. Flux modulations of the X-ray light curve are interpreted in the literature as due to the orbital period of the binary system.
The only non-X-ray measurement of the orbital period, based on optical observations of H$\alpha$ radial velocity profiles, yielded $P_{orb} = 308^{+26}_{-23}$\,days \citep{Moritani-2018}, again compatible with previous results (see Table \ref{tab:appendix:orbitalPeriod} for a summary of the orbital period measurements). Although \cite{Casares-2012} did not estimate the orbital period, they showed that the main parameters of the H$\alpha$ emission line (equivalent width, full width at half maximum and centroid velocity) are modulated within the 321-day X-ray period and calculated an orbital solution of the binary. 

The analysis of the most recent available  \emph{Swift}-XRT data set including the re-analysis of older data (264 data points for MJD 54857--58466; see  Figure \ref{fig:light curves} (top)), considered in this study using a method based on Pearson's correlation coefficient (PCC method; see Appendix \ref{appendix:orbitalPeriod}), results in an orbital period of $P_{orb} = 317.3\pm 0.7 (stat) \pm 1.5(sys)$\,days (see Table \ref{tab:appendix:orbitalPeriod}).
Systematic uncertainties are derived from the application of the different methods for the period derivation applied to 1000 Monte Carlo-generated data sets. 
The description of the application of additional methods, such as discrete correlation functions (DCF) or phase dispersion minimisation (PDM), together with details of the period analysis and derivation of the associated systematic uncertainties can be found in Appendix \ref{appendix:orbitalPeriod}.
Periodic signals on timescales different from the orbital period have been observed in high-mass X-ray binaries, e.g., the super-orbital period of 1667\,days detected in radio, optical, X-ray and gamma-ray observations of the binary \lsi \citep{2002ApJ...575..427G,Li_2011,Ackermann:2013d,2015A&A...575L...6P,2016A&A...591A..76A,2020A&A...643A.170K}.
A signal at periods other than the  orbital period has not been found for periods between 50 and 1000 days in an analysis of the X-ray observations of \h using the PCC method (see Appendix \ref{appendix:orbitalPeriod} and Figure \ref{fig:orbitalAnalysis} d).
The analysis of MC-generated light curves with similar time structures shows that the PCC periodogram obtained from the analysis of the measured light curve is consistent with expectations from statistical fluctuations.


The large gamma-ray dataset, spanning almost 15 years of gamma-ray observations (Figure \ref{fig:light curves} bottom), \replaced{allows for the first time the orbital period to be determined from gamma-ray data only}{allows the orbital period of this system to be determined from gamma-ray data alone for the first time}.
The analysis approach is equivalent to the X-ray analysis discussed above (PCC method), resulting in an orbital period $P_{\gamma}=316.7\pm4.4 (stat)\pm2.5(sys)$\,days, consistent with the value obtained at X-ray energies (see again Appendix \ref{appendix:orbitalPeriod} for details on the analysis).
Given the known strong correlation of X-ray and \added{VHE} gamma-ray fluxes observed in the past (\cite{Aliu-2014}, see also subsection \ref{sec:correlations}), this result is not unexpected.

%
\begin{deluxetable*}{lcccc}[!htb]
\centering
\tablecolumns{5}
\tablecaption{
Orbital period derivation results from literature in comparison with 
the outcome of this work.
The applied methods are:
Peak fitting \citep{Bongiorno-2011},
Z-transformed discrete correlation functions \citep[ZDCF]{Alexander-1997},
phase dispersion minimisation method \citep[PDM]{Stellingwerf-1978},
discrete correlation functions \citep[DCF]{Edelson-1988}, and
correlation analysis comparing the light curves with a binned-average light curve \citep[PCC]{2019AN....340..465M}.
Errors indicate $1\sigma$ statistical uncertainties.
\label{tab:appendix:orbitalPeriod}}
\tablehead{
\colhead{Publication} &
\colhead{Energy Range} &
\colhead{MJD Range} &
\colhead{Method} &
\colhead{Orbital period} \\
\colhead{} &
\colhead{} &
\colhead{} &
\colhead{} &
\colhead{(days)} 
}
\startdata
%
\cite{Bongiorno-2011} & \emph{Swift}-XRT & 54857--55647 & Peak fitting & $321\pm5$ \\
\cite{Aliu-2014}  & \emph{Swift}-XRT & 54857--55972 & ZDCF & $315^{+6}_{-4}$ \\
\cite{Moritani-2018} &  \emph{Swift}-XRT & 54857--57052 & ZDCF & $313^{+11}_{-8}$  \\
\cite{Moritani-2018} &  H$\alpha$ & 56597--57856 & Fourier Analysis & $308^{+26}_{-23}$ \\
\cite{2019AN....340..465M} & \emph{Swift}-XRT & 54857--57860 & PCC & $316.2^{+1.8}_{-2.0}$ \\
%
\hline
%
%
This work & X-ray & 54857--58466 & PCC & $317.3\pm0.7(stat)\pm1.5(sys)$ \\
This work & X-ray & 54857--58466 & PDM & $316.7\pm1.5(stat)\pm1.5(sys)$ \\
This work & X-ray & 54857--58466 & DCF & $318.9\pm2.2(stat)\pm1.5(sys)$\\
This work & Gamma ray & 53087--58490 & PCC & $316.7\pm 4.4(stat) \pm 2.5(sys)$ \\
This work & Gamma ray & 53087--58490 & PDM & $319.0\pm 3.4(stat) \pm 2.5(sys)$\\
%
\if\fordiscussion2
\textbf{for discussion VTS/HESS} & & & \\
This work & VERITAS & 54830--58490 & PDM & $319.0\pm 6.6$\\
This work & VERITAS & 54830--58490 & PCC & $316.4\pm 6.4$ \\
This work & H.E.S.S. & 53087--58143 & PDM & $317.4\pm 14.8$\\
This work & H.E.S.S. & 53087--58143 & PCC & $319.8\pm 17.6$ \\
\fi
\enddata

\end{deluxetable*}

Table \ref{tab:appendix:orbitalPeriod} summarises the outcome of the different orbital period analyses in comparison with results from the literature.
For period-folding, in this paper the result from the PCC method was used. It was applied to the \emph{Swift}-XRT data with the following values: 
\mbox{MJD$_0$ = 54857} (arbitrarily set to the date of the first observations of the source with \textit{Swift}) and period \mbox{$P = 317.3\pm0.7_{stat}\pm1.5_{sys}$\,days}.
It should be stressed that none of the conclusions on the physical properties of the binary system depend on this particular choice (see also Appendix \ref{appendix:phaseFolding} and Figures \ref{fig:lc-phaseFolded-uncertaintiesX}, \ref{fig:lc-phaseFolded-uncertaintiesG}).

\subsection{Phase-folded light curves}

\begin{figure}
\plotone{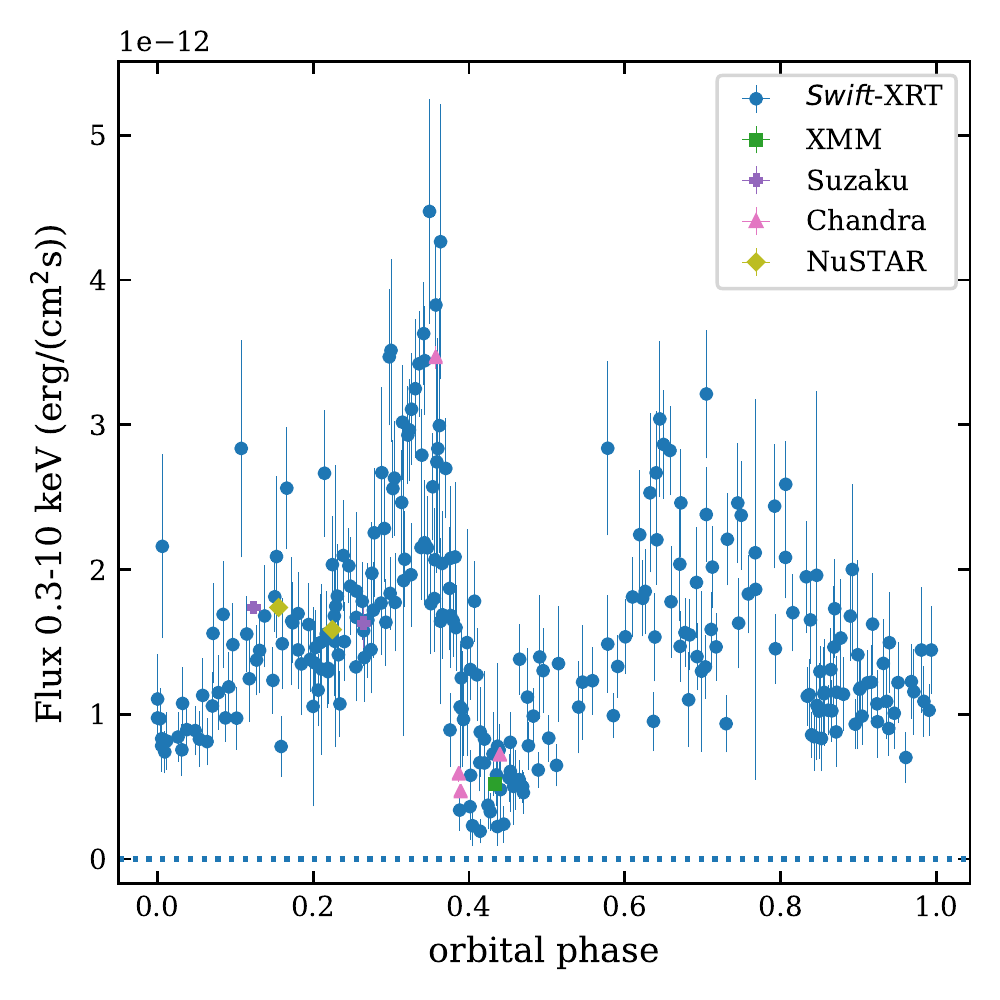}
\plotone{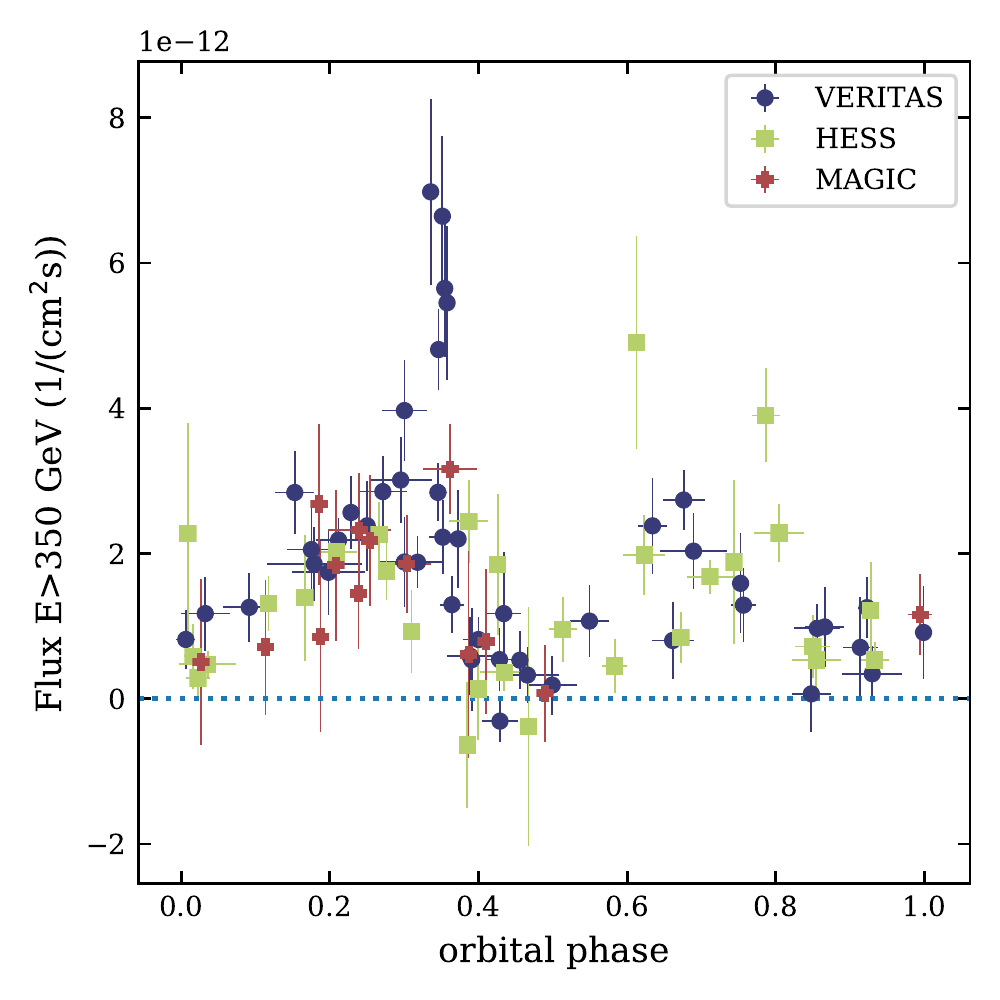}
\caption{\label{fig:lc-phaseFolded}
X-ray  (0.3--10\,keV; top) and gamma-ray ($>$350\,GeV; bottom) light curves as a function of the orbital phase, assuming an orbital period of 317.3\,days (MJD$_0$=54857.0; \citet{Bongiorno-2011}).
Vertical error bars indicate statistical uncertainties; note that these are smaller than the marker size for all X-ray instruments except for \emph{Swift}-XRT.
Fluxes are averaged over time intervals indicated by the horizontal lines (for most flux points these are smaller than the marker size).
}
\end{figure}



\begin{figure*}
\gridline{
\fig{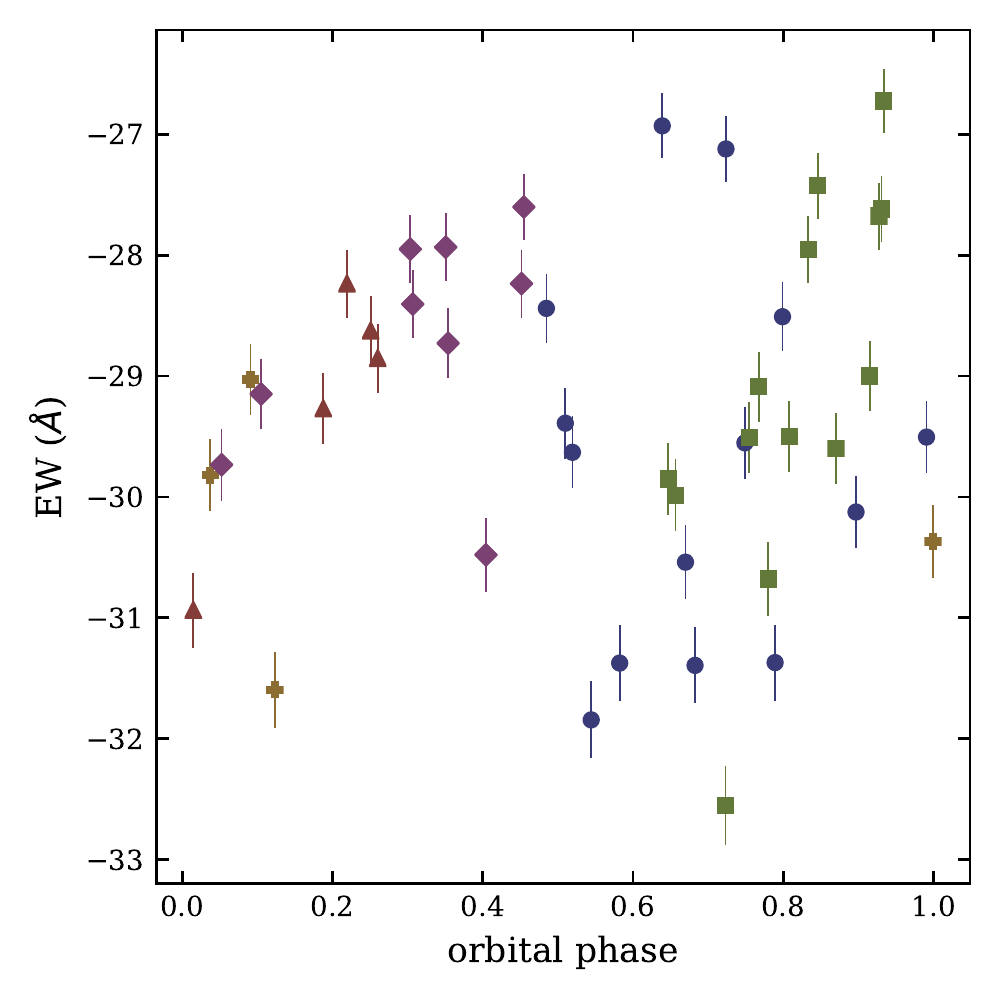}{0.33\textwidth}{(a)}
\fig{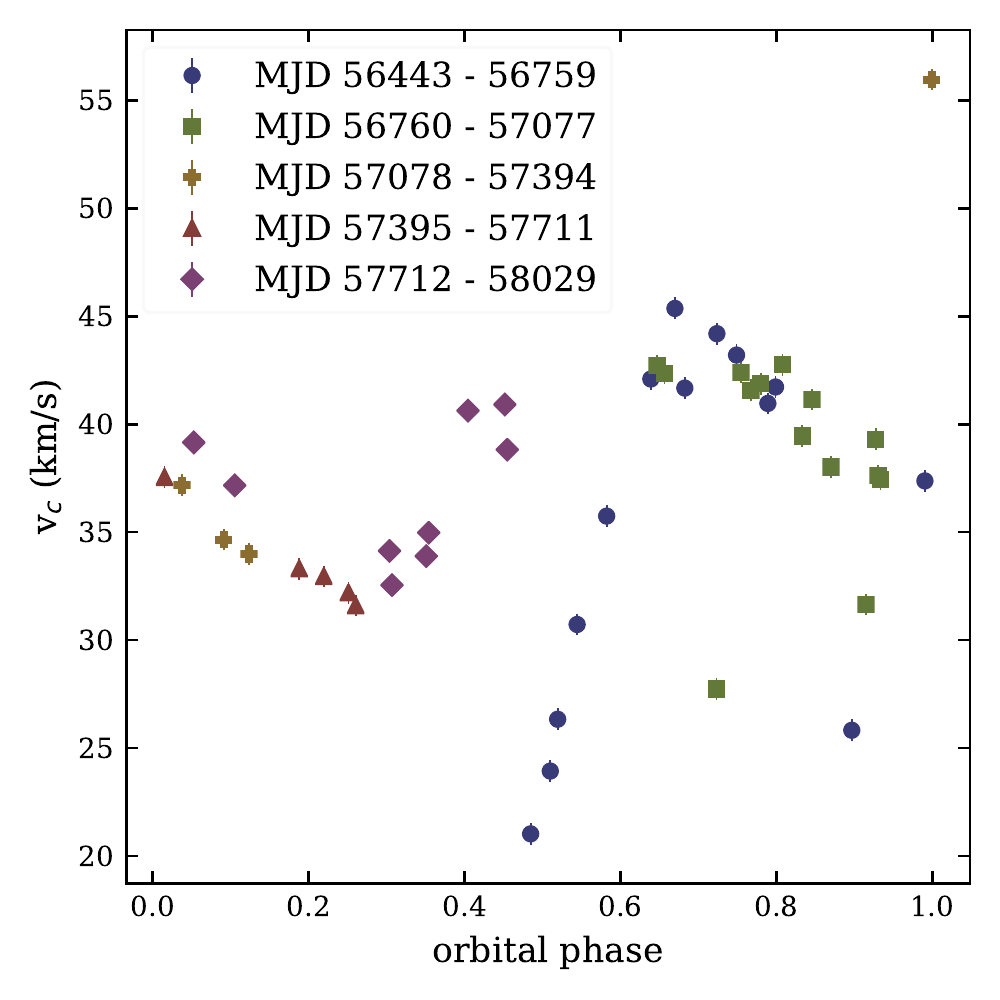}{0.33\textwidth}{(b)}
\fig{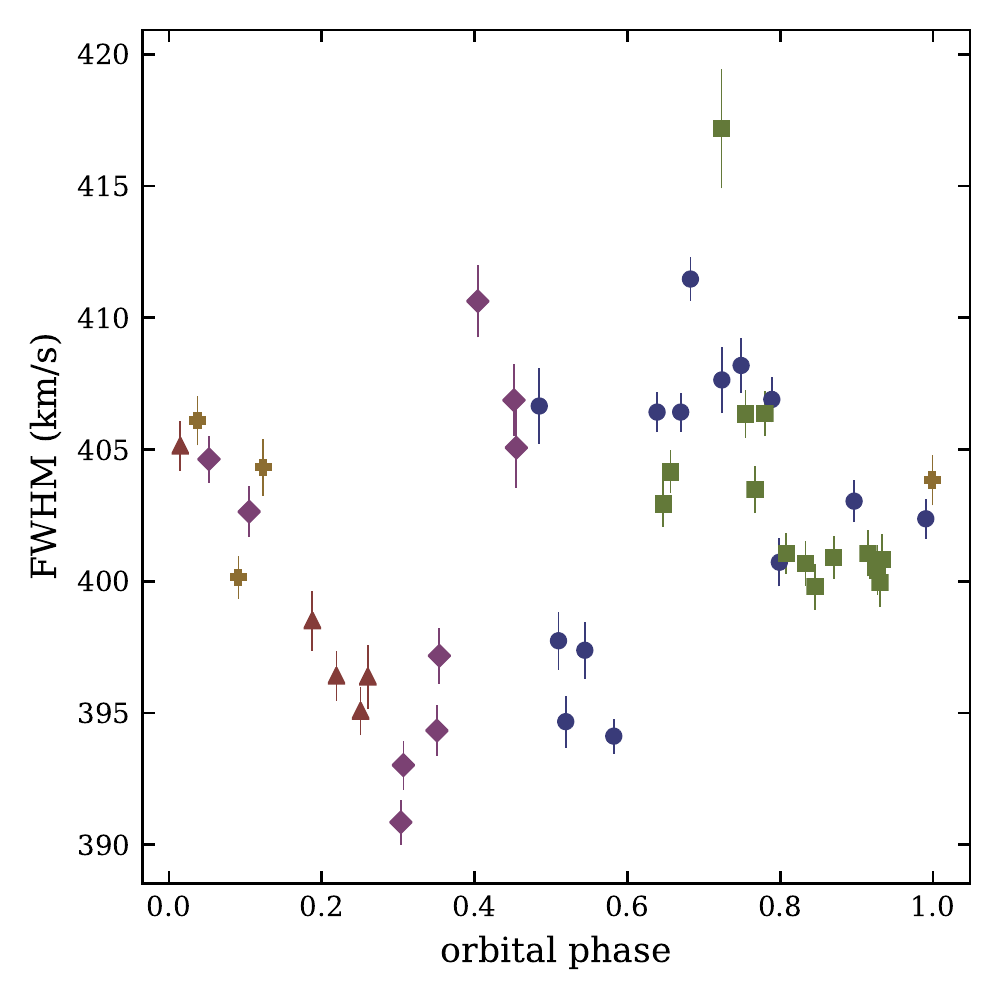}{0.33\textwidth}{(c)}}
\caption{\label{fig:opticalPhaseFolded}
Results from H$\alpha$-measurements of MWC 148 vs. orbital phase, assuming an orbital period of 317.3\,days.
Different markers indicate individual orbits, as indicated by the intervals in MJD in the figure legend in the central panel.
(a) H$\alpha$ equivalent width EW;
(b) H$\alpha$ radial velocity;
(c) H$\alpha$ FWHM.
See section \ref{sec:optical} for definitions of EW, FWHM.}
\end{figure*}

The light curves at X-ray and gamma-ray energies were both folded with the adopted orbital period of 317.3\,days using $\phi=0$ as MJD=54857.0, 
arbitrarily set to the date of the first \textit{Swift}-XRT observations \citep{Bongiorno-2011}. These folded light curves reveal several  prominent features (see Figure \ref{fig:lc-phaseFolded}):
a maximum around phases 0.2-0.4, a  minimum at phases 0.4-0.6, a second (lower) peak around phases 0.6-0.8, and a broader plateau around 0.8-0.2. 
\h is now detected at energies $> 350$\,GeV with high statistical significance at all orbital phases with the exception of the region around the first minimum, at phases 0.4-0.6 (Table \ref{table:EnergySpectrum}). According to the orbital solution of \cite{Casares-2012}, the periastron is located at phase $\phi$=0.967, coincident with the gamma- and X-ray flux plateau, while apastron takes place during the dip/minimum, at phase $\phi$=0.467. For the orbital solution proposed by \cite{Moritani-2018}, periastron takes place at $\phi$=0.663, during the second peak, while apastron happens at $\phi$=0.163 (see Figure \ref{fig:OrbitalSolutions} for a representation of the suggested solutions).
The phase-folded parameters of the H$\alpha$ measurement results are shown in Figure \ref{fig:opticalPhaseFolded}.
No significant modulation is seen in EW, whereas the radial velocity and FWHM show small variations as function of orbital phase. 
Note that \cite{Moritani-2018} proposed an orbital solution using the radial velocity of H$\alpha$.

\h has been observed over 14 orbits at X-ray and 18 orbits at gamma-ray energies (see Appendix \ref{appendix:detailedLCs} and especially Figures \ref{fig:lc-phaseFolded-perOrbit-Xray} and \ref{fig:lc-phaseFolded-perOrbit-GammaRay}).
Despite more than 270 X-ray pointings and $\sim 450$\,h of gamma-ray observations, these data cover most orbits relatively sparsely.
Only five orbits (orbits 7-11 in Figure \ref{fig:lc-phaseFolded-perOrbit-Xray}) have been regularly monitored at X-ray energies over more than 50\% of an individual orbit. 
Ground-based gamma-ray instruments covered an even smaller range, as they cannot observe the binary for a large fraction of its orbit due to visibility constraints.
Four orbits have been observed in the phase range 0.2-0.4 with notable coverage in X-ray and gamma-ray observations (Figure \ref{fig:LC-detail}) and are discussed in more detail in section \ref{sec:outburstJan2018}.


\subsection{Correlation analysis}
\label{sec:correlations}

\begin{figure}
\centering
\plotone{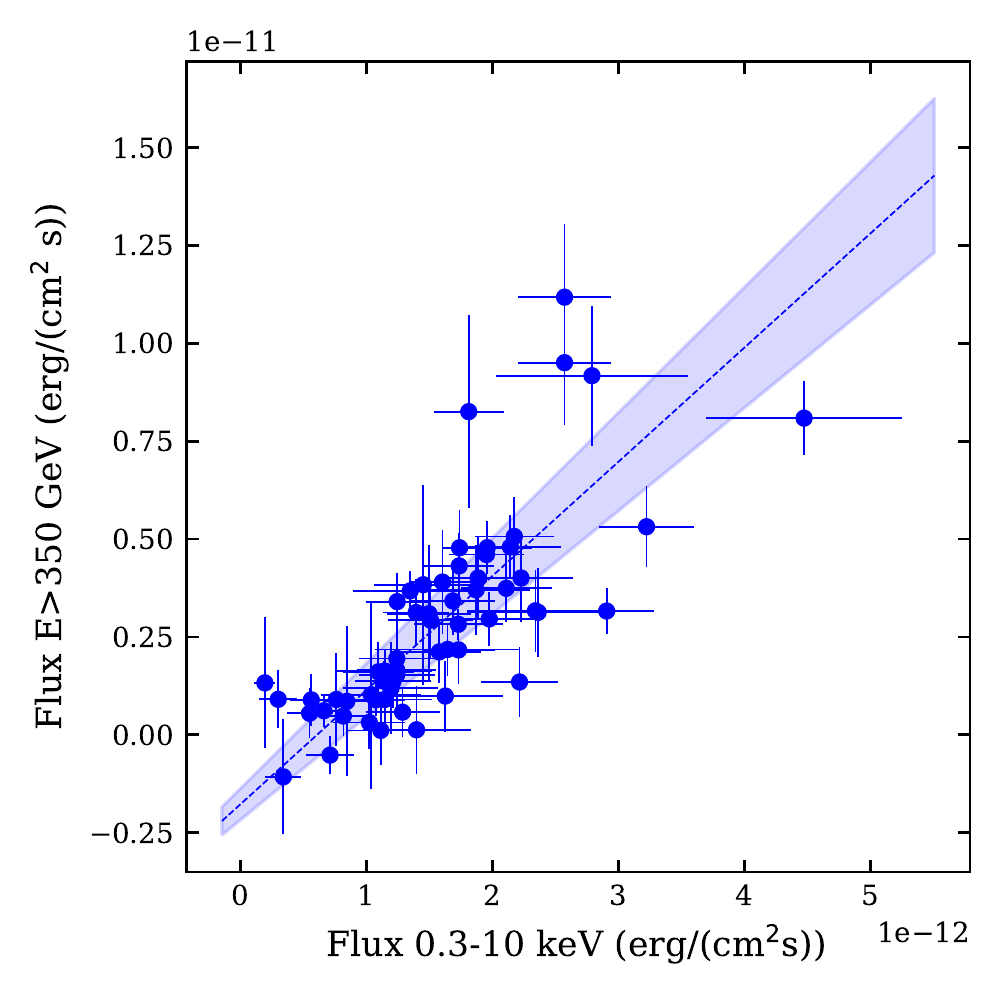}
\plotone{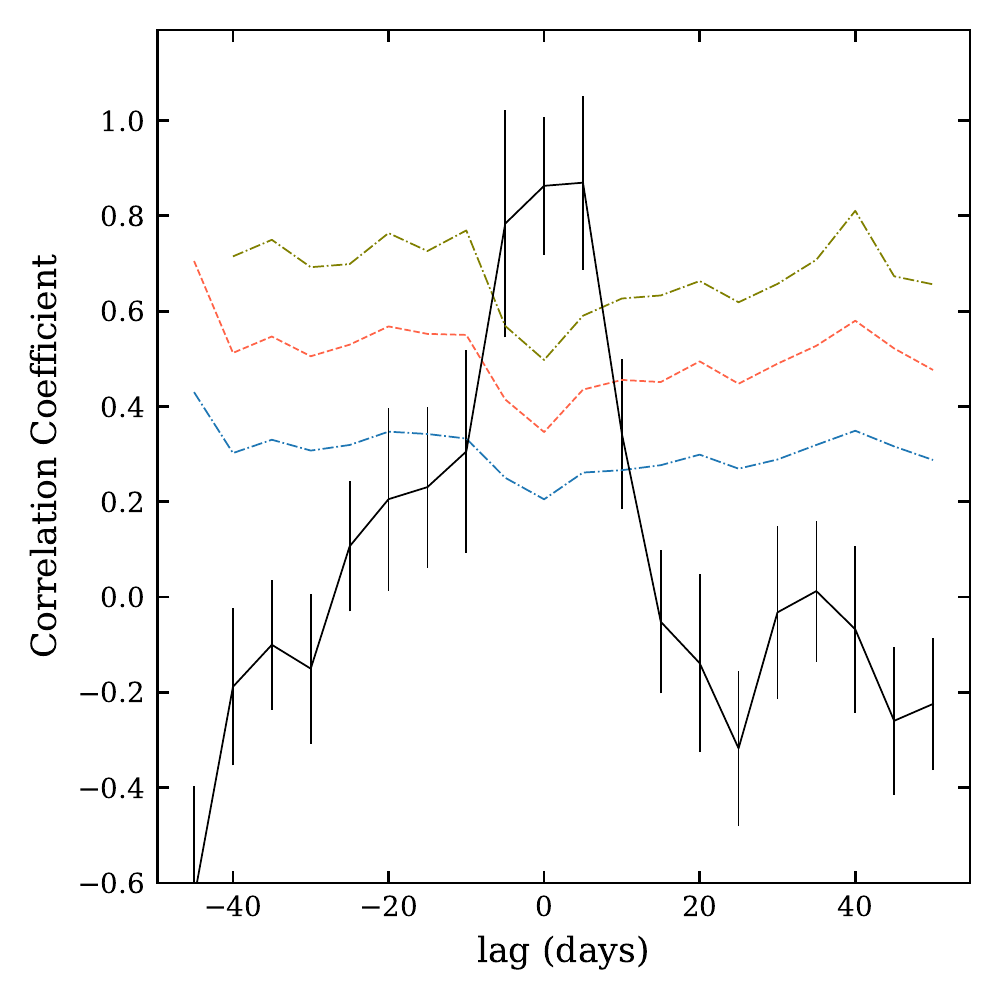}
\caption{\label{fig:XrayGrayCorrelation}
Top: Contemporaneous gamma-ray ($>$350\,GeV) vs X-ray (0.3--10\,keV) integral fluxes.
In total, 59 measurements were selected with X-ray measurements taking place during the period of gamma-ray observations.
The dashed line and shaded area indicate the results of a linear fit to the data and the one sigma error range, respectively.
Bottom: Discrete cross-correlation function (DCF) between gamma-ray and X-ray data assuming a bin width for the time lag $\Delta \tau=5$\,days.
The dashed lines indicate the 2, 3, and 4\,$\sigma$ confidence levels (from bottom to top) derived from 100,000\,toy Monte Carlo light curves.
Only correlation coefficients with time lags consistent with zero are significant.
}
\end{figure}

Correlations between X-ray and gamma-ray flux measurements are shown in Figure \ref{fig:XrayGrayCorrelation} (top).
Gamma-ray observations are averaged over several nights in order to achieve sufficient statistical detection significance, while X-ray observations are nightly averages. 
Measurements are considered to be contemporaneous in this context for X-ray observations taking place during the time period of the grouped gamma-ray observations (the maximum interval length for the grouped gamma-ray observations is $\sim$10\% of the orbital period).
Average X-ray fluxes are used for periods of gamma-ray observations including several X-ray data points.
It is worth stating that both the X-ray and gamma-ray fluxes are measured to be variable on timescales as short as hours, and the total X-ray exposure associated with each point is typically just a small fraction of the gamma-ray exposure.
This might contribute to some of the scattering between fluxes in both energy bands as observed in Figure \ref{fig:XrayGrayCorrelation} (top).
The discrete correlation function method \citep[DCF]{Edelson-1988} is used to quantify the correlation and to search for possible time lags between X-ray and gamma-ray emission patterns.
Figure \ref{fig:XrayGrayCorrelation} (bottom) shows the DCF using a binning of 5\,days (different binnings have been tried with no significant changes in the results) and with confidence limits calculated using the toy Monte Carlo technique as recommended in \cite{2003ApJ...584L..53U}.
The correlation coefficient is $0.82$ at time lag $\tau=0$ and the p-value for non-correlation is $4\times 10^{-15}$, calculated from 59 flux pairs.
The ratio of gamma-ray ($>$ 350 GeV) to X-ray flux (0.3-10 keV) is on average $2.9\pm 0.3$, estimated from a linear fit as shown in  Figure \ref{fig:XrayGrayCorrelation} (top) for the chosen integration ranges in energy.
This underlines the equality or even dominance of the gamma-ray energy range for the emission of the binary system with respect to the X-ray regime (see also Figure \ref{fig:SEDphaseRange}).
The fit also indicates a non-zero X-ray flux (F$_{\mathrm{0.3-10 keV}} = (6.1\pm 1.5)\times10^{-13}$\,erg/cm$^2$/s) for a vanishing gamma-ray component, suggesting an X-ray source partially unrelated to the gamma-ray emission.  
No indication of a significant time lag between X-ray and gamma-ray emission is observed (see Figure \ref{fig:XrayGrayCorrelation} bottom), although this measurement is not sensitive to time lags $\tau  \lessapprox 10$\,days due to the required grouping of the gamma-ray observations.
In general, this confirms the strong correlation between gamma-ray and X-ray emission, as already shown in \cite{Aliu-2014}.
Ranked and un-ranked correlation analyses result in similar correlation coefficients between 0.75 and 0.82.
This  points towards a close-to-linear correlation, although statistical uncertainties of the photon fluxes observed in both energy bands prevent us from making any strong statements regarding possible non-linearities between gamma-ray and X-ray emission.

\begin{figure*}
\gridline{
\fig{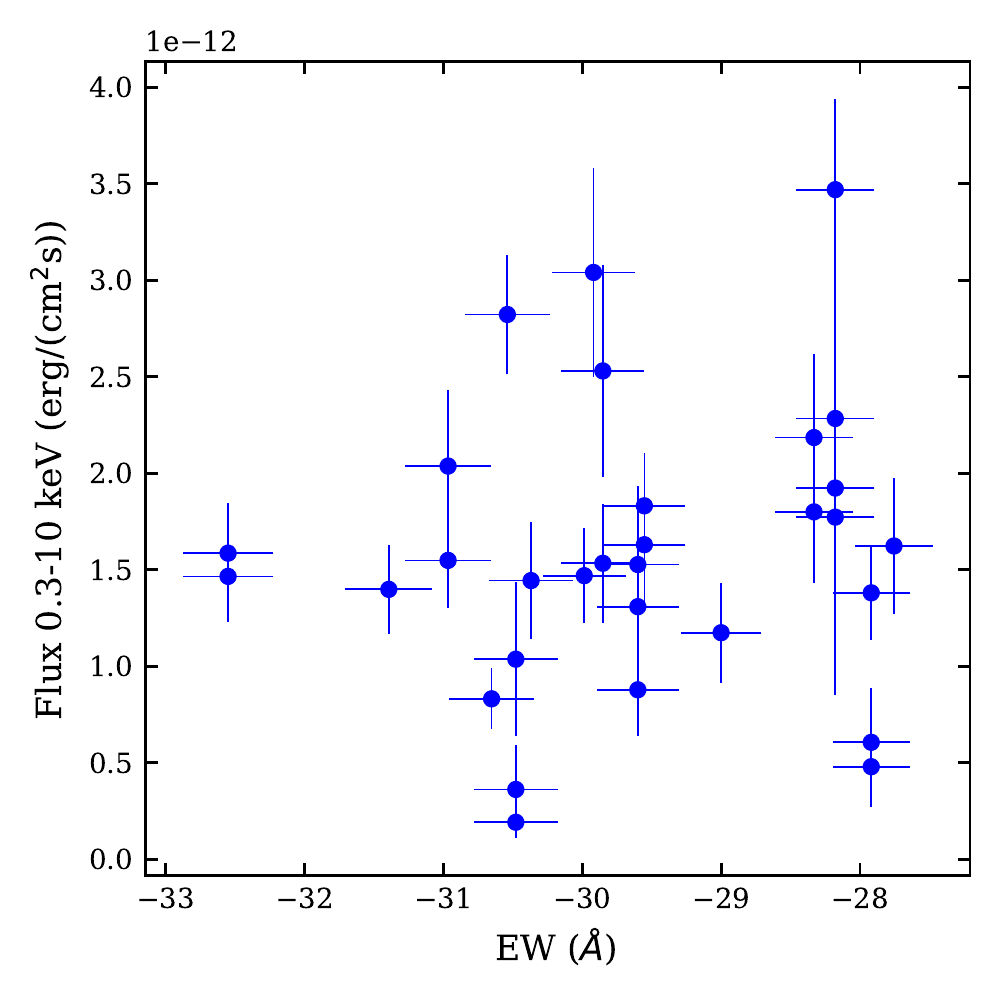}{0.33\textwidth}{(a)}
\fig{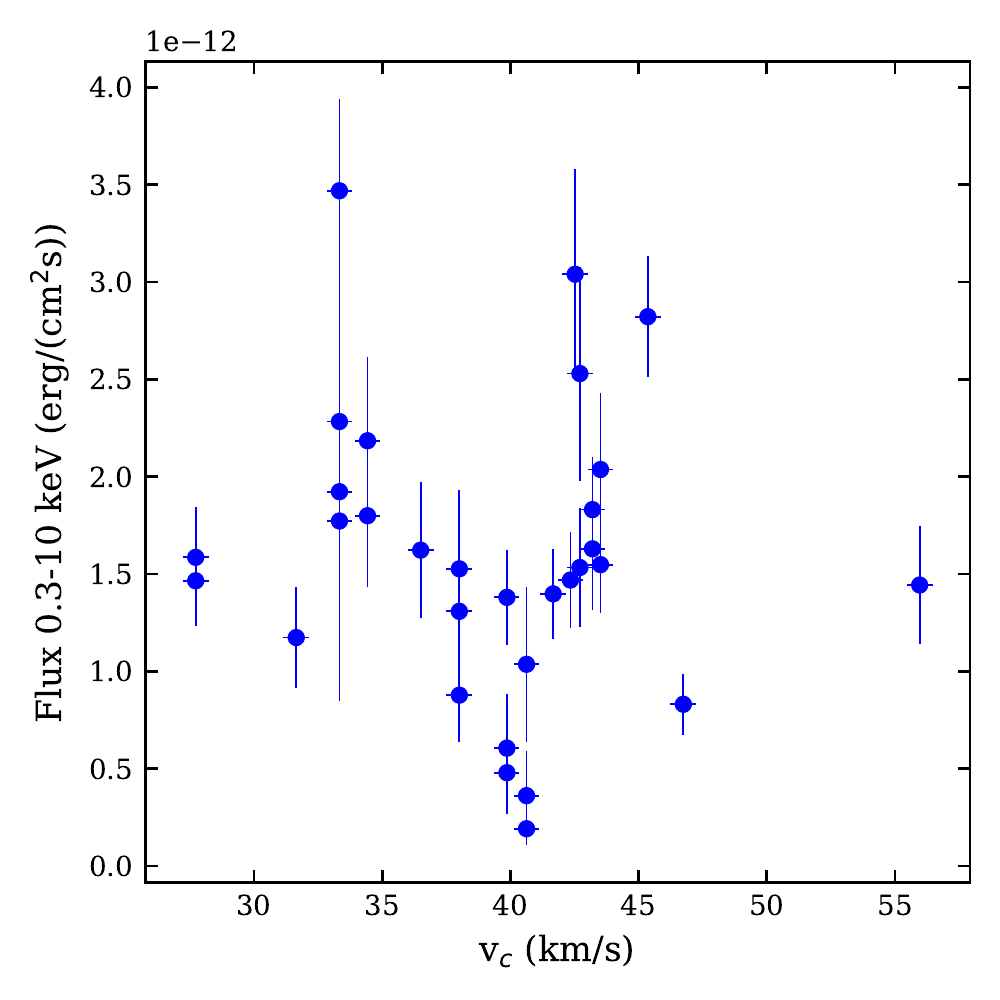}{0.33\textwidth}{(b)}
\fig{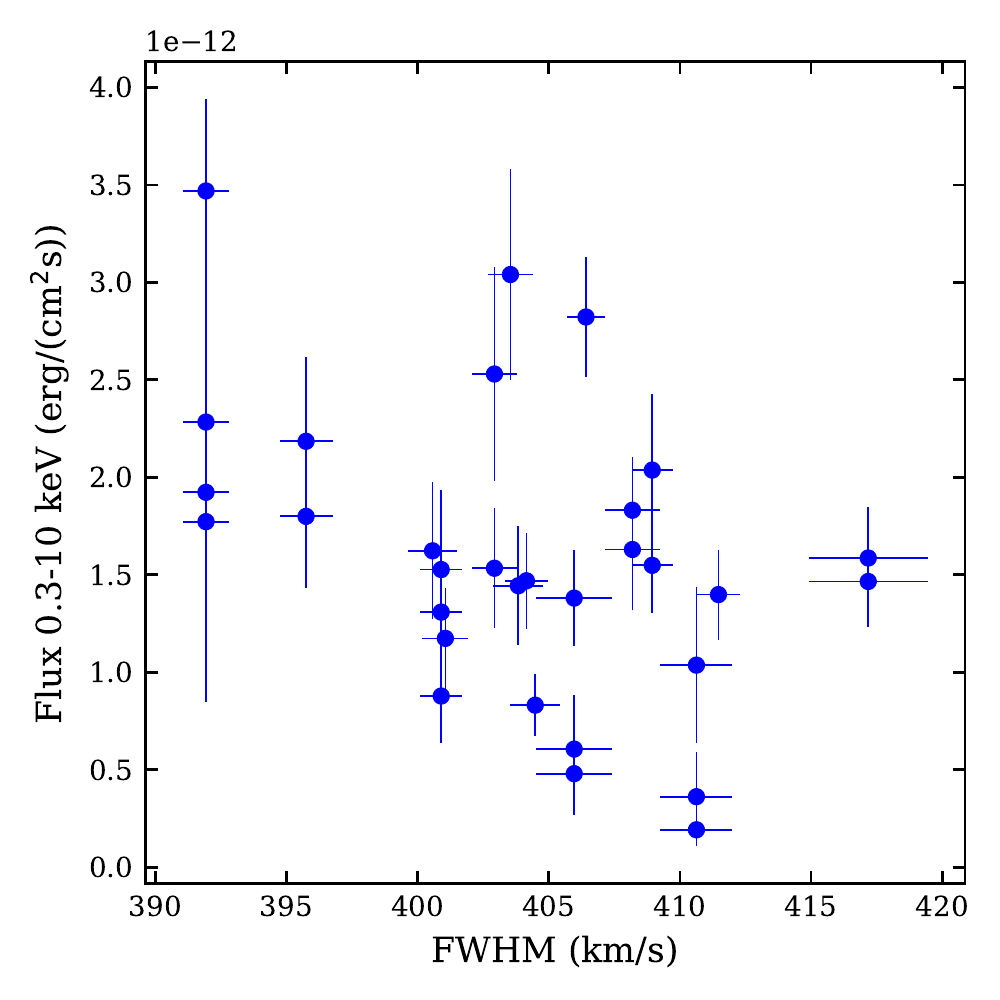}{0.33\textwidth}{(c)}}
\gridline{
\fig{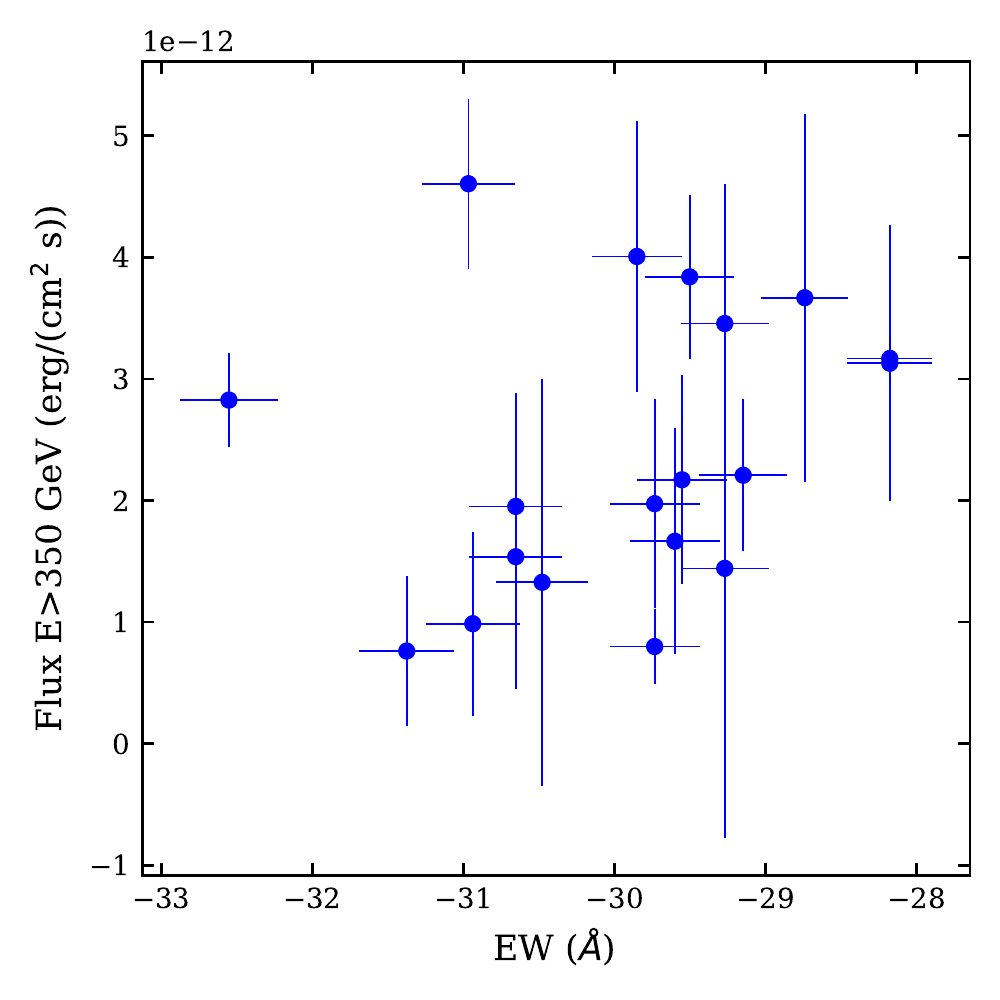}{0.33\textwidth}{(d)}
\fig{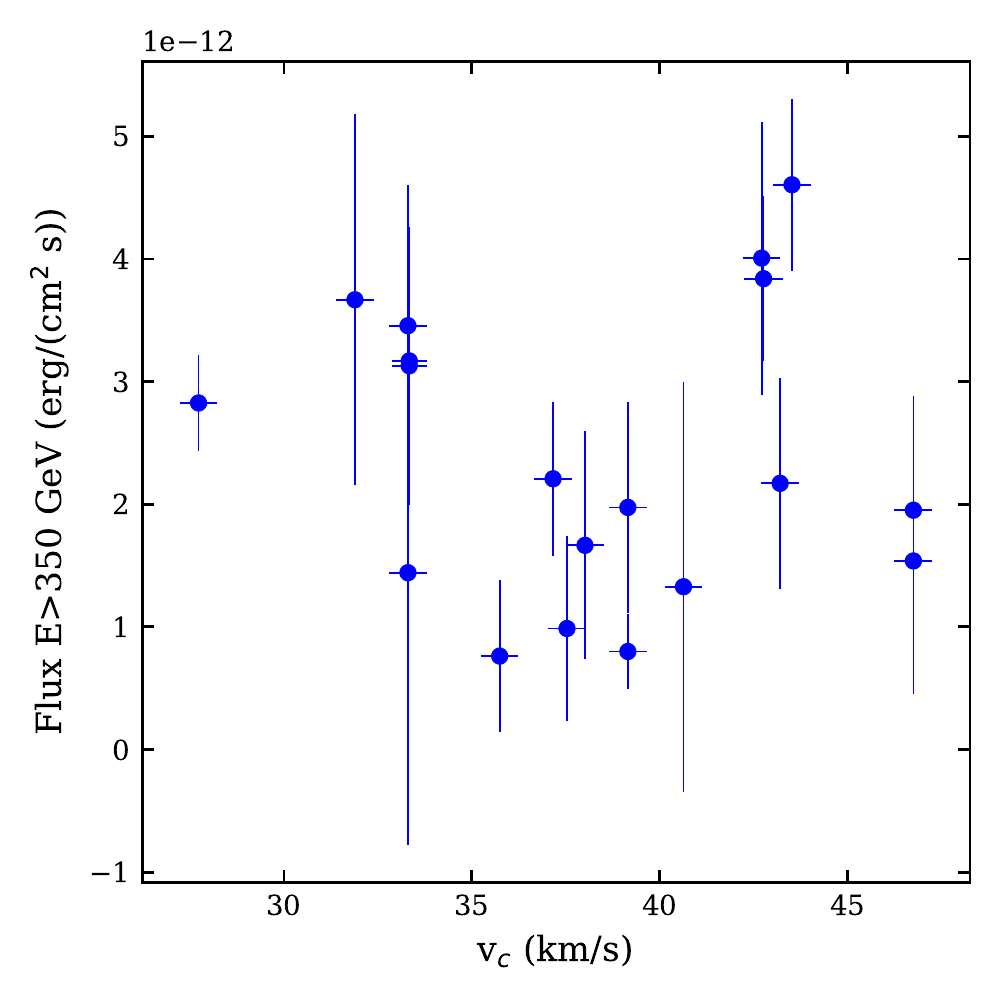}{0.33\textwidth}{(e)}
\fig{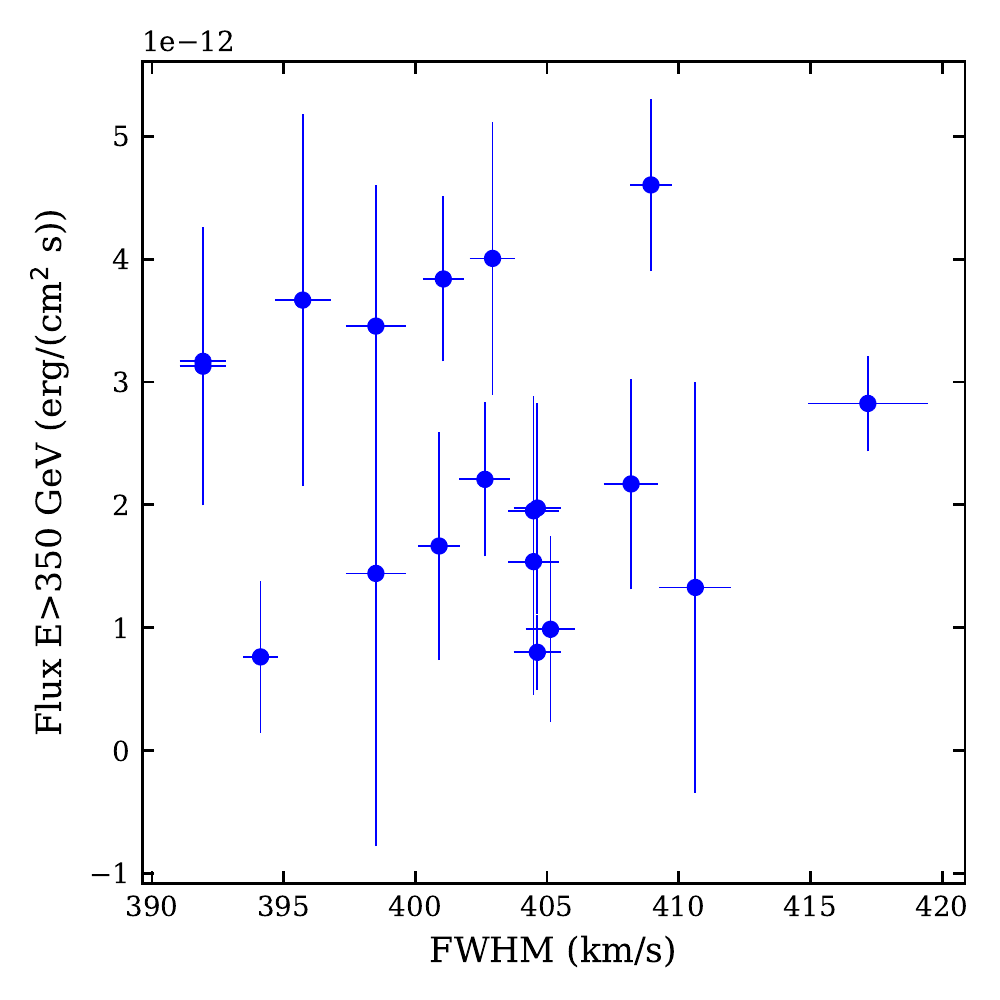}{0.33\textwidth}{(f)}}
\caption{\label{fig:XGvsOptical}
H$\alpha$-measurements of MWC 148 vs X-ray (top; panels a-c) and gamma-ray (bottom; panels d-f) measurements.
In total 31 (19) measurements were selected with a maximal difference of observation dates of 5\,days for the X-ray/H$\alpha$ (gamma-ray/H$\alpha$) correlation analysis.
}
\end{figure*}

No correlation is found between any parameters obtained from the optical H$\alpha$ measurements and X-ray or gamma-ray fluxes (Figure \ref{fig:XGvsOptical}).
The Spearman correlation coefficients are between -0.4 and 0.33 for time lags $\tau=0$, corresponding to p-values for non-correlation between 0.02 and 0.85 (calculated from 18 gamma-ray/H$\alpha$ and 31 X-ray/H$\alpha$ pairs).
A relaxed definition for contemporaneous data is used for the H$\alpha$ correlation analysis with a time span of $\pm 5$ days, well below the shortest timescales of $\approx50$ days observed for H$\alpha$ profile variability \citep{2015ApJ...804L..32M}. 
This relaxation of the contemporaneity criteria is necessary to obtain a reasonable number of pairs of observations, but might obscure variability on shorter timescales. The low gamma-ray fluxes and large uncertainties might hide possible correlations. 
It should also be noted that the optical H$\alpha$ data were obtained during five orbital periods (MJD 56597--57857; orbits 12--16 in Figures \ref{fig:lc-phaseFolded-perOrbit-Xray}, \ref{fig:lc-phaseFolded-perOrbit-GammaRay}) with generally sparse coverage of X-ray and gamma-ray observations.  
This results in no contemporaneous observation pairs at TeV/ H$\alpha$ around orbital phases 0.3 -- 0.6 which correspond to the first maximum and the dip in the folded light curve. In case of the contemporaneous observation pairs at X-ray/ H$\alpha$, there is no coverage at orbital phases 0.0 -- 0.2 and 0.5 -- 0.6. This imperfect coverage of orbital phases might also obscure possible correlations.

%
%
%


\subsection{Flux states and spectral analysis}
\label{sec:spectra}

\begin{figure*}[!htb]
\gridline{
\fig{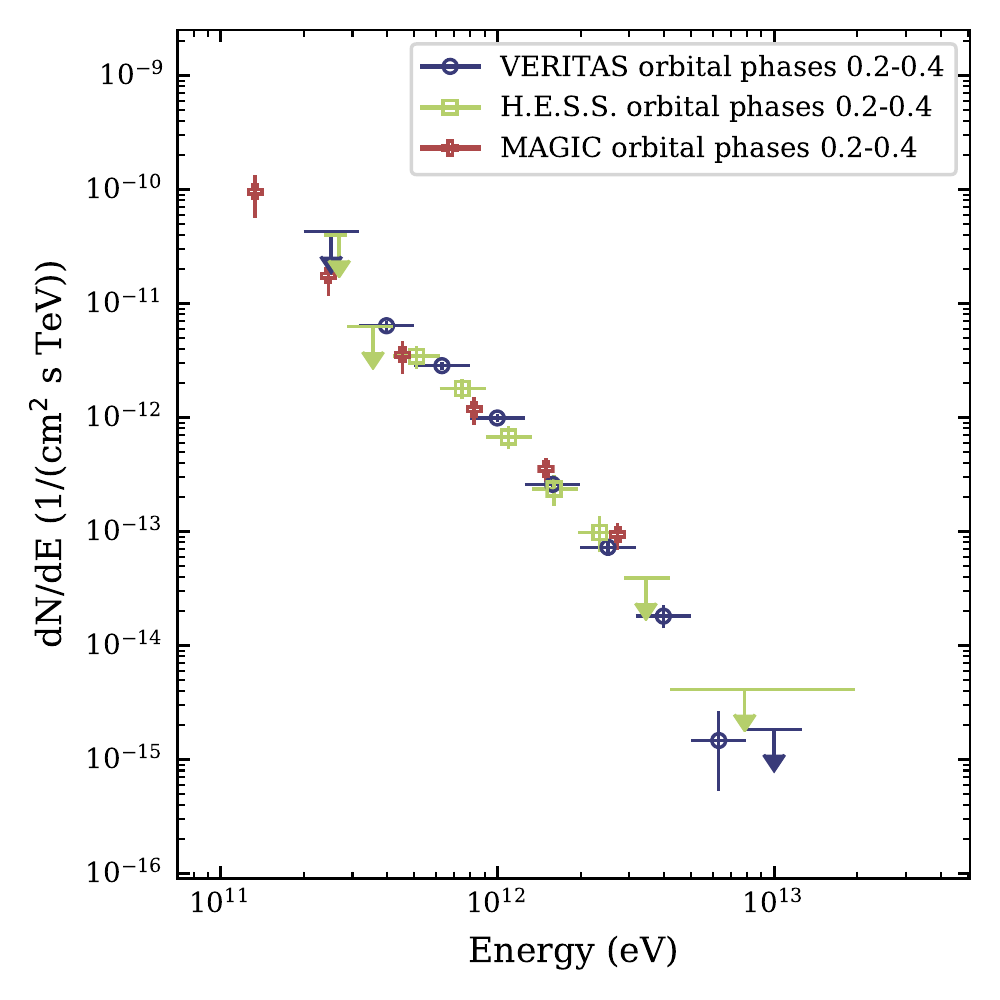}{0.5\textwidth}{(a)}
\fig{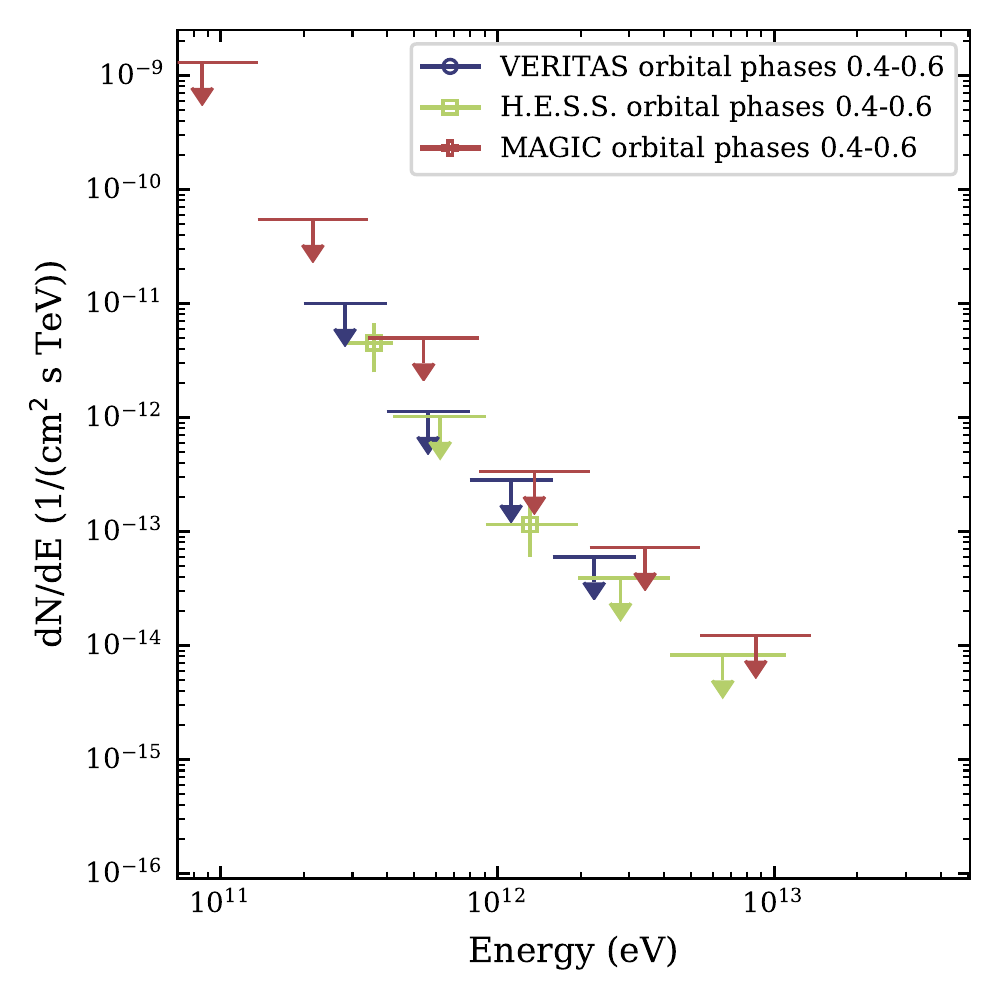}{0.5\textwidth}{(b)}}
\gridline{
\fig{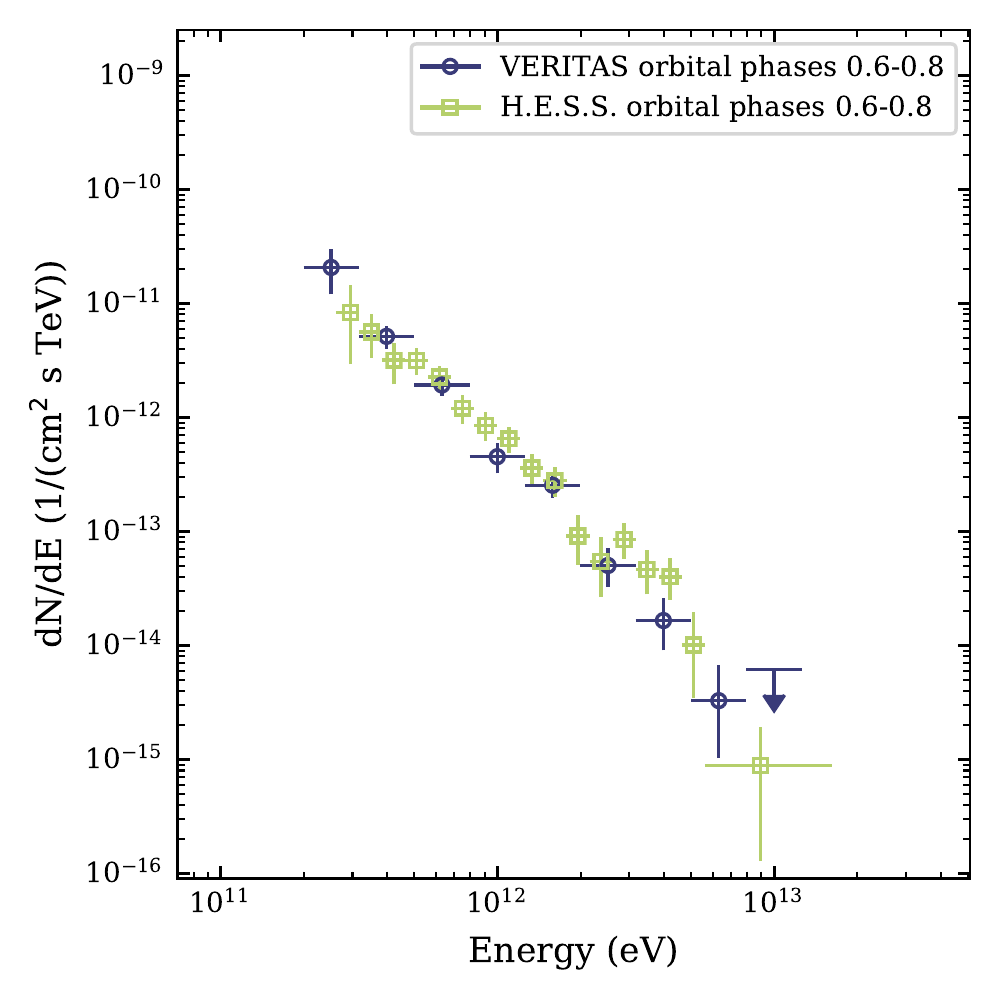}{0.5\textwidth}{(c)}
\fig{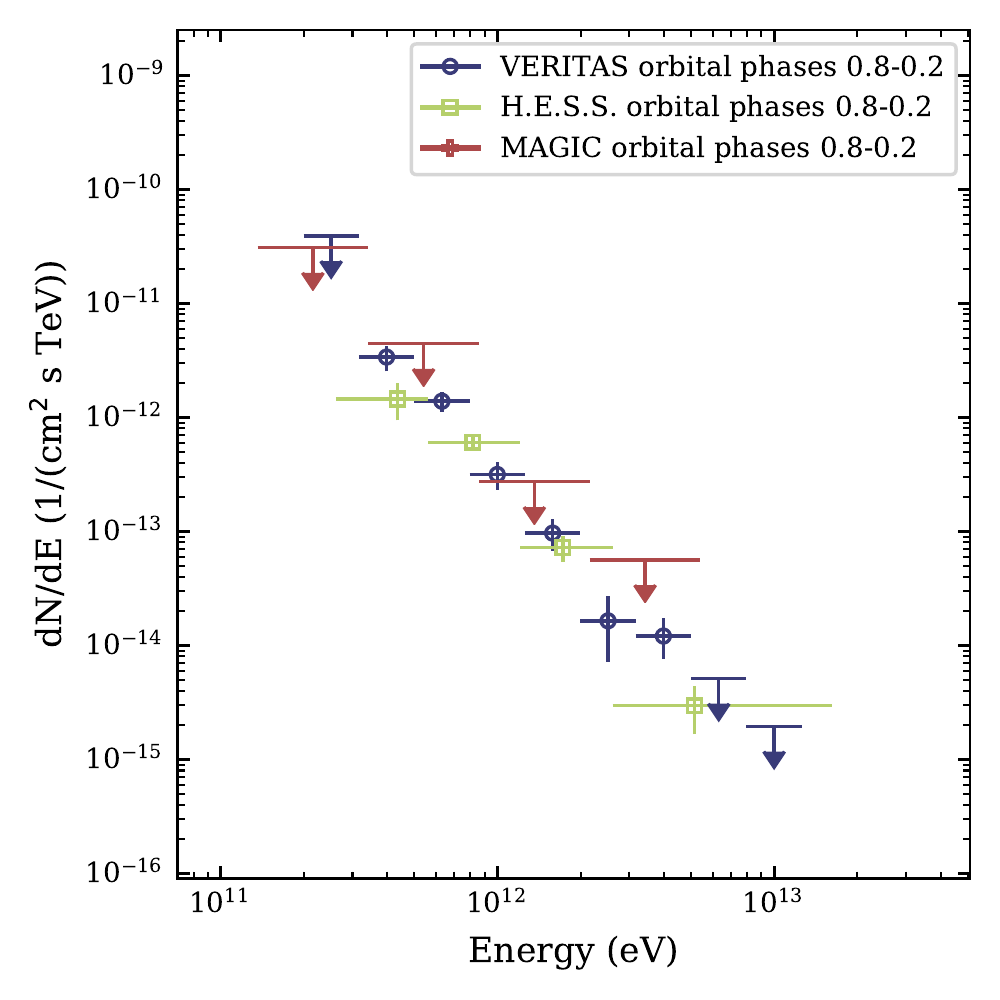}{0.5\textwidth}{(d)}}
\caption{\label{fig:TeVSpectra}
Differential energy spectra of photons above 200 GeV obtained by \hess, MAGIC and VERITAS  averaged over all available orbits. 
The figure shows the results for four different orbital phase bins:
(a) orbital phases 0.2--0.4; (b) orbital phases 0.4--0.6; 
(c) orbital phases 0.6--0.8; (d) orbital phases 0.8--0.2.
Vertical error bars show $1\sigma$ uncertainties; downwards pointing arrows indicate upper limits at the 95\% confidence level.
}
\end{figure*}

%
%
%
%
%
%
%
\begin{deluxetable*}{lcccccc}
\centering
\tablecolumns{7}
\tablewidth{0pt}
\tablecaption{Summary of observations and spectral analyses at gamma-ray energies averaged over different ranges in orbital phase assuming an orbital period of 317.3 days.
The table lists the  results of the  power-law fits ($dN/dE = N_0(E/0.5\mathrm{TeV})^{-\Gamma}$) to the differential energy spectra obtained separately for \hess, MAGIC, and VERITAS.
For one measurement, results of a fit using a power-law with exponential cut off ($dN/dE = N_0 (E/0.5\mathrm{TeV})^{-\Gamma} e^{-E/E_c}$; marked as \textit{epl}) is listed.
Errors are $1\sigma$ statistical errors only. 
The systematic errors are 11-20\% on the flux constant and 0.10-0.15 on the spectral index.
\label{table:EnergySpectrum}}
\tablehead{
\colhead{Observatory} &
\colhead{Exposure} &
\colhead{Significance\tablenotemark{a}} &
\colhead{flux normalization} &
\colhead{photon index} &
\colhead{cut-off} &
\colhead{$\chi^2/N$\tablenotemark{c}} \\
\colhead{} &
\colhead{(h)} &
\colhead{($\sigma$)} &
\colhead{constant at 0.5 TeV\tablenotemark{b}} & 
\colhead{} &
\colhead{energy} &
\colhead{} \\
\colhead{} &
\colhead{} &
\colhead{(pre-trial)} &
\colhead{[cm$^{-2}$s$^{-1}$TeV$^{-1}$]} &
\colhead{} &
\colhead{[TeV]} &
\colhead{}
}
\startdata
\hline
\multicolumn{7}{c}{Orbital phase range 0.2-0.4 (high state)} \\
\hline
VERITAS & 119.3 & 25.5 & $(45.6 \pm 2.4) \times 10^{-13}$ & $2.67 \pm 0.04$ & -- &$45.17/7$ \\
VERITAS (epl) & &  & $(57.2 \pm 4.2) \times 10^{-13}$ & $1.79 \pm 0.16$ & $1.75\pm0.38$ &$4.5/6$ \\
H.E.S.S.   &  17.3  & 9.9 & $(34.0\pm 4.4) \times 10^{-13}$ & $2.45\pm0.12$ & -- & $4.6/7$ \\
MAGIC   &  35.5  & 9.6 & $(34.8\pm4.6) \times 10^{-13}$  & $2.48\pm0.11$ & -- & $13.5/8$ \\
\hline
\hline
\multicolumn{7}{c}{Orbital phase range 0.4-0.6 (low state)} \\
\hline
VERITAS & 40.0 & 2.4 & -- & -- & -- & -- \\
H.E.S.S.   &  18.7  & 2.9 &  --  & --  & -- & --  \\
MAGIC   &  5.2  & 0.6 &  --  & --  & -- & --  \\
\hline
\hline
\multicolumn{7}{c}{Orbital phase range 0.6-0.8 (high state)} \\
\hline
VERITAS & 28.4 & 11.0 & $(31.6 \pm 3.9) \times 10^{-13}$ & $2.52 \pm 0.14$ & -- & $3.3/6$ \\
H.E.S.S.   & 30.4  & 13.8 & $(32.0\pm3.5) \times 10^{-13}$ & $2.34\pm0.1$ & -- & $15.25/15$ \\
MAGIC   &  -- & -- & -- & -- & -- & -- \\
\hline
\hline
\multicolumn{7}{c}{Orbital phase range 0.8-0.2} \\
\hline
VERITAS & 72.7 & 8.9 & $(20.8 \pm 3.1) \times 10^{-13}$ & $2.67 \pm 0.17$ & -- & $3.2/4$ \\
\hess   &  53.6  & 8.3 & $(16.2\pm2.6) \times 10^{-13}$ & $2.55\pm0.15$ & -- & $5.3/2$ \\
MAGIC   &  27.4 & 3.4 & --  & -- & -- & -- \\
\enddata
\tablenotetext{a}{Significances are calculated following the maximum likelihood ratio test method proposed in \citet{Li:1983}.}
\tablenotetext{b}{Flux normalisation constants are calculated for all data sets for easier comparison at the same energy (0.5 TeV) and not at the de-correlation energy. }
\tablenotetext{c}{\textit{N} comes from number of spectral bins in estimated energy used to fit the data. The spectrum itself is obtained via forward folding to those bins.}
\end{deluxetable*}

The data sets are divided into four different bins in orbital phase for the spectral analysis: the two higher flux states with orbital phases 0.2--0.4 and 0.6--0.8, the low state at orbital phases 0.4-0.6, and the medium (plateau) state for the remaining phases 0.8--0.2.
This approach ignores the variability of the source on timescales shorter than these bins, as well as possible orbit-to-orbit variability, but allows a sufficient
number of excess events to be collected for high-statistics spectra.

%
\begin{figure*}[!htb]
\gridline{
\fig{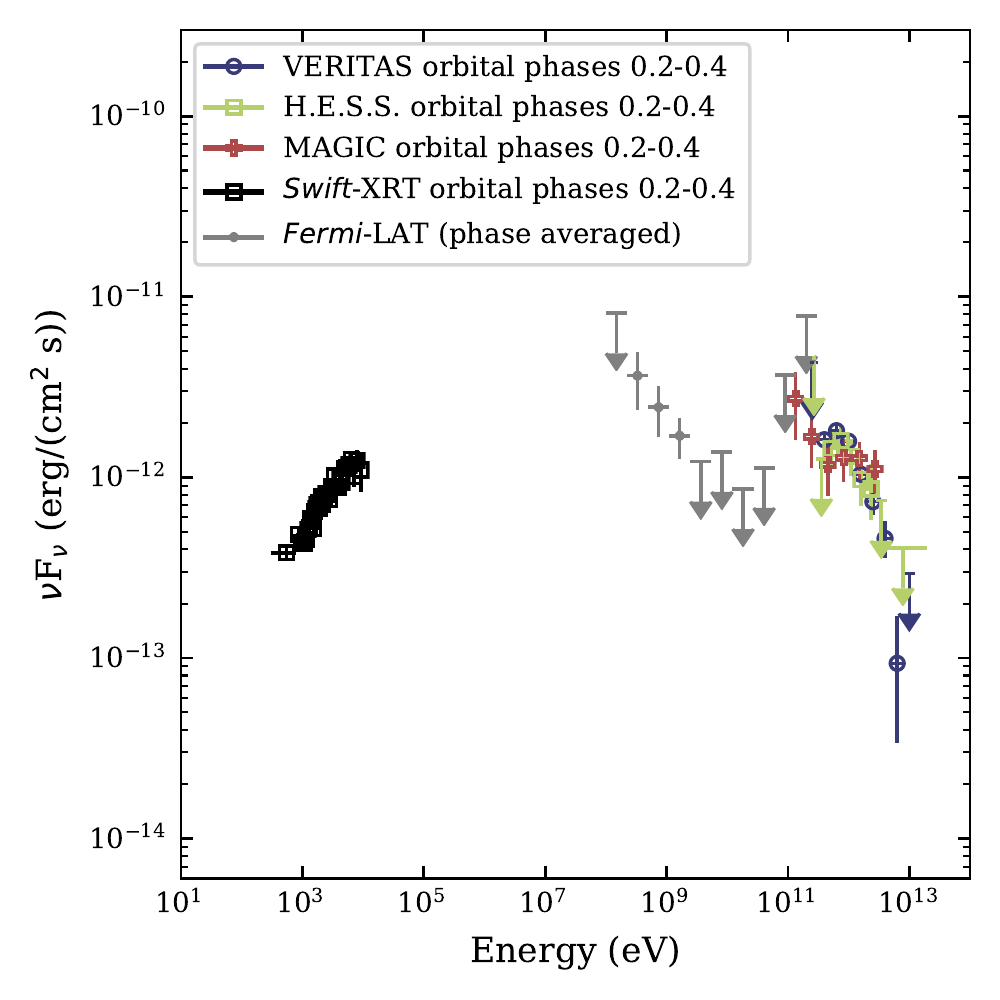}{0.5\textwidth}{(a)}
\fig{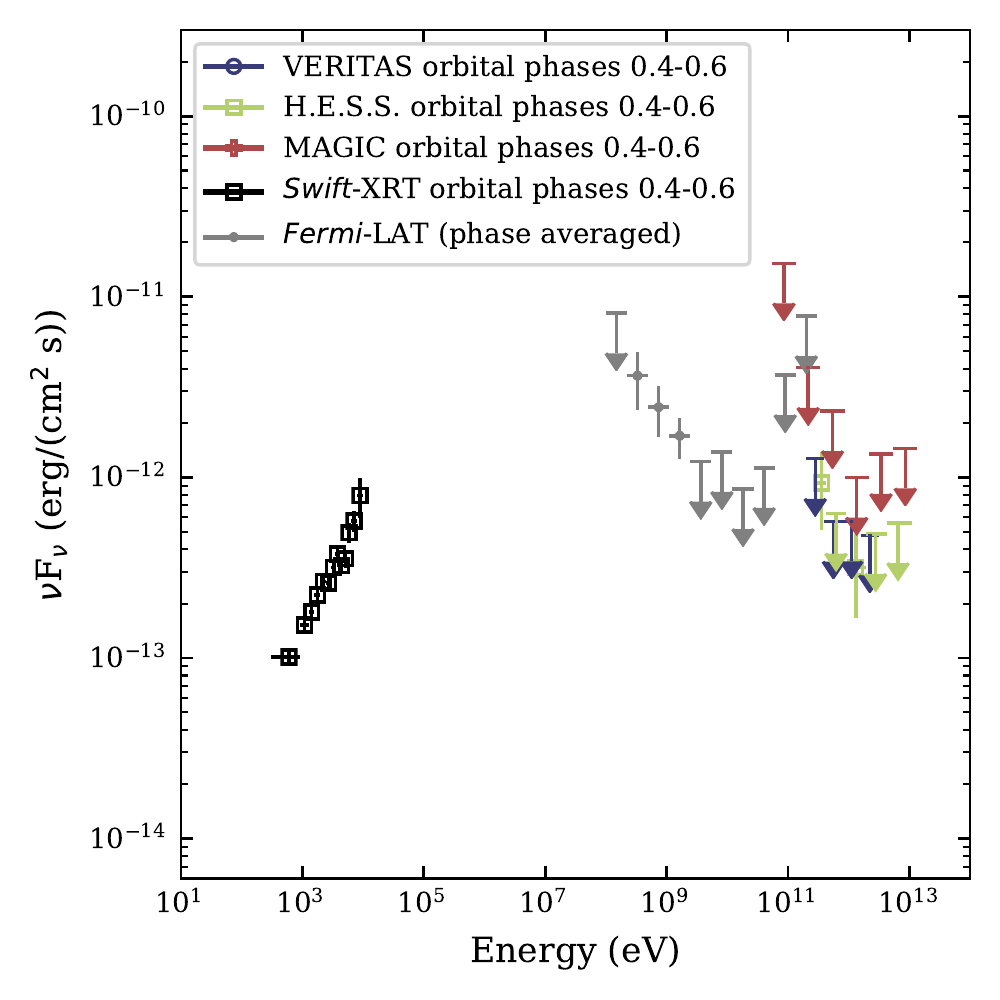}{0.5\textwidth}{(b)}}
\gridline{
\fig{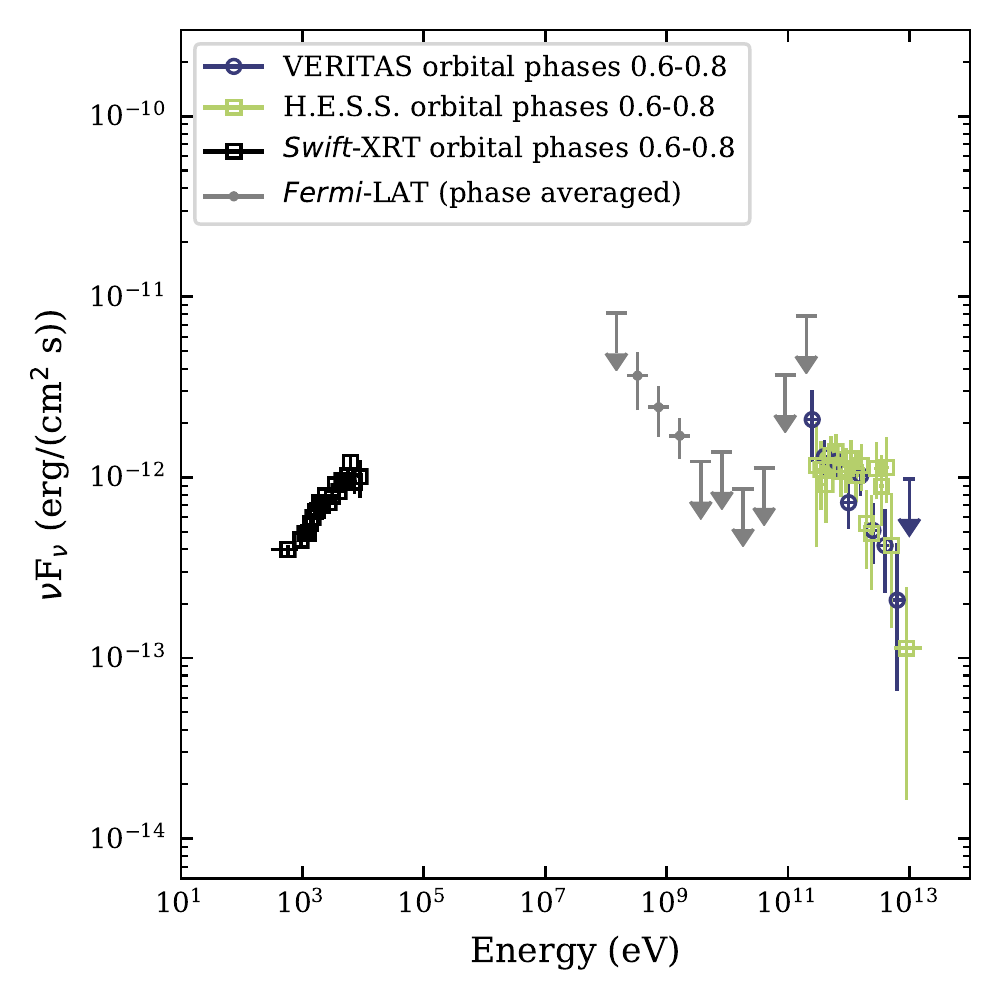}{0.5\textwidth}{(c)}
\fig{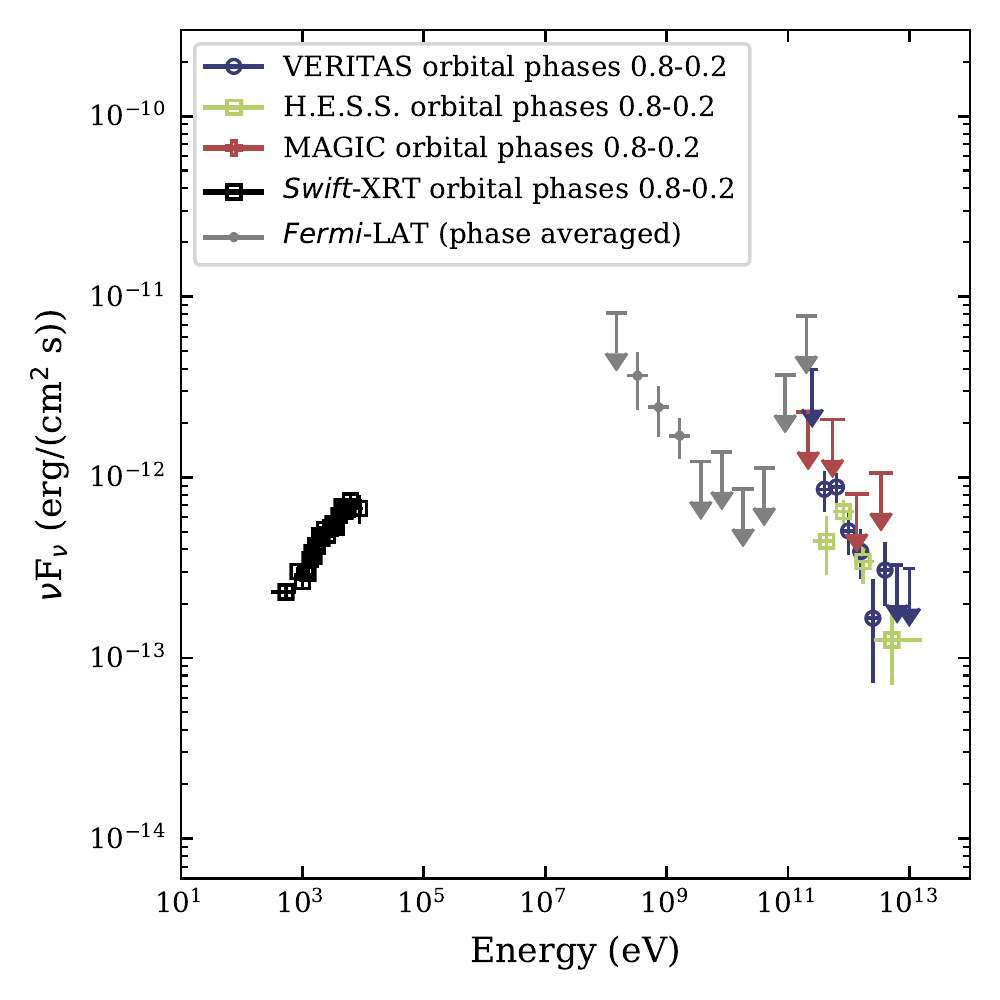}{0.5\textwidth}{(d)}}
\caption{\label{fig:SEDphaseRange}
Spectral energy distribution for \h averaged over all available orbits.
The figure shows the result for four different phase bins:
(a) orbital phases 0.2--0.4;
(b) orbital phases 0.4--0.6;
(c) orbital phases 0.6--0.8;
(d) orbital phases 0.8--0.2.
Downwards pointing arrows indicate upper limits at the 95\% confidence level.
Gray points show the phase-averaged spectrum measured by the \textit{Fermi}-LAT \citep{2017ApJ...846..169L}.
}
\end{figure*}

Results from the spectral analysis are summarised in Figure \ref{fig:TeVSpectra} and Table \ref{table:EnergySpectrum}. The differential gamma-ray energy spectra in the four phase ranges are consistent within statistical uncertainties with simple power-laws $dN/dE = N_0(E/(0.5\,\mathrm{TeV)})^{-\Gamma}$ with the exception of the long-exposure data set obtained with VERITAS for phase bin 0.2--0.4.
Here, a spectral fit assuming a power law with exponential cut off ($dN/dE = N_0(E/(0.5\,\mathrm{TeV)})^{-\Gamma} e^{-E/E_c}$) with $E_c=1.75\pm 0.38$\,TeV provides a significantly better reduced $\chi^2$ compared to a power-law fit without cut off.
Shorter exposure times at other phase bins of \hess and MAGIC observations prevent us from measuring or refuting similar changes in the spectral shapes with those data sets. 
All measurements of the differential energy spectra by \hess, MAGIC, and VERITAS are compatible, within statistical and systematic uncertainties.
No variability of the spectral index $\Gamma$ is observed among spectra in different orbital phase bins, considering the statistical uncertainties.
The spectral energy distributions for the four different phase bins are shown in Figure \ref{fig:SEDphaseRange}, including the X-ray spectra, which have been obtained in the same way by averaging over the discussed phase bins.
\h is only weakly detected in the energy range covered by the \textit{Fermi}-LAT, 
therefore only a phase-averaged spectrum is shown, taken from  \cite{2017ApJ...846..169L}.


\subsubsection{Detailed examination of orbits 9, 10, 16, and 17}
\label{sec:outburstJan2018}

%

%
\begin{figure*}[!htb]
\gridline{
\fig{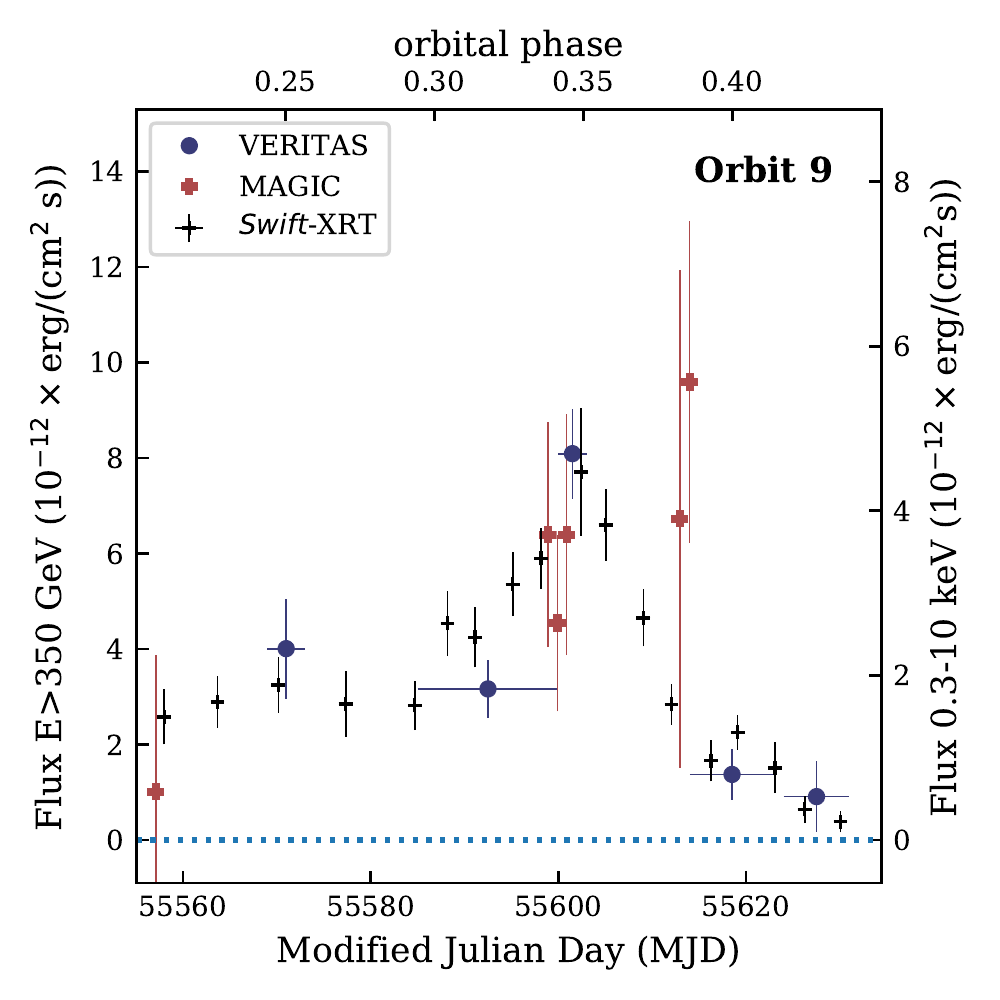}{0.5\textwidth}{}
\fig{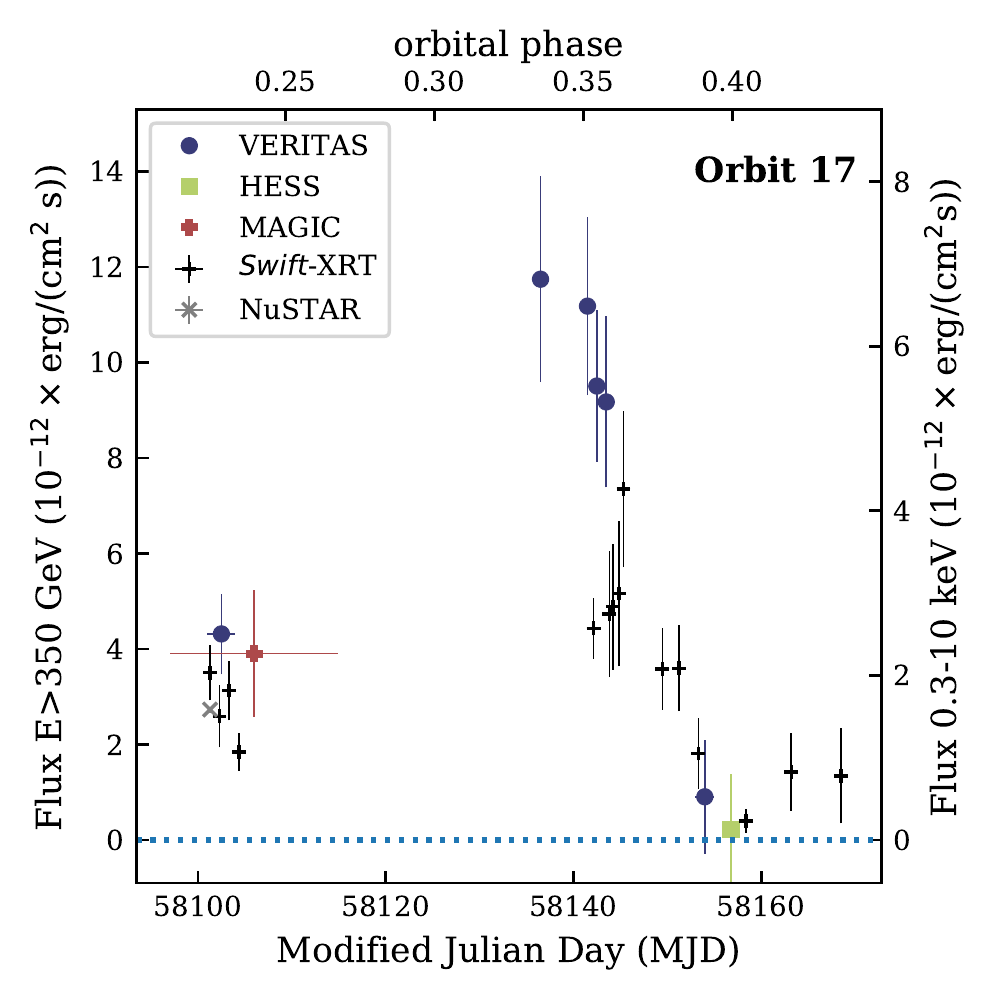}{0.5\textwidth}{}}
\gridline{
\fig{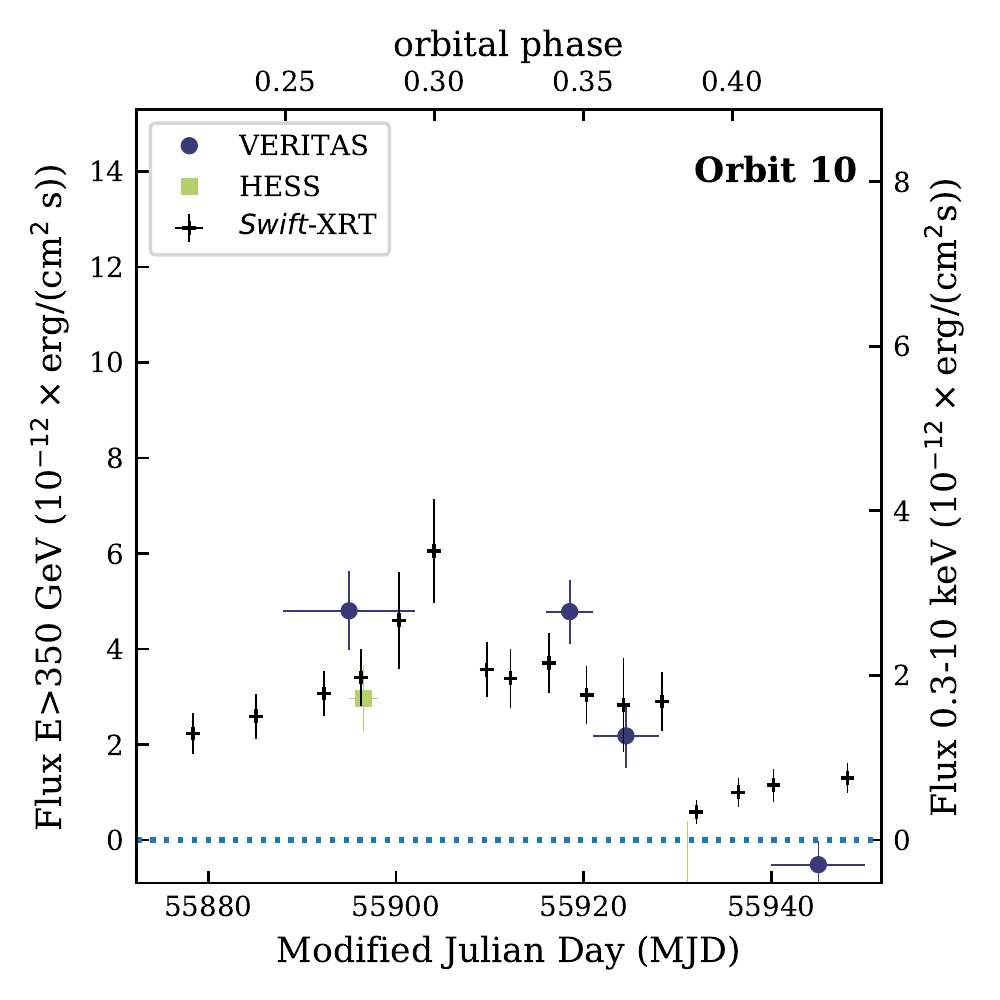}{0.5\textwidth}{}
\fig{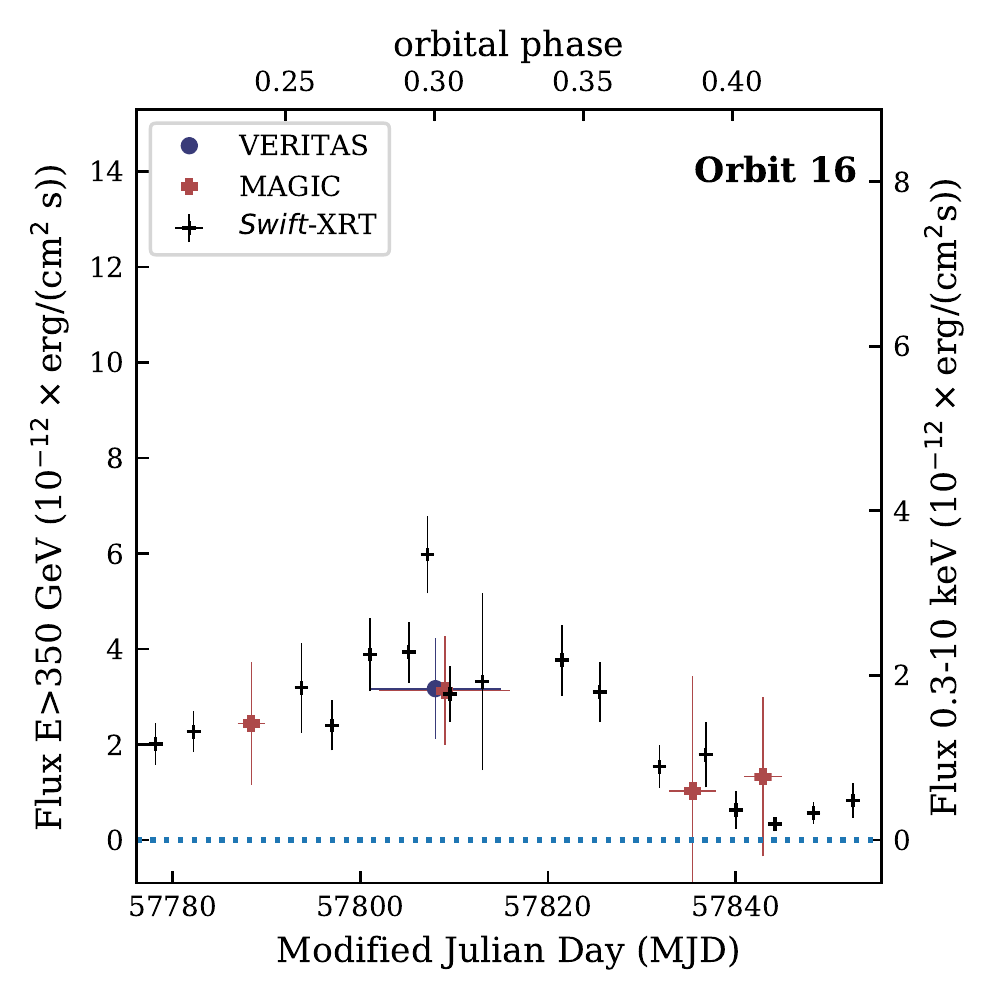}{0.5\textwidth}{}}
\caption{\label{fig:LC-detail}
Detailed light curves for four orbits with deep X-ray and gamma-ray coverage in gamma rays (H.E.S.S., MAGIC, VERITAS, $>$350\,GeV; left axis) and X-rays (NuSTAR, \emph{Swift}-XRT, 0.3-10\,keV; right axis).
Orbits are numbered following the start of the gamma-ray observations (phase zero of the Orbit 1 is at MJD 52953).
The orbits 9 and 17 (upper two panels) represent high states of the source while the orbits 10 and 16 (lower two panels) represent low states.
Vertical lines indicate statistical uncertainties.
Fluxes are averaged over time intervals indicated by the horizontal lines (for most flux points smaller than the marker size).
An orbital period of 317.3\,days is assumed.
\explain{Changed units of left y-axis to erg/cm2/s}}
\end{figure*}

Four orbits in the datasets presented provide reasonably good X-ray and gamma-ray coverage around orbital phase values of 0.2--0.45, which includes the first maximum and subsequent flux decay. 
Figure \ref{fig:LC-detail} shows the light curves for the four orbits with gamma-ray and X-ray coverage: 9, 10, 16, and 17.
The shapes of the light curves of these four orbits differ notably and suggest the existence of orbit-to-orbit variability.
The short-time scale variability on time scales of 20 days and less has been seen at X-ray energies (see also Figure \ref{fig:light curves}), the observed differences in fluxes between orbits might be a biased view due to observations taking place at slightly different orbital phases.
Observations during orbits 9 (2011, Jan) and 17 (2018, Jan) 
 reveal gamma-ray and X-ray emission
in bright states, comparable to the highest ever observed from this binary. 
The X-ray flux in orbits 10 and 16 is a factor of $\sim1.5$ lower compared to orbits 9 and 17 at orbital phases of 0.35. 
The maximum gamma-ray flux also appears to be lower, although the low cadence of observations in this energy range does not allow a firm statement to be made.

 The bright state in orbit 9 has been reported in previous work by the MAGIC and VERITAS collaborations \citep{Aleksic-2012,Aliu-2014}. 
The gamma-ray flux for orbit 17 reached a level of $(7.0\pm1.3)\times 10^{-12}$\,photons cm$^{-2}$ s$^{-1}$ above 350\,GeV (corresponding to $\approx 7$\% of the flux of the Crab Nebula above the same energy). 
In this same orbit, \h was detected in three individual nights within one week of observations with $>7$\,$\sigma$ statistical significance per night  (see Table \ref{table:EnergySpectrumHighStates}). 

The determination of variability time scales is limited by the cadence of the observations.
For orbit 17, the flux changes from the highest state on MJD 58136 to a flux below the detection limit on MJD 58153 (flux upper limit for MJD 58153: F$_{99\%  \mathrm{CL}} (>350\,\mathrm{GeV})$ $< 2.6\times 10^{-12}$\,photons cm$^{-2}$ s$^{-1}$), indicating a flux decay time shorter than 17\,days.
A similar time scale of roughly 20 days or less, again limited by the cadence and detection statistics of the observations, is observed for orbit 9.

%
%
%
%
%
%
\begin{deluxetable*}{cccccc}
\centering
\tablecolumns{6}
\tablewidth{0pt}
\tablecaption{Summary of observations and spectral analyses at gamma-ray energies for orbits 9 and 17. 
The table lists the results of the  power-law fits to the differential energy spectra obtained for VERITAS and MAGIC.
Errors are 1\,$\sigma$ statistical errors only. 
The systematic errors are 11-20\% on the flux constant and 0.10-0.15 on the spectral index.
\label{table:EnergySpectrumHighStates}}
\tablehead{
\colhead{Time range} &
\colhead{Exposure} &
\colhead{Significance\tablenotemark{a}} &
\colhead{flux normalization} &
\colhead{photon index} &
\colhead{$\chi^2/N$} \\
\colhead{MJD} &
\colhead{(h)} &
\colhead{($\sigma$)} &
\colhead{constant at 0.5 TeV} & 
\colhead{} &
\colhead{} \\
\colhead{} &
\colhead{} &
\colhead{(pre-trial)} &
\colhead{[cm$^{-2}$s$^{-1}$TeV$^{-1}$]} &
\colhead{} &
\colhead{}
}
\startdata
\hline
\multicolumn{6}{c}{VERITAS observations of orbit 9 (Jan 2011)} \\
\hline
55585-55600 & 11.3  & 6.1 & $(3.8\pm0.9)\times 10^{-12}$ & $2.71\pm0.30$ & $2.7/3$ \\
55600-55603 & 8.7 & 11.6 & $(8.3\pm1.2)\times 10^{-12}$ & $2.69\pm0.17$ & $4.5/4$\\
\hline
\multicolumn{6}{c}{MAGIC observations of orbit 9 (Jan 2011)} \\
\hline
55585-55600 & 4.5  & 5.9 & $(4.9\pm0.9)\times 10^{-12}$ & $2.48\pm0.20$ & $4.2/3$ \\
\hline
\multicolumn{6}{c}{VERITAS observations of orbit 17 (Jan 2018)} \\
\hline
58136 & 1.8 & 7.5 & $(11.9 \pm 2.6) \times 10^{-12}$ & $2.22 \pm 0.24$ & $3.3/4$ \\
58141 & 2.5 & 7.7 & $(9.2 \pm 2.2) \times 10^{-12}$  & $2.20 \pm 0.24$ & $2.0/4$ \\
58142 & 3.6 & 7.8 & $(11.6 \pm 2.4) \times 10^{-12}$  & $2.56 \pm 0.28$ & $1.5/3$ \\
58143 & 3.1 & 5.8 & $(7.3 \pm 2.0) \times 10^{-12}$ &  $2.15 \pm 0.22$ & $3.2/2$ \\
\enddata
\tablenotetext{a}{Significances are calculated following the maximum likelihood ratio test method proposed in \citet{Li:1983}.}
\end{deluxetable*}

\begin{figure}
\plotone{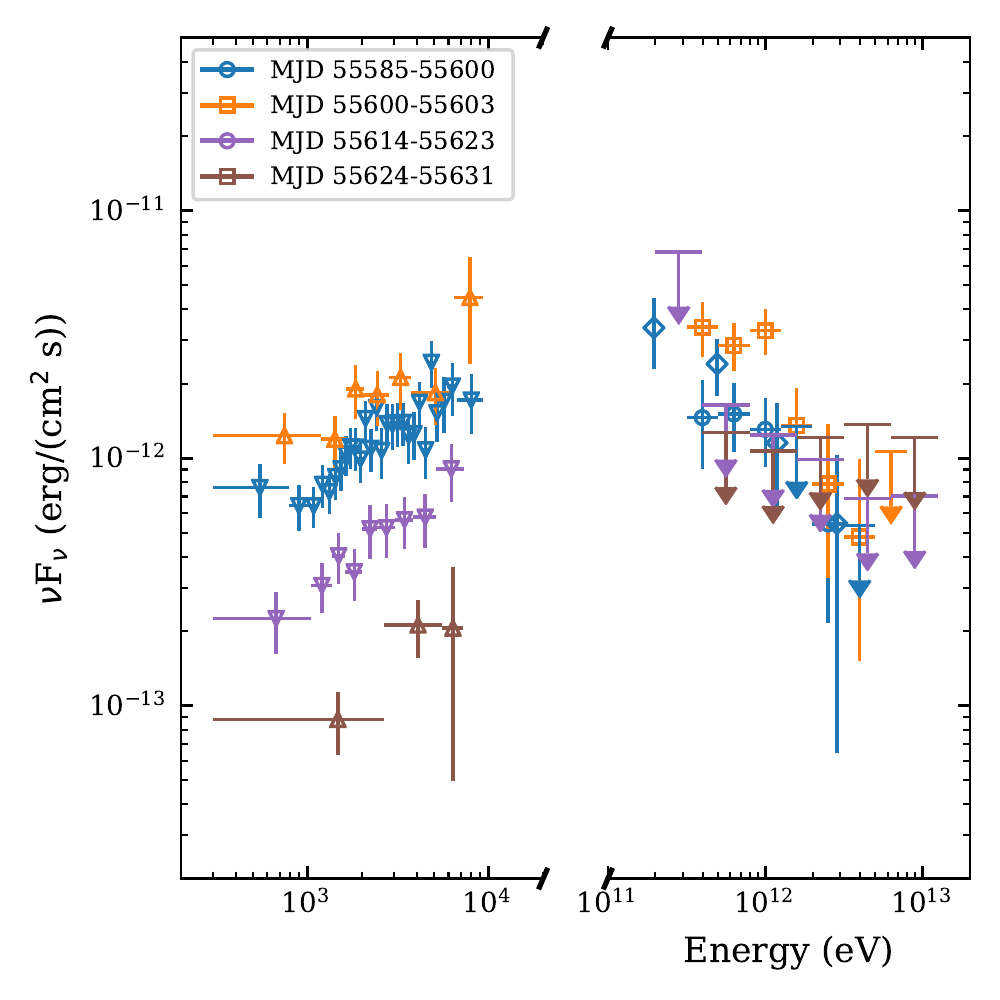}
\plotone{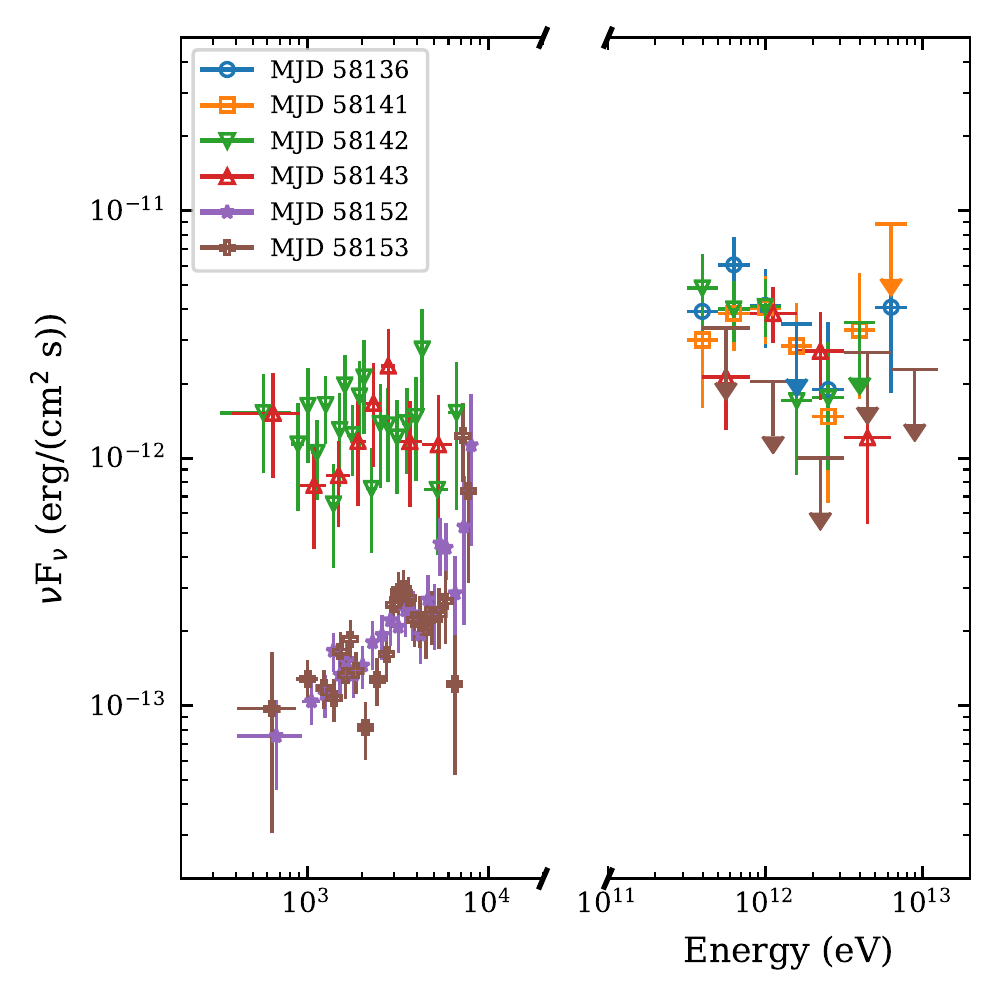}
\caption{\label{fig:SED2-detail}
Spectral energy distribution for \h during two orbital period ranges (Orbit 9: top panel; Orbit 17: bottom panel) with deep coverage in X-ray (\emph{Swift}-XRT) and gamma-ray observations.
Gamma-ray spectra obtained by VERITAS are shown in all figures. 
Orbit 9 also shows results from MAGIC \citep{Aleksic-2012}, indicated by diamond-shaped markers.
Downwards pointing arrows indicate upper limits at the 95\% confidence level.
\explain{Removed white space in figures and changed axes scale}
}
\end{figure}

The spectral energy distributions for orbits 9 and 17 are shown in Figure \ref{fig:SED2-detail} 
and results for spectral fits to the gamma-ray data, assuming power-law functions, are given in Table  \ref{table:EnergySpectrumHighStates}.
Despite the dramatic changes of the overall flux levels, no evidence for variability of the spectral index is detected, within statistical errors, for all six periods of gamma-ray observations.

%
%
%


\section{Discussion}
\label{sec:discussion}

\subsection{VHE results \& Acceleration mechanisms}
\h has been observed at VHE with irregular coverage across 18 orbits, spanning an observational period from 2004 to 2019.
These observations result in a statistically significant detection of non-thermal emission at all orbital phases, except phases
0.4--0.6.
The phase-averaged luminosity of \h above 1\,TeV is $\approx 10^{32}$\,erg s$^{-1}$ (assuming a distance of 1.4\,kpc), making it one of the faintest gamma-ray binaries known to date \citep[see][]{2013A&ARv..21...64D}.
%
The VHE gamma-ray emission extends beyond several TeV (see Figure \ref{fig:TeVSpectra}), indicating the existence of radiating particles with at least similar energies.
The energy range for the detected emission and the spectral shapes of individual spectra are, within statistical uncertainties, independent of the orbital phase of the measurement.
The spectra follow approximately power-law shapes with spectral indeces in the range of 2.3--2.6 (Table \ref{table:EnergySpectrum}), in line with generic expectations from diffusive shock acceleration.
%

In general and independent of the nature of the compact object (black hole or neutron star), the VHE emission can be produced through interactions of accelerated hadrons or leptons with surrounding matter, magnetic, or radiation fields. 
%
The literature (e.g., \cite{2007MNRAS.380..320K}) provides detailed discussions of mechanisms for acceleration and losses (e.g.,  synchrotron radiation, anisotropic inverse Compton scattering, or adiabatic losses) as functions of density and magnetic field strengths in binary systems.
The discussion of the non-thermal processes at work in \h, including particle acceleration and X- and gamma-ray emission and absorption processes, is limited by the uncertainties present in some of the fundamental characteristics of the system. In particular, the unknown nature of the compact object allows different scenarios for the required power source for particle acceleration to be considered (e.g., the interaction of the stellar wind with an accretion-driven relativistic jet or with a pulsar wind).
Equally important are the difficulties to determine the orbital geometry, given that two discrepant solutions with large uncertainties are found in the literature (Figure \ref{fig:OrbitalSolutions}). 
Further uncertainties include the size and orientation of the equatorial disk of the Be star; the mass-loss rates and velocities of both equatorial and isotropic stellar winds; the properties of the accreting object (in the case of a black hole), or the pulsar wind (in the case of a neutron star); and the magnetic field configuration in the acceleration region.
%


\subsection{Flux variability studies}
The 15\,years of \hess, MAGIC, and VERITAS observations of \h, combined with measurements at optical and X-ray energies, suggest multiple patterns of variability at different time scales in the binary system.
The most prominent modulation, interpreted as the orbital period, is consistently seen, within statistical uncertainties, in X-ray and gamma-ray measurements ($P_X=317.3\pm 0.7(stat)\pm 1.5(syst)$; $P_\gamma=316.7\pm 4.4(stat)\pm2.5(syst)$, see Sec.~\ref{sec:period}). 
In both bands, the orbital-phase-folded light curves (Figure \ref{fig:lc-phaseFolded} top and bottom) show two maxima at phases $\Phi\approx 0.3$ and $\Phi\approx 0.6-0.8$, separated by a deep flux-minimum at orbital phases $\Phi\approx 0.4$.
In the VHE band, the first maximum is more pronounced and exhibits, at least for some orbits, a sharp peak with a duration of a few days. 
The locations of the extrema in the X-ray and gamma-ray light curves along the binary orbit are not known, due to the large uncertainties in the orbital solutions, as discussed above.
Assuming orbital parameters as presented in \citet{Casares-2012} (Figure \ref{fig:OrbitalSolutions}, top), the first maximum and the second maximum occur before and after apastron, where the environment around the compact object could likely be the 
least disturbed by the winds of the massive star.
For the  orbital solution proposed by \citet{Moritani-2018}, instead, the first maximum is after apastron and the second maximum shortly after periastron. 
A direct comparison with other gamma-ray binaries is affected by these uncertainties. 
However, it is worth pointing out that the two other long-period systems known (\hbin and \tpsr) show their emission maxima close to or around the periastron passages. 

The analysis of a total of 59 X-ray -- VHE flux pairs reveals a strong correlation between these two energy bands (Figure \ref{fig:XrayGrayCorrelation}, top). 
No time lag is observed (Figure \ref{fig:XrayGrayCorrelation}, bottom), although the grouping of the gamma-ray observations (to gain statistics) prevents us from probing time lags $\tau \leq$10 days.
The X-ray -- VHE correlation suggests that emission in both bands is produced by the same population of accelerated particles. 
One-zone leptonic models can describe the X-ray to TeV spectral distribution and the correlation between these two energy bands \added{in HESS J0632+057}, as seen in \cite{2009ApJ...690L.101H,Aleksic-2012,Aliu-2014}. 
In this scenario, a population of electrons produces X-rays via synchrotron emission and VHE gamma-rays through inverse Compton scattering off thermal photons from the companion Be star. 
It has been shown in~\cite{2020ApJ...888..115A} that a one zone leptonic model can describe well a small set of contemporaneous SED data at hard X-rays and VHE gamma-rays observed with NuSTAR and with VERITAS, respectively. 

One of the main difficulties in modelling \h's SED data lies in accounting for the observed orbit-to-orbit variability. 
\added{Detailed hydrodynamical simulations of \h by \cite{2017MNRAS.471L.150B} and \cite{2018MNRAS.479.1320B} propose a trapping of parts of the shocked pulsar wind by the massive star as a cause of orbit-to-orbit variability. These authors suggest that the peaks of the X-ray and VHE gamma-ray emission are a consequence of the accumulation of the hot shocked plasma injected by the shocked pulsar wind and the consequent disruption of the spiral arm in the periastron–apastron direction. 
It is proposed that the observed drop in the X-ray and gamma-ray fluxes is caused by the disruption of the stellar wind by the pulsar wind.}
In \added{the simpler model of} of~\cite{2020ApJ...888..115A}, this variability is assumed to be due to fluctuations of the electron density at the emission zone, which is treated as a free parameter of the fit. 
This kind of approach can only be applied to contemporaneous X-ray and VHE gamma-ray observations and it is only possible because of the large overlap between the electron energies responsible for the X-rays and VHE gamma-rays in this leptonic scenario. 


For the first time, contemporaneous X-ray and gamma-ray observations have been presented for orbital phases 0.2-0.45 (around first maximum and minimum) for four different orbits (Figure \ref{fig:LC-detail},  \ref{fig:SED2-detail}).
The measurements suggest that flux levels are different between these periods, indicating the existence of orbit-to-orbit variability in the emission pattern.
%
No evidence for a super-orbital period, as observed e.g., in the gamma-ray binary \lsi \citep{2016A&A...591A..76A}, is found.
On shorter integration times, strong variability of the VHE flux on time scales of a few days was seen during 2011 (Orbit 9) and 2018 (Orbit 17) observations at orbital phase $\sim 0.35$ (prior to the local flux minimum; see Fig.~\ref{fig:LC-detail}). 
Similar day-scale variability has been detected during VERITAS observations of the apastron passage of the compact object in the \lsi binary system~\citep{lsi_flares}.

In the GeV band, the weak flux of \h hinders the detection of orbital variability with the \textit{Fermi}-LAT \citep{2017ApJ...846..169L}. 
%
No variability pattern is seen in H$\alpha$ equivalent width, whereas the radial velocity and the H$\alpha$ FWHM increase in the orbital phase range 0.3-0.6 and decrease between 0.7 and 0.3.
The data demonstrate (with low statistical significance) a somewhat higher scatter at $\phi>0.5$, which may indicate a perturbed state of the stellar disk at these orbital phases. 
The correlation coefficients obtained for the different H$\alpha$ parameters (EW, FWHM and centroid velocity) and X-ray or VHE emission do not show significant values for the observed time lags (Figure \ref{fig:XGvsOptical}).
Optical spectroscopy measurements of MWC 148 during and after the time of the high state around 2018, Jan 25 with the RCC telescope at Rozhen NAO, Bulgaria, indicate changes in the structure or size of the equatorial disk of the Be star \citep{2018ATel11233....1S}, although comparable measurements have  been carried out only during  different orbital phases (2018, Jan 25: phase 0.36 and 2014, Jan 23: phase 0.71).
Equivalent width measurements by the Southern African Large Telescope \citep{SALT2018} confirm this: pointings with a measured $|$EW$|$ of $>40$\,\AA \ in Feb 2018 indicate a larger circumstellar disk compared to that at the same phase in earlier cycles (e.g., $|$EW$|$ of 28 March 2017 was $\approx 30$\,\AA; see also measurements by \cite{Casares-2012,2015ApJ...804L..32M}).

\subsection{Comparison with other binaries and model descriptions}
Similarities of the emission pattern observed for \h with those of other gamma-ray binaries are notable.
Similar correlations between X-rays and VHE emission have been found during an outburst of \lsi \citep{2009ApJ...706L..27A}, but not consistently across different orbits \citep{2011ApJ...738....3A}.
However, many short-period systems like \lsi are characterised by single-maximum light curves~\citep{2013A&ARv..21...64D}, while long-period gamma-ray binaries (e.g., \hbin, \tpsr) show a similar double-peaked structure in X- and gamma-ray emission (see e.g., \citet{2013A&ARv..21...64D,2018ApJ...867L..19A, 2020A&A...633A.102H}) as observed in \h. 
%
\cite{Moritani-2018} and \cite{2019AN....340..465M} compare the emission patterns of \h with those of \hbin and relate the double-peak structure of the orbital-phase-folded light curves in the keV-VHE bands in both systems to the inclined disk geometry and the change of environmental parameters during the first and second crossing of the circumstellar disk by the compact object.
The higher ambient density and increased magnetic field strengths in the stellar equatorial wind zone lead to enhanced particle acceleration via wind-wind interaction and possibly increased radiation efficiency due to an increased energy loss timescale for the electron population. Close to periastron, when the compact object leaves the disk, acceleration and radiation efficiency decrease leading to the dip in the keV and VHE light curves.

\replaced{\cite{Tokayer-2021} obtain a description of the orbital variability pattern in X-rays by using a similar leptonic model based on intra-binary shock emission, but with the following modifications: 
two electron populations are assumed, where one moves slowly in the shock and the other is accelerated along the shock flow; beaming effects of the accelerated component in the cone-shaped shock front lead to the enhanced flux state around orbital phases of 0.7-0.9 (when the flow direction and the line-of-sight aligns); 
the inclined stellar disk with respect to the orbital plane leads to enhanced absorption of the X-ray and gamma-ray emission; the change of the magnetic fields at the intra-binary shock position due to the eccentric orbit and the resulting change in acceleration efficiency provide favorable conditions for the high emission state around phases 0.3-0.4. 
Note that the scenario of~\cite{Tokayer-2021} relies on the assumption of a particular orbital geometry  which is derived from the shape of the X-ray light curve and is independent from the ones derived by~\cite{Casares-2012} and~\cite{Moritani-2018}.}
{\cite{An_2017} obtain a quantitative description of the orbital variability pattern in X-rays of the binary \fgl, which shows a similar double-peaked structure as observed in \h. 
The model used a simple leptonic model based on intra-binary shock emission as discussed in the previous section, but with the following modifications: 
two electron populations are assumed, where one moves slowly in the shock and the other is accelerated along the shock flow; beaming effects, as proposed by \cite{2010A&A...516A..18D} are added to  the accelerated component in the cone-shaped shock front. Applying this model to \h, these components would lead to the enhanced flux state around orbital phases of 0.7-0.9 (when the flow direction and the line-of-sight aligns). 
The stellar disk of the Be star in \h is the main difference compared to the diskless O-star in \fgl and might lead to enhanced absorption of the X-ray and gamma-ray emission. Additional changes of the magnetic fields at the intra-binary shock position due to the eccentric orbit and the resulting change in acceleration efficiency might be important to provide favourable conditions for the high emission state around phases 0.3-0.4 \citep{Tokayer-2021}.}

An alternative scenario to explain the variability pattern of \h could be provided
by flip-flop models. 
\citet{2012ApJ...744..106T} and \citet{2012ApJ...756..188P} proposed this scenario to explain the emission from \lsi.
%
\replaced{In this scenario, the compact object is a pulsar (see also Mochol, Moritani)}{In this scenario, the compact object is a pulsar. Different authors support the pulsar nature of the compact object in \h \citep[see ][]{Mochol2014, 2015ApJ...804L..32M}} as might be the case for the binaries \lsi, 
\lsf~\citep{ls5039_pulsar} or, as in the confirmed pulsar binaries \hbin~\replaced{\citep{Johnston1992}}{\citep{Johnston1992,1999APh....10...31K}} and \tpsr~\citep{Abdo2009,2009ApJ...705....1C}.
%
In the
model proposed by \citet{2012ApJ...744..106T} for \lsi, the pulsar switches from a propeller regime
at periastron to an ejector regime at apastron, which is triggered by changes in the mass-loss rate
of the companion star. Hence, for periods when the disk is at its largest, the propeller regime could
be active also at apastron, reducing or impeding VHE emission and explaining the long-term variability in the
binary. 
It should be noted that a recent Astronomer's Telegram \citep{2021ATel14297....1W} reported on the detection of a periodicity from \lsi
at a period 269.196 ms.
This period is almost exactly that marked as the preferred one for the flip-flop model to work
(see Figure 4 of \citet{2012ApJ...756..188P}).
In this model, optical and VHE emission are predicted to be anti-correlated. \citet{2015ApJ...804L..32M} argued that a similar mechanism could work in \h, although the orbital phases
in which the gamma-ray emission occurs are different due to the geometry of the system. In the
case of \h, the strong gas pressure will overcome the pulsar-wind pressure at periastron, suppressing VHE emission (assuming the ephemerides of \citet{Casares-2012}). The second
minimum will take place at apastron, where the photon field and magnetic field density are low.
In the case of \lsi, \cite{2016A&A...591A..76A} found the VHE periodicity as predicted, but no
hints for (anti-)correlation between optical and VHE are observed. The latter could be the result
of the strong, short-timescale intra-day variation displayed by the H$\alpha$ fluxes. In \h,
a similar situation can be at work, since both the contemporaneity of the observations (5-day time
span) and the integration times are large, even more than for the case of \lsi --which were already problematic.
The lack
of (anti-)correlation between optical and high-energy emission (X- and gamma rays) observed in
\h can also be explained by the difference in the integration times required in
these frequencies, which is in agreement with the results found by \cite{2016A&A...591A..76A} for the case
of \lsi.

Regardless of the assumed model and the nature of the compact object, two additional processes can affect the orbital modulation of VHE emission -- gamma-gamma absorption processes and the anisotropic nature of inverse-Compton radiation.
%
Close to periastron, the increased soft photon density can produce severe gamma-gamma absorption losses which are potentially able to explain dips in the lightcurves of several gamma-ray binary systems~\citep[see e.g.][]{2005ApJ...634L..81B,2006ApJ...643.1081D,2017ApJ...837..175S}. 
Anisotropic inverse-Compton effects are important for close to edge-on orientations, as suggested by~\citet{2006MNRAS.368..579B}. 
In the context of \lsf, \citet{khangulian08,dubus08a} have shown that a model taking into account both effects can reproduce well the observed orbital light curve, producing somewhat different results to the isotropic inverse-Compton case spectrum at inferior and superior conjunctions.
The observed cut-off in the energy spectrum for orbital phase bins 0.2-0.4 at $E_c = 1.75\pm0.38$\,TeV can have multiple, possibly correlated, origins:
limiting properties of the shock determining the acceleration efficiency and the maximum energy of accelerated particles;
photon fields determining the absorption of multi-TeV gamma rays; or the  effective temperature distribution of the low-energy target photons  (e.g., due to substantial contribution of the stellar disk), which impact the energy of the gamma rays obtained from inverse-Compton processes.
%

%
An additional channel of information which can allow the proposed models to be discriminated against each other is the observation of flux variability patterns and outbursts at shorter time scales than the orbital period, as shown in Figure~\ref{fig:LC-detail}.
Such variability suggests a rapid change of the environmental properties leading to particle acceleration, or of the surrounding medium impacting the VHE emission, possibly because of a peculiar geometry in the system (e.g., beaming, as proposed for \hbin in e.g., \citet{khangulian11_psrb,chernyakova20_psrb}).
In terms of the discussed models, however, short-term variability can be accommodated for by several theoretical explanations. 
In the flip-flop scenario these timescales could be interpreted as the timescales on which the pulsar wind is quenched due to increased circumstellar disk density. 
In the \hbin-like model, the same timescales would correspond to the time on which the compact object intersects the disk's region of influence and exits from it close to periastron.
Additionally, stellar winds of massive stars can exhibit strong density variations. 
Flux variability can be connected to the clumpiness of the equatorial or isotropic stellar wind, which can lead to keV-VHE flux changes, see e.g.~\citet{clumpy_wind, lsi_flares}. 
In this case the variability time can allow us to estimate the characteristic size of the corresponding high-density structure.
Deeper, simultaneous multiwavelength (keV-GeV-VHE) observations of \h with current or future facilities can help to clarify these points. 
Such observations may allow one to find the shortest timescales of variability in each band and more firmly establish the similarities and differences of the variability patterns. 
Additional future H$\alpha$ observations used as tracer for the radius of the circum-stellar disk are also helpful, since the data set shown in this paper is rather sparse.

\section{Summary and conclusions}
\label{sec:summary}

The results of the deepest study of the gamma-ray binary \h at TeV energies with H.E.S.S., MAGIC and VERITAS,  comprising a total of 450~\,h of data spanning almost 15\,years, are presented. This multi-year campaign is embedded in a multi-wavelength context, which includes X-ray (\textit{Swift}-XRT, Chandra, \textit{XMM}-Newton, NuStar and Suzaku) and optical H$\alpha$ observations. The results of the spectral and temporal analyses are summarized as follows:
\begin{itemize}
    \item For the first time, the orbital period at TeV energies was determined, yielding a value of $316.7\pm4.4$\,days. This solution is in agreement with the $317.3\pm0.7$-day period  derived from the latest \emph{Swift}-XRT X-ray data set.  
    
    \item The light curve and spectral energy distribution along the orbit was characterized. In the phase-folded light curve, two well-differentiated peaks are visible, a dip phase and a broader plateau phase. The VHE SEDs for all of these phases (except the dip phase, where only upper limits could be derived) are generally characterized as power-laws, showing no variability in the spectral slope within statistical errors. Only the spectrum measured with VERITAS during the phases 0.2--0.4 favors a power-law with exponential cutoff at 1.75\,TeV.
    
    \item The strong correlation between X-rays and gamma rays suggest a common origin of the radiation, indicating the existence of a single population of particles. An indication for an X-ray source partially not related to the gamma-ray emission was however found.
    
    \item The lack of correlation between H$\alpha$ and X-ray or gamma-rays might point towards a negligible role of the disk of the Be star in the modulation of the non-thermal emission, but is possibly an effect of the fast variability of H$\alpha$ compared with the sparse overlap of the datasets at different energies. If the H$\alpha$ spectra change on a time scale of days, much shorter than the grouped gamma-ray time scales, one can not measure a possible correlation with the sensitivity of current VHE/TeV instruments.

    \item The ratio of gamma-ray to X-ray flux underlines the equality or even dominance
    of the gamma-ray energy range for the emission of \h.
    
    \item Two outbursts during orbits 9 (2011, January) and 17 (2018, January) revealed enhanced gamma-ray and X-ray emission comparable to the highest flux ever observed from this binary. 
    Furthermore, a flux decay \added{time} of roughly 20\,days or less was detected for two orbits. Contemporaneous H$\alpha$ data taken on 2018, January 25 indicate that the size of the circumstellar disk had increased during those days, suggesting that the decretion disk was larger and its structure had changed.  

\end{itemize}

Looking forward, it is obvious that deeper, simultaneous multi-wavelength (H$\alpha$ and keV-GeV-TeV) observations of \h with current and/or future more-sensitive instruments are required to characterise its emission. The determination of the orbital geometry of the system is of utmost importance and requires a coordinated multi-year optical campaign. Finally, the wealth of data presented for \h 
awaits theoretical modelling taking consistently all aspects of the spectral and temporal measurements into account.

\section*{Availability of Data}

The data sets generated and analysed during the current study are available through the websites of the \h\footnote{\url{https://www.mpi-hd.mpg.de/hfm/HESS/pages/publications/auxiliary/auxinfo_hessj0632_HMVdata.html}}, MAGIC\footnote{\url{http://vobs.magic.pic.es/fits/}}, and VERITAS\footnote{\url{https://github.com/VERITAS-Observatory/VERITAS-VTSCat}} Instruments and from the Zenodo data repository\footnote{\url{https://doi.org/10.5281/zenodo.5157848}}.

\acknowledgments

The support of the Namibian authorities and of the University of Namibia in facilitating the construction and operation of \hess is gratefully acknowledged, as is the support by the German Ministry for Education
and Research (BMBF), the Max Planck Society, the German Research Foundation (DFG), the Helmholtz Association,
the Alexander von Humboldt Foundation, the French Ministry of Higher Education, Research and Innovation, the Centre
National de la Recherche Scientifique (CNRS/IN2P3 and CNRS/INSU), the Commissariat \`a l'\'energie atomique et aux \'energies alternatives (CEA), the U.K. Science and Technology Facilities Council (STFC), the Knut and Alice Wallenberg
Foundation, the National Science Centre, Poland grant no. 2016/22/M/ST9/00382, the South African Department of Science
and Technology and National Research Foundation, the University of Namibia, the National Commission on Research,
Science \& Technology of Namibia (NCRST), the Austrian Federal Ministry of Education, Science and Research and the
Austrian Science Fund (FWF), the Australian Research Council (ARC), the Japan Society for the Promotion of Science
and by the University of Amsterdam. We appreciate the excellent work of the technical support staff in Berlin, Zeuthen,
Heidelberg, Palaiseau, Paris, Saclay, T\"ubingen and in Namibia in the construction and operation of the equipment. This work
benefited from services provided by the \hess Virtual Organisation, supported by the national resource providers of the
EGI Federation.

The MAGIC collaboration would like to thank the Instituto de Astrof\'{\i}sica de Canarias for the excellent working conditions at the Observatorio del Roque de los Muchachos in La Palma. The financial support of the German BMBF, MPG and HGF; the Italian INFN and INAF; the Swiss National Fund SNF; the ERDF under the Spanish Ministerio de Ciencia e Innovaci\'{o}n (MICINN) (PID2019-104114RB-C31, PID2019-104114RB-C32, PID2019-104114RB-C33, PID2019-105510GB-C31,PID2019-107847RB-C41, PID2019-107847RB-C42, PID2019-107988GB-C22); the Indian Department of Atomic Energy; the Japanese ICRR, the University of Tokyo, JSPS, and MEXT; the Bulgarian Ministry of Education and Science, National RI Roadmap Project DO1-268/16.12.2019 and the Academy of Finland grant nr. 320045 is gratefully acknowledged. This work was also supported by the Spanish Centro de Excelencia ``Severo Ochoa'' (SEV-2016-0588, CEX2019-000920-S), the Unidad de Excelencia ``Mar\'{\i}a de Maeztu'' (CEX2019-000918-M, MDM-2015-0509-18-2) and by the CERCA program of the Generalitat de Catalunya; by the Croatian Science Foundation (HrZZ) Project IP-2016-06-9782 and the University of Rijeka Project 13.12.1.3.02; by the DFG Collaborative Research Centers SFB823/C4 and SFB876/C3; the Polish National Research Centre grant UMO-2016/22/M/ST9/00382; and by the Brazilian MCTIC, CNPq and FAPERJ.


VERITAS is supported by grants from the U.S. Department of Energy Office of Science, the U.S. National Science Foundation and the Smithsonian Institution,  by NSERC in Canada, and by the Helmholtz Association in Germany. We acknowledge the excellent work of the technical support staff at the Fred Lawrence Whipple Observatory and at the collaborating institutions in the construction and operation of the instrument.
This research used resources provided by the Open Science Grid, which is supported by the National Science Foundation and the U.S. Department of Energy's Office of Science, and resources of the National Energy Research Scientific Computing Center (NERSC), a U.S. Department of Energy Office of Science User Facility operated under Contract No. DE-AC02-05CH11231. 

DFT acnowledges support by grants PGC2018-095512-B-I00, SGR2017-1383, and AYA2017-92402-EXP.

This research has made use of data obtained from the Chandra Data Archive and the Chandra Source Catalog; it has made use of data obtained from the Suzaku satellite, a collaborative mission between the space agencies of Japan (JAXA) and the USA (NASA); it is based on observations obtained with XMM-Newton, an ESA science mission with instruments and contributions directly funded by ESA Member States and NASA; it made use of data from the NuSTAR mission, a project led by the California Institute of Technology, managed by the Jet Propulsion Laboratory, and funded by the NASA.

\software{Astropy \citep[version 3.01,][]{astropy-2013, astropy-2018},
               CIAO \citep[v.4.9,][]{2006SPIE.6270E..1VF}
               Eventdisplay \citep[v4.83,][]{Maier-2017},
               HEAsoft \citep[v6.22,][]{2014ascl.soft08004N},
               IRAF \citep{1986SPIE..627..733T},
               PyAstronomy \citep[version 0.13,][]{pya},
               xrtpipeline (v.0.13.4),
               XSPEC \citep[v12.9.1m,][]{1996ASPC..101...17A}}

\vspace{5mm}
\facilities{Swift,  XMM-Newton, CXO, Suzaku, NuSTAR, MAGIC, H.E.S.S., VERITAS, CFHT, OAO:1.88m}

\def\aj{AJ}%
\def\actaa{Acta Astron.}%
\def\araa{ARA\&A}%
\def\apj{ApJ}%
\def\apjl{ApJ}%
\def\apjs{ApJS}%
\def\ao{Appl.~Opt.}%
\def\apss{Ap\&SS}%
\def\aap{A\&A}%
\def\aapr{A\&A~Rev.}%
\def\aaps{A\&AS}%
\def\azh{AZh}%
\def\baas{BAAS}%
\def\bac{Bull. astr. Inst. Czechosl.}%
\def\caa{Chinese Astron. Astrophys.}%
\def\cjaa{Chinese J. Astron. Astrophys.}%
\def\icarus{Icarus}%
\def\jcap{J. Cosmology Astropart. Phys.}%
\def\jrasc{JRASC}%
\def\mnras{MNRAS}%
\def\memras{MmRAS}%
\def\na{New A}%
\def\nar{New A Rev.}%
\def\pasa{PASA}%
\def\pra{Phys.~Rev.~A}%
\def\prb{Phys.~Rev.~B}%
\def\prc{Phys.~Rev.~C}%
\def\prd{Phys.~Rev.~D}%
\def\pre{Phys.~Rev.~E}%
\def\prl{Phys.~Rev.~Lett.}%
\def\pasp{PASP}%
\def\pasj{PASJ}%
\def\qjras{QJRAS}%
\def\rmxaa{Rev. Mexicana Astron. Astrofis.}%
\def\skytel{S\&T}%
\def\solphys{Sol.~Phys.}%
\def\sovast{Soviet~Ast.}%
\def\ssr{Space~Sci.~Rev.}%
\def\zap{ZAp}%
\def\nat{Nature}%
\def\iaucirc{IAU~Circ.}%
\def\aplett{Astrophys.~Lett.}%
\def\apspr{Astrophys.~Space~Phys.~Res.}%
\def\bain{Bull.~Astron.~Inst.~Netherlands}%
\def\fcp{Fund.~Cosmic~Phys.}%
\def\gca{Geochim.~Cosmochim.~Acta}%
\def\grl{Geophys.~Res.~Lett.}%
\def\jcp{J.~Chem.~Phys.}%
\def\jgr{J.~Geophys.~Res.}%
\def\jqsrt{J.~Quant.~Spec.~Radiat.~Transf.}%
\def\memsai{Mem.~Soc.~Astron.~Italiana}%
\def\nphysa{Nucl.~Phys.~A}%
\def\physrep{Phys.~Rep.}%
\def\physscr{Phys.~Scr}%
\def\planss{Planet.~Space~Sci.}%
\def\procspie{Proc.~SPIE}%
\let\astap=\aap
\let\apjlett=\apjl
\let\apjsupp=\apjs
\let\applopt=\ao
\bibliographystyle{aasjournal}
\bibliography{bibliography}

\begin{appendix}
\label{appendix:appendix}

\section{Orbital period determination}
\label{appendix:orbitalPeriod}

Literature provides a variety of methods for periodicity analysis of sparse astronomical data \citep[e.g.][]{Graham-2013}.
The performance of the different techniques depend on quality, coverage, and shape of the given light curves.
There is no clear guidance which technique is best for a given data set.
For these reasons, the following methods are evaluated using Monte Carlo-generated light curves:
discrete correlation functions 
\citep[DCF,][]{Edelson-1988,Robertson-2015}\footnote{The implementation of the DCF method in the python package pyDCF (\url{https://github.com/astronomerdamo/pydcf}) is used.}, 
correlation analysis comparing the light curves with a binned-average light curve \citep[PCC,][]{2019AN....340..465M}, 
Lomb-Scargle periodograms \citep{1976Ap&SS..39..447L, 1982ApJ...263..835S}\footnote{The implementation of the Lomb-Scargle method in astropy is used (\url{https://docs.astropy.org/en/stable/timeseries/lombscargle.html})}, and
phase dispersion minimisation method 
\citep[PDM,][]{Stellingwerf-1978}\footnote{The implementation of the PDM method in the python package PyAstronomy (\url{https://github.com/sczesla/PyAstronomy}) is used.}.

%
\begin{figure}[!htb]
\plotone{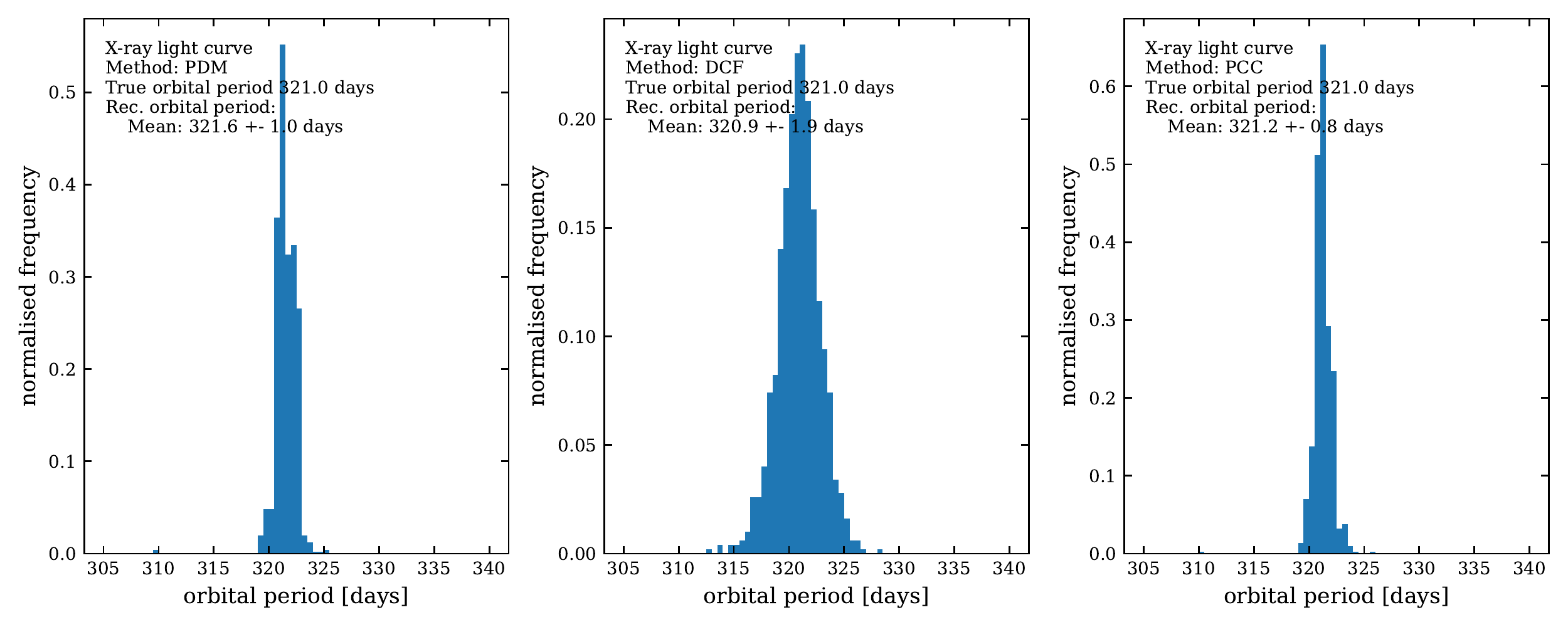}
\plotone{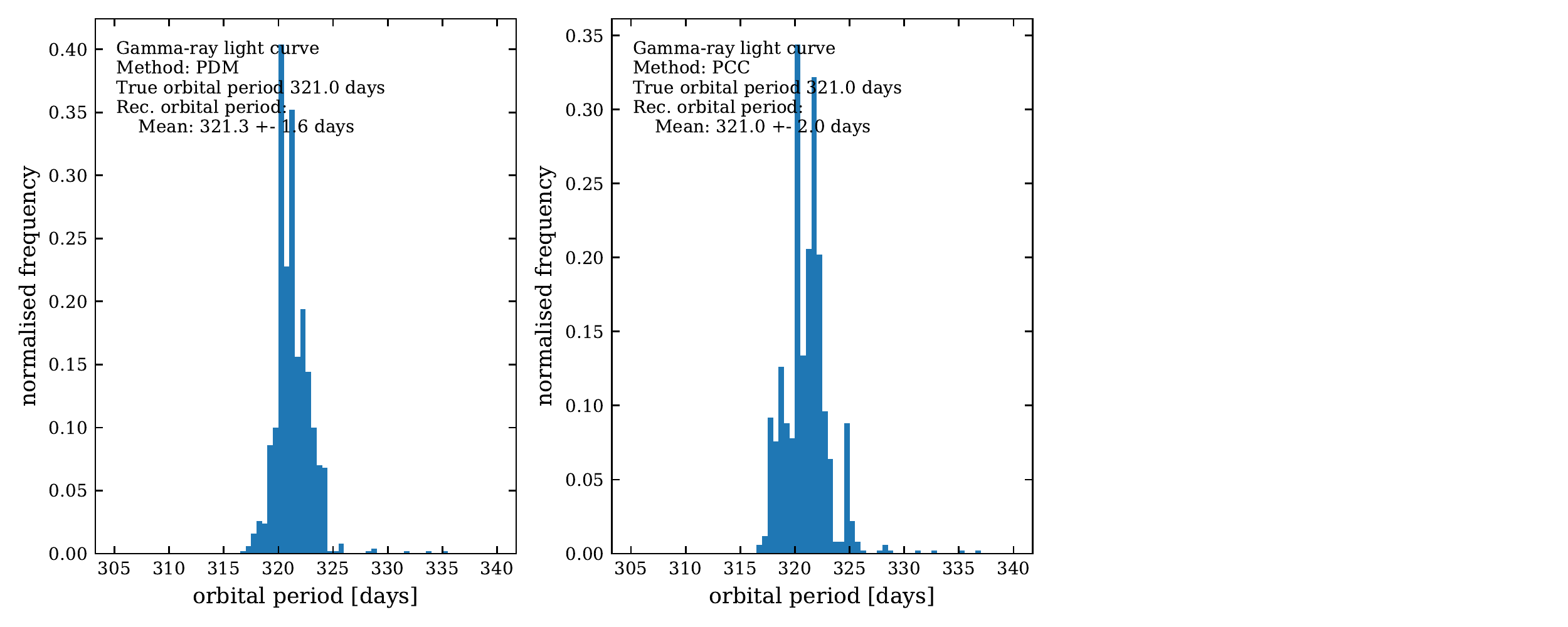}
\caption{\label{fig:toyMCresults}
Outcome of the analysis of 1000 Monte Carlo-generated light curves using the following data sets as templates for the toy MC:
Top: \emph{Swift}-XRT;
Bottom: VERITAS and H.E.S.S..
In all cases a true orbital period of 321\,days is assumed.
Different techniques for periodicity analysis, as indicated in the figure panels, have been applied to the same MC-generated light curves.
Results are given both as mean and 68\% fiducial interval for all methods.}
\end{figure}

The Monte Carlo-generated light curves are based on the phase-binned average profiles of the observed fluxes in X- or gamma rays and are generated assuming the same distribution in time as the measurements, with flux values randomly altered according to the corresponding uncertainties.
As the phase-binned averaged profile removes variability which is not due to statistical uncertainties of the measurement, additional power-law noise following a $(1/f)^{1.2}$ spectrum is added \citep{Timmer-1995}.

%
\begin{figure*}
\gridline{
\fig{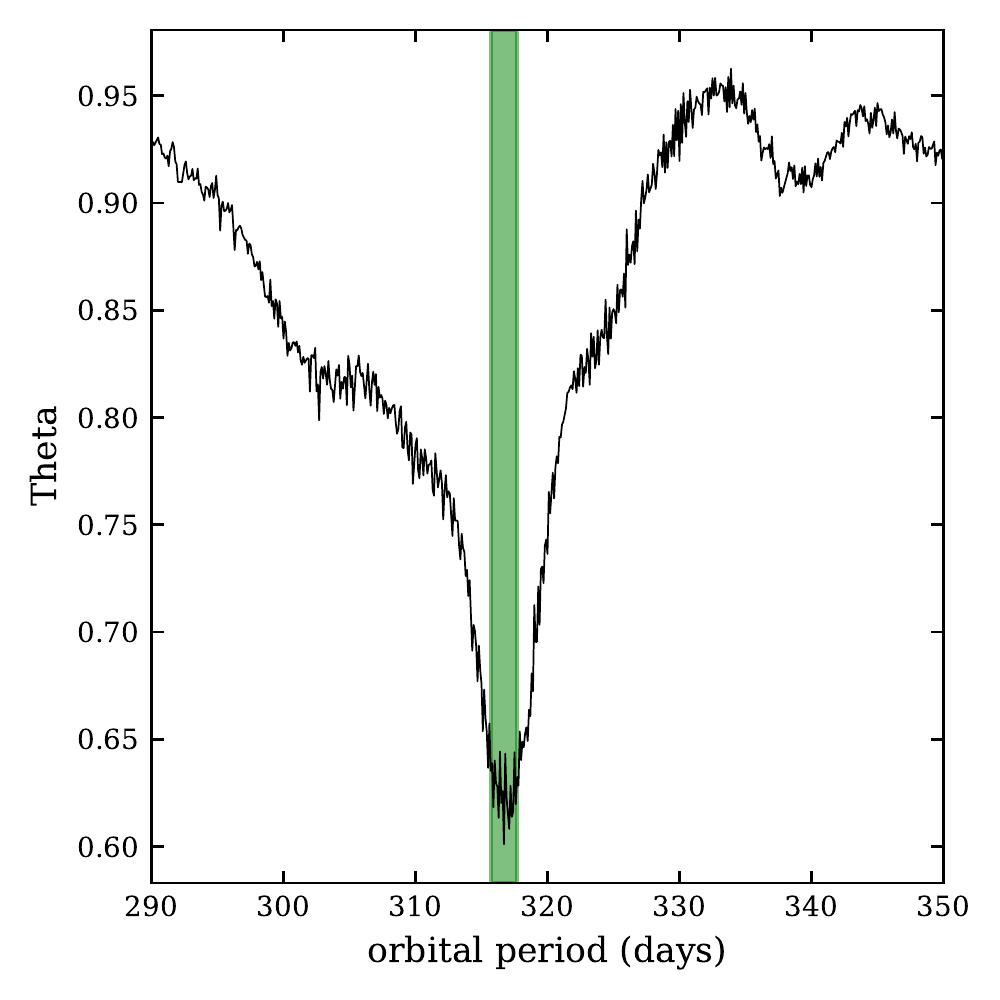}{0.333\textwidth}{(a)}
\fig{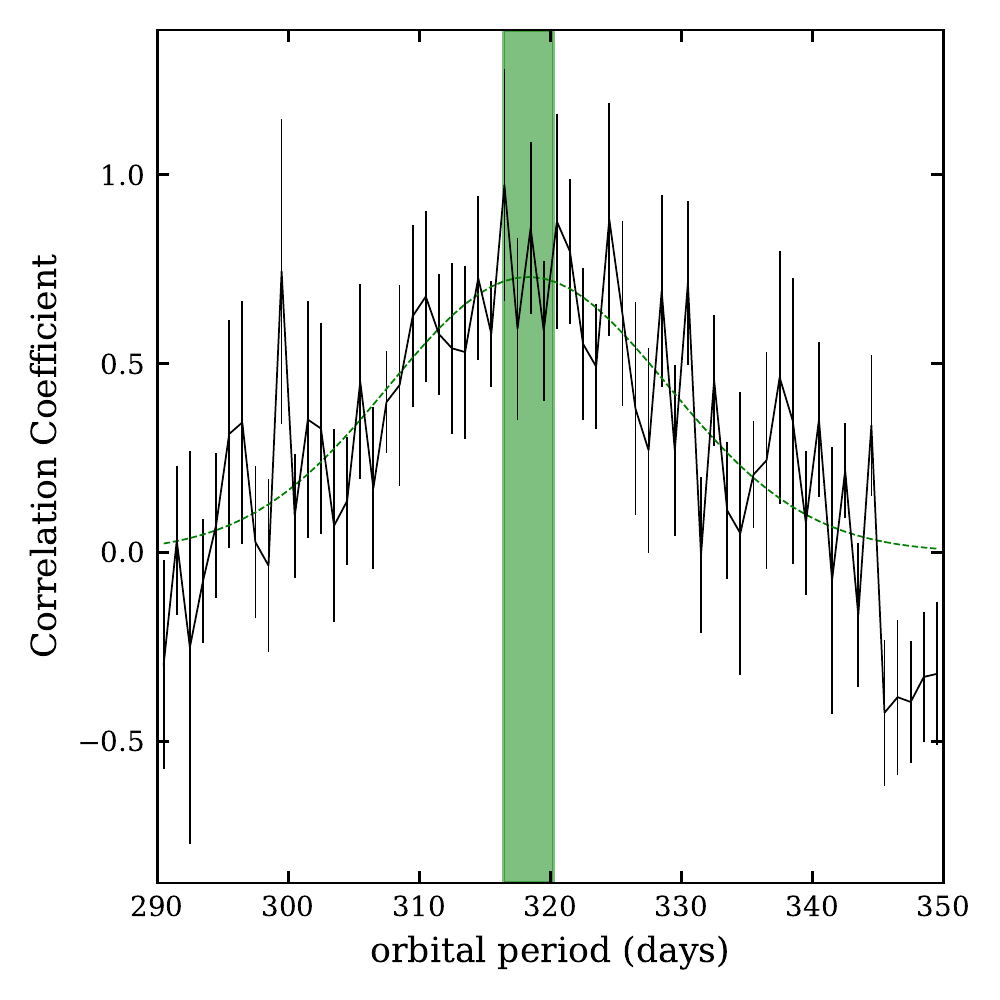}{0.333\textwidth}{(b)}
\fig{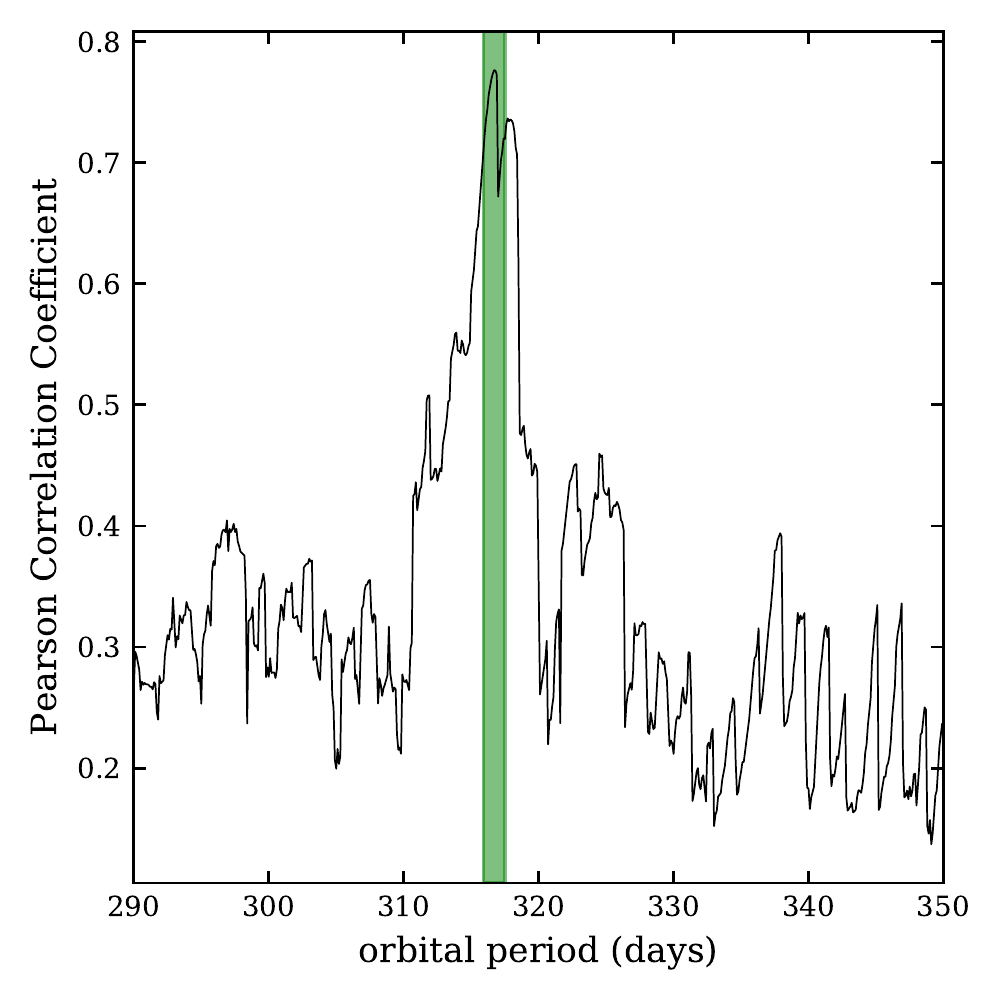}{0.333\textwidth}{(c)}}
\gridline{
\fig{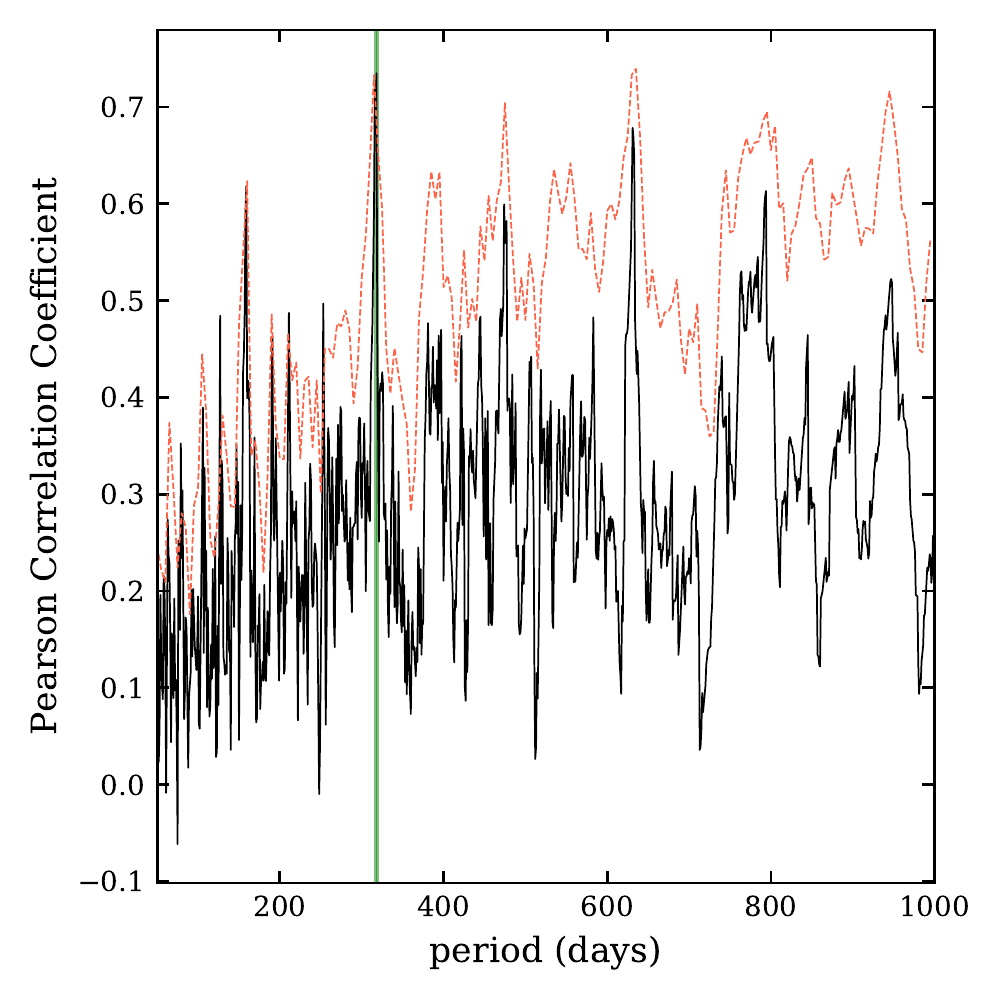}{0.333\textwidth}{(d)}
\fig{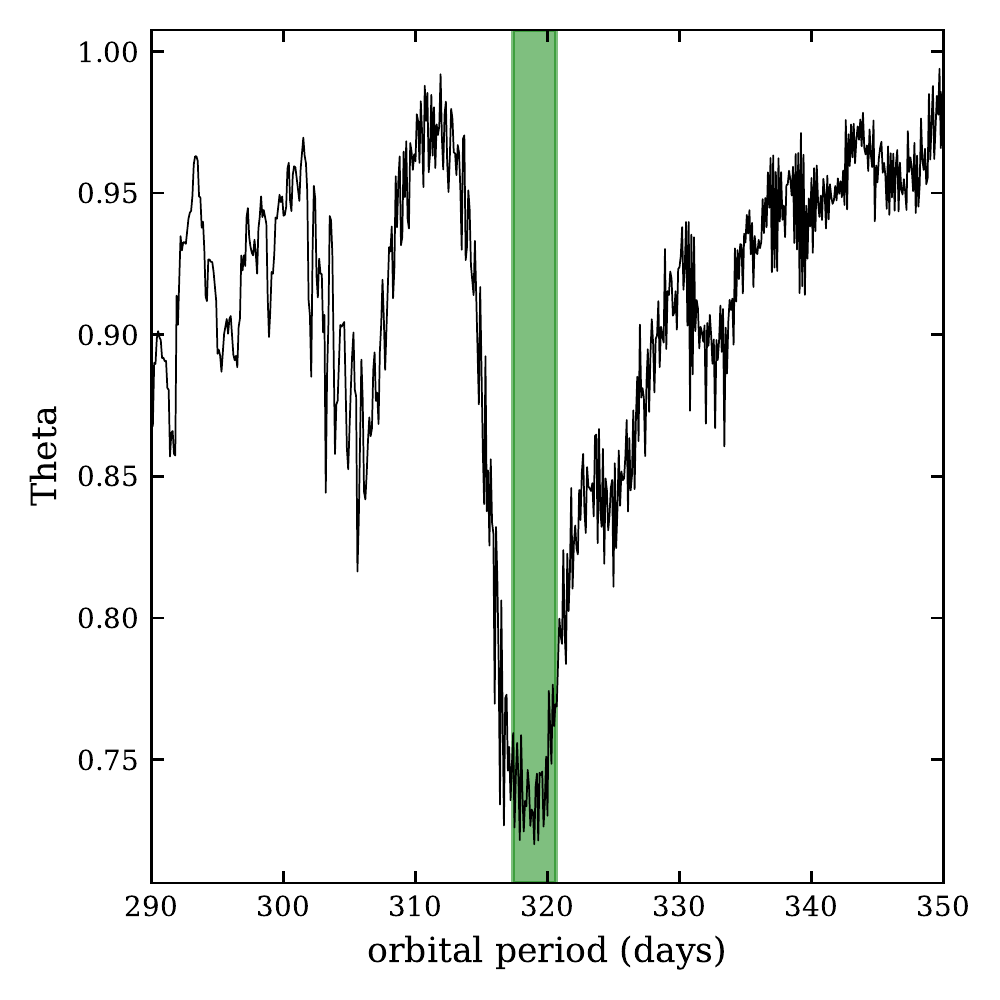}{0.333\textwidth}{(e)}
\fig{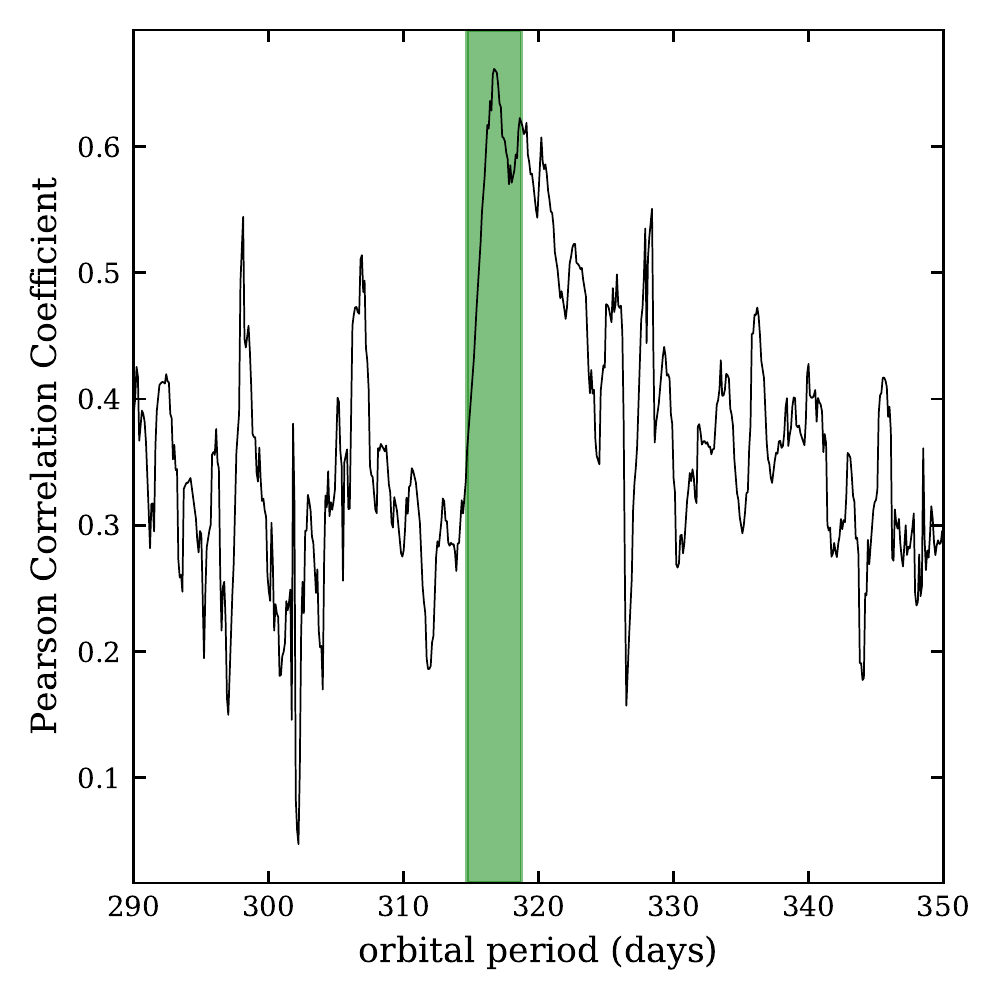}{0.333\textwidth}{(f)}}
\caption{\label{fig:orbitalAnalysis}
Results of the periodicity analysis of \emph{Swift}-XRT (a-d) and gamma-ray (e,f) measurements 
applying different methods:
a) Phase dispersion (PDM; \emph{Swift}-XRT);
b) Discrete correlation coefficient (DCF; \emph{Swift}-XRT);
c) Pearson correlation coefficient (PCC; \emph{Swift}-XRT); 
d) Search for periods other than the orbital period in the range 50 to 1000 days; Pearson correlation coefficient (PCC; \emph{Swift}-XRT); the red-dashed line indicates the 99\% containment level obtained from Monte Carlo sets of the XRT light curve (to take spectral leakage into account, an orbital periodicity of 317.3\,days is assumed in the Monte Carlo sets);
e) Phase dispersion (PDM; gamma-ray energies);
f) Pearson correlation coefficient (PCC; gamma-ray energies).
All coefficients are plotted as a function of orbital period.
The green shaded areas indicate the 68\% fiducial interval obtained by the analysis of MC-generated light curves (see text for details).
}
\end{figure*}

Figure \ref{fig:toyMCresults} shows the results of the different techniques applied to 1000 Monte Carlo-generated light curves based on the \emph{Swift}-XRT and gamma-ray data sets assuming a true orbital period of 321 days.
The search region is restricted to a $\pm 30$\,day-interval around the true orbital period.
This prior is necessary, as the PDM method tends towards reconstructing from a significant fraction ($>20$\%) of light curves an orbital period roughly half the true orbital period.
Applying the method of Lomb-Scargle to the generated light curves leads to inconsistent results, as the largest peak in the periodograms  for most light curves is at half of the true period.
This result is also obtained when applying this method to the \emph{Swift}-XRT data.
Folding the light curve with such short periods leads to inconsistent solutions, as both the PDM and Lomb-Scargle methods ignore its non-sinusoidal shape and in particular the very low fluxes respectively non-detections of the system visible at orbital phases of $\approx 0.4$ (assuming 317.3\,days of orbital period).

All shown methods are able to reconstruct, for the majority of the MC light curves, the true orbital period with an uncertainty of less than  2 days for the \emph{Swift}-XRT and gamma-ray data set.
The DCF method does not provide reliable estimations when applied to the gamma-ray measurements or on the toy MC based on the gamma-ray data.
This is probably due to the sparsity of the gamma-ray dataset and the larger uncertainties of the flux measurements.
Statistical uncertainties for the given datasets are derived from the 68\% fiducial intervals of the corresponding toy MC analysis.
Systematic uncertainties are derived from the largest difference to the expected values (0.6\,days) and the impact of the choice of the bin width on the calculation of the averaged light curves.
Bin widths from 0.025 to 0.1 in orbital phase are tested and the largest difference is used to estimate the contribution to the systematic uncertainties.
The total systematic uncertainty for the orbital period determination is estimated to be $\delta_{sys}^{Swift \ \mathrm{XRT}} = 1.5$\,days and $\delta_{sys}^{\mathrm{Gamma}} = 2.5$\,days.

Confidence limits on the correlation coefficients are obtained in a similar fashion: 1000 toy MC light curves with similar data structure to the one obtained by observations are used to calculate the 95 and 99\% quantiles.
Two  null hypotheses are distinguished: a constant flux is assumed for the determination of the statistical significance of the correlation coefficients for the orbital period analysis (see Figure \ref{fig:orbitalAnalysis}, c)).
For the search for modulation periods other than the orbital period, toy MC light curves as described above, including the average observed orbital modulation pattern, are generated and analysed with the different period determination methods (see Figure \ref{fig:orbitalAnalysis}, d)).


Figure \ref{fig:orbitalAnalysis} shows the results of application of all three techniques on the \emph{Swift}-XRT and gamma-ray light curve measurements (see Figure \ref{fig:light curves}).
A summary of all obtained orbital periods together with those reported in the literature is given in Table \ref{tab:appendix:orbitalPeriod}.


\section{Impact of orbital period uncertainty on phase-folded light curves}
\label{appendix:phaseFolding}

Uncertainties on the orbital period determination might lead to significant differences in the shape of the phase-folded light curves given the long total observation time of $\approx$15\,years for gamma-ray and $\approx$10\,years for X-ray measurements.
To test this, four different orbital periods are assumed: $P_{+}=319.5$\,days, $P_{-}=315.1$, $P_{M}=313$\,days, and $P_{H\alpha}=308$\,days.
$P_{+}$ and $P_{-}$ correspond to a change of the orbital period by 1\,$\sigma$ statistical error plus the  systematic uncertainty;
$P_{M}$ and $P_{H\alpha}$ correspond to the solutions presented in \cite{Moritani-2018}.

\begin{figure}
\gridline{
\fig{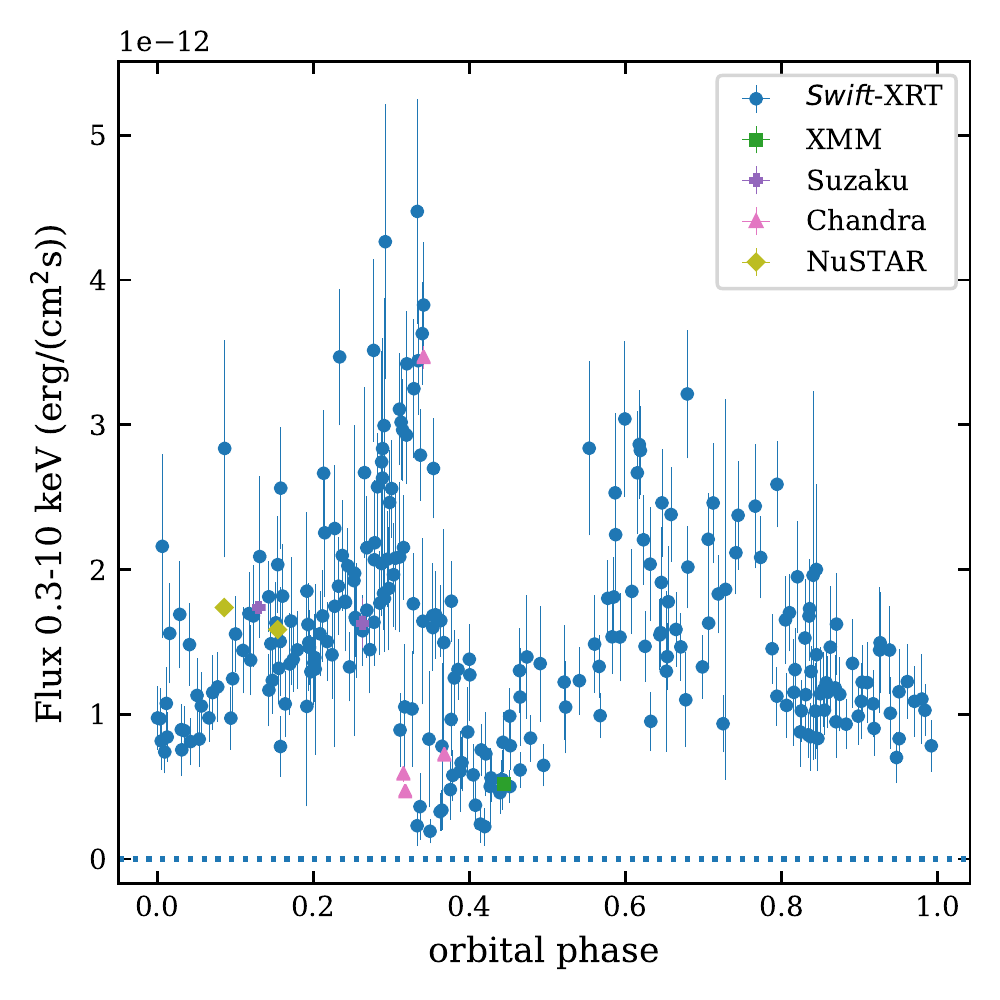}{0.45\textwidth}{(a) X-rays ($P=319.5$ days)}
\fig{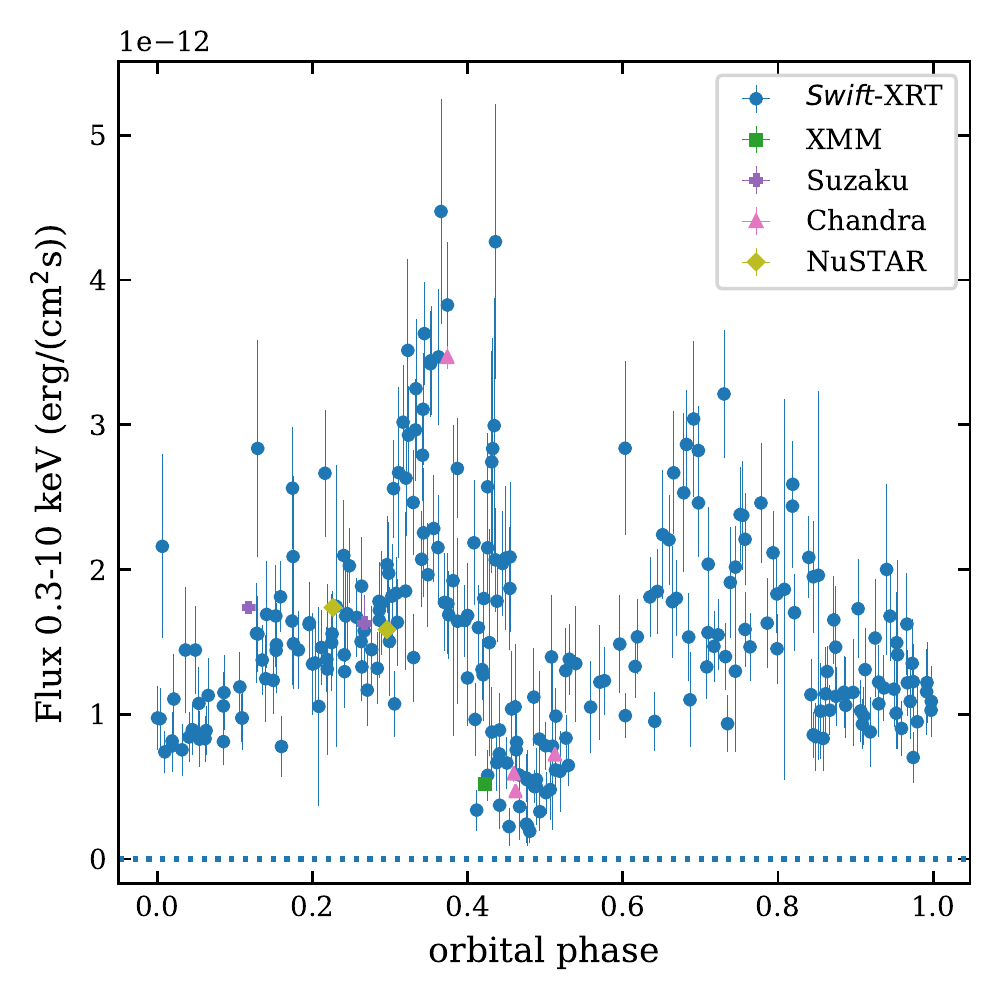}{0.45\textwidth}{(b) X-rays ($P=315.1$ days)}}
\gridline{
\fig{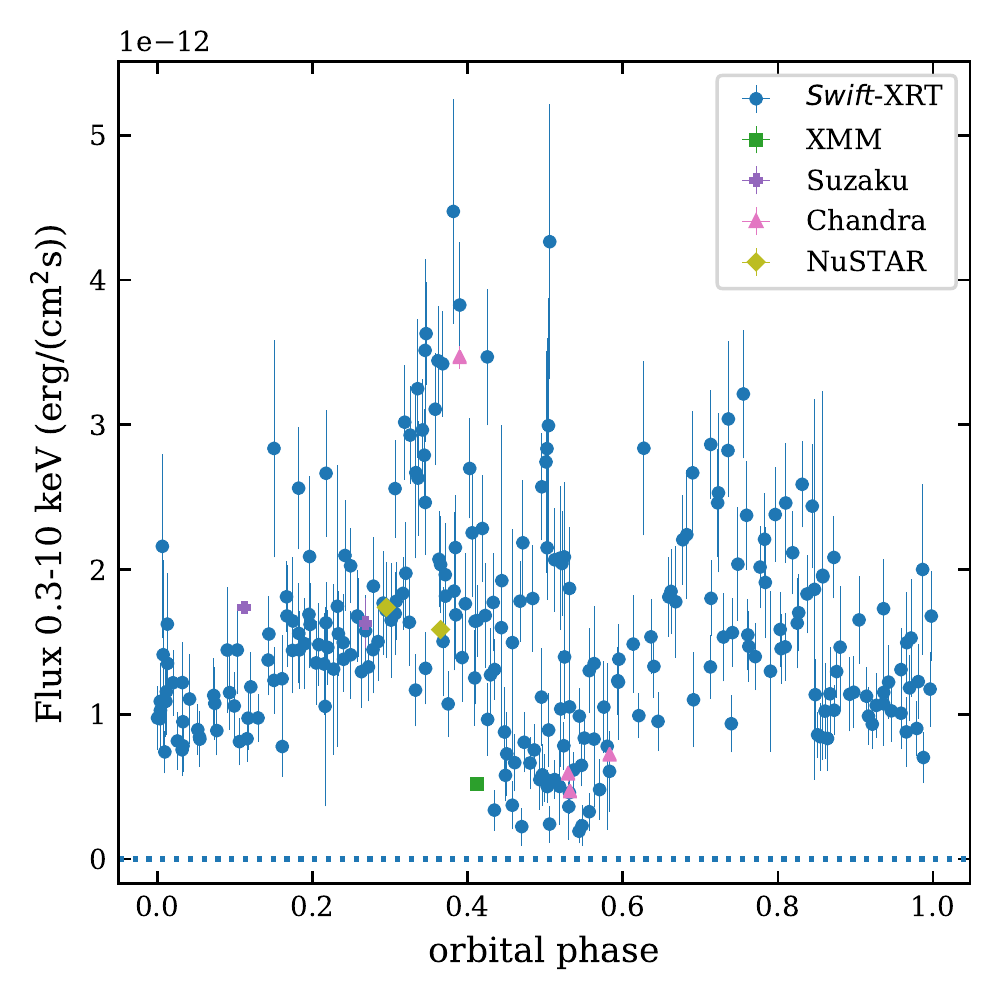}{0.45\textwidth}{(c) X-rays ($P=313$ days)}
\fig{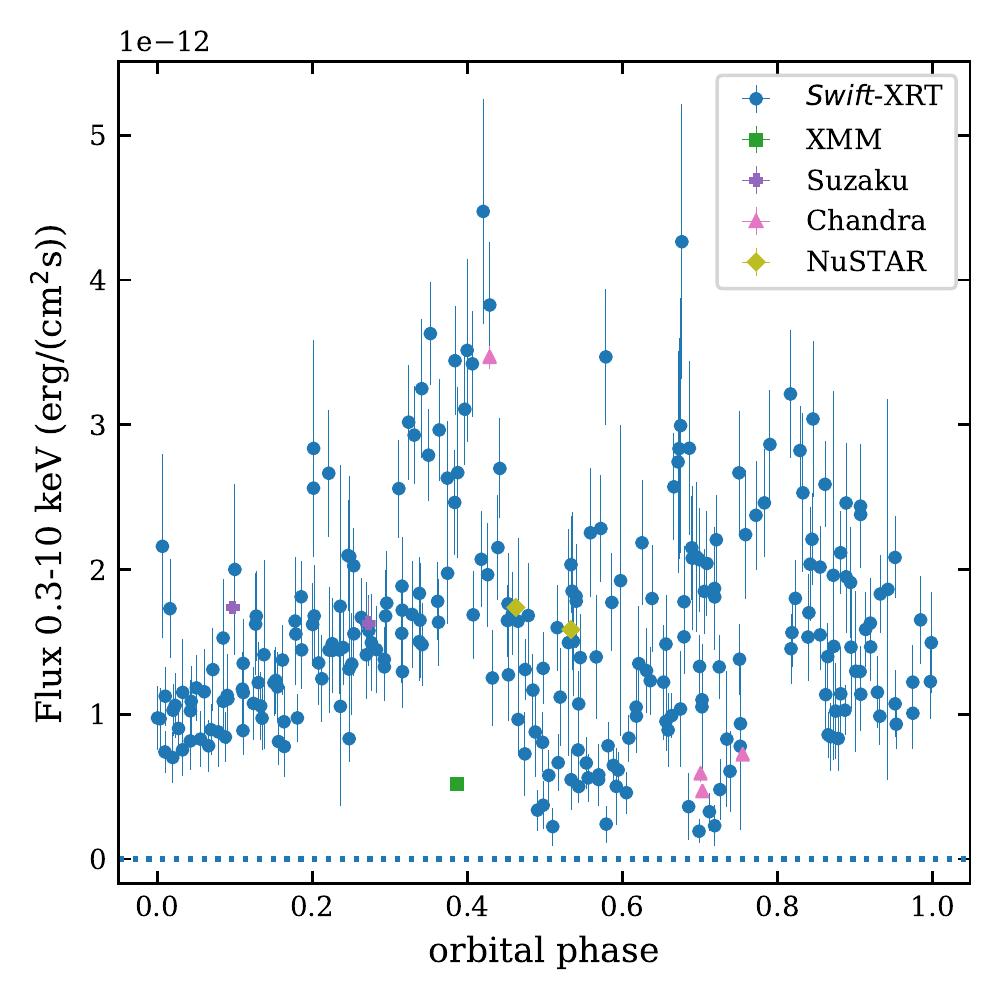}{0.45\textwidth}{(d) X-rays ($P=308$ days)}
}
\caption{\label{fig:lc-phaseFolded-uncertaintiesX}
X-ray  (0.3--10\,keV) light curves as function of orbital phase assuming orbital periods as indicated below the figures.
For further details, see Figure \ref{fig:lc-phaseFolded} and text.}
\end{figure}

\begin{figure}
\gridline{
\fig{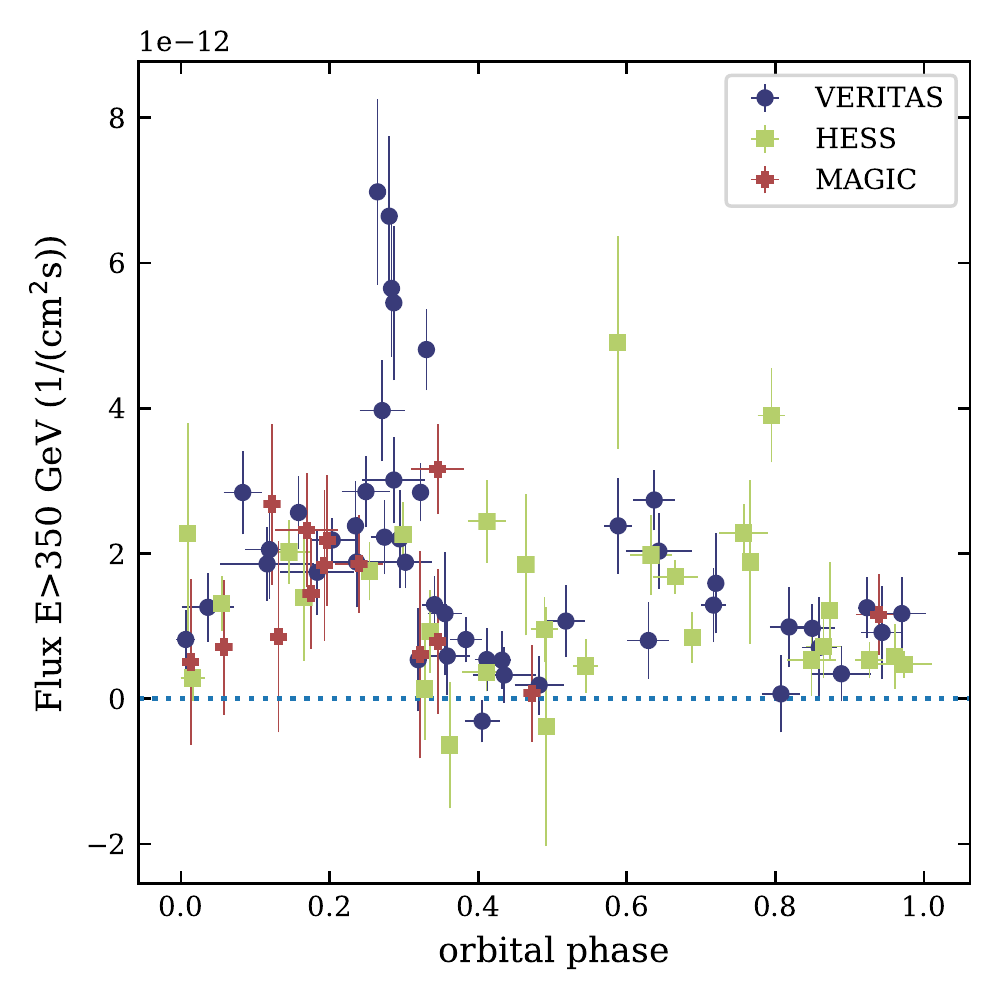}{0.45\textwidth}{(a) Gamma-rays ($P=319.5$ days)}
\fig{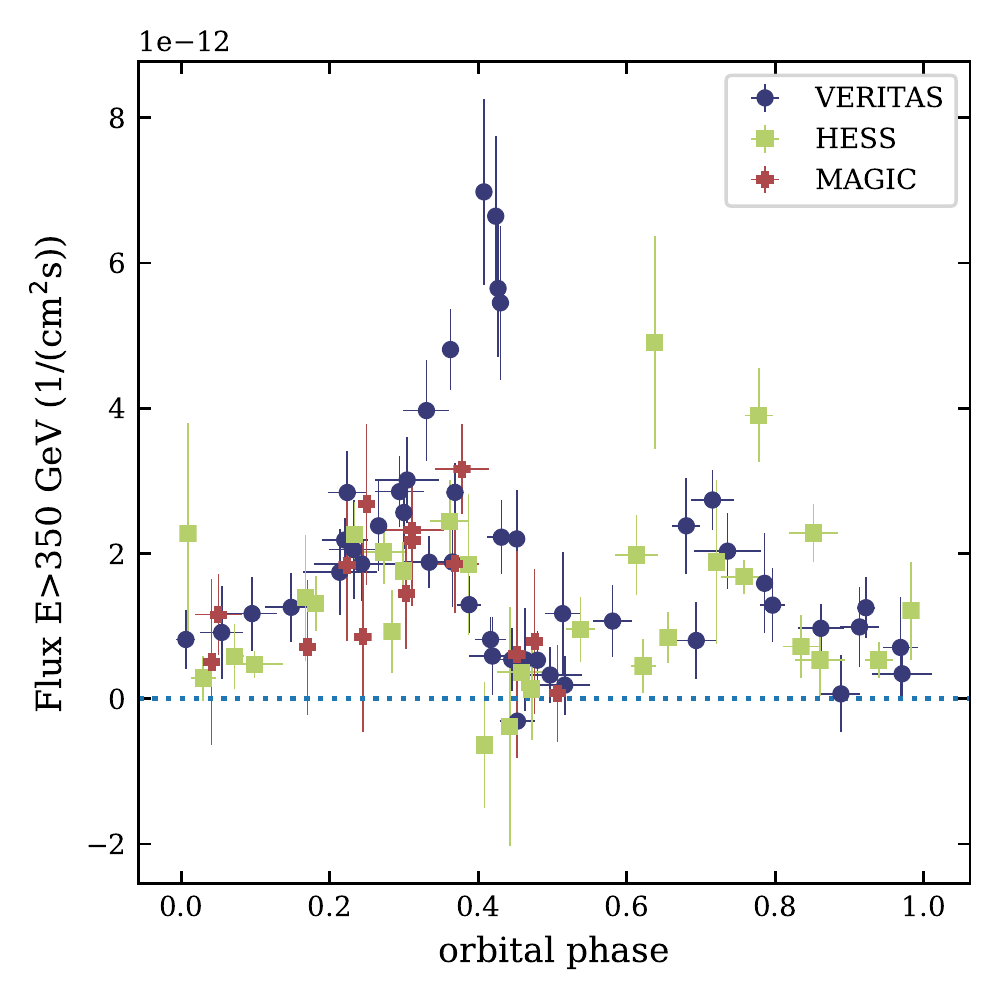}{0.45\textwidth}{(b) Gamma-rays ($P=315.1$ days)}}
\gridline{
\fig{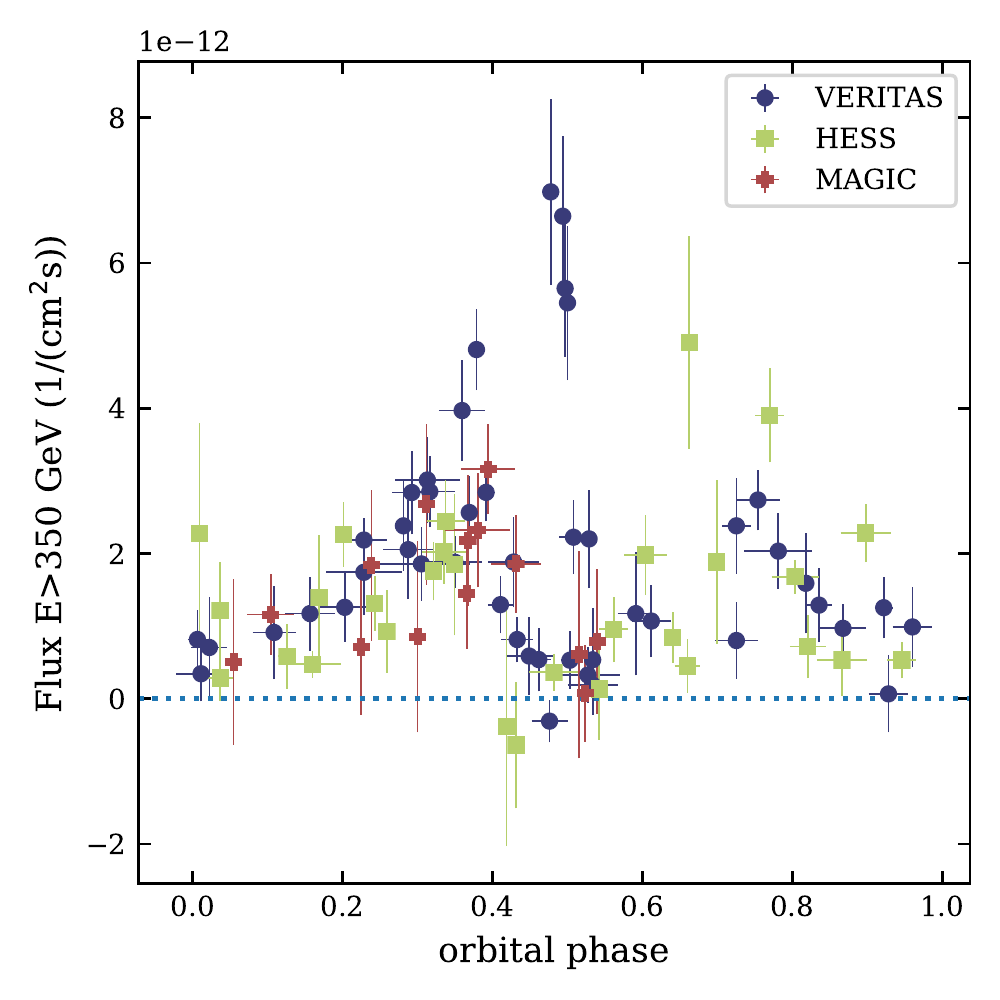}{0.45\textwidth}{(c) Gamma-rays ($P=313$ days)}
\fig{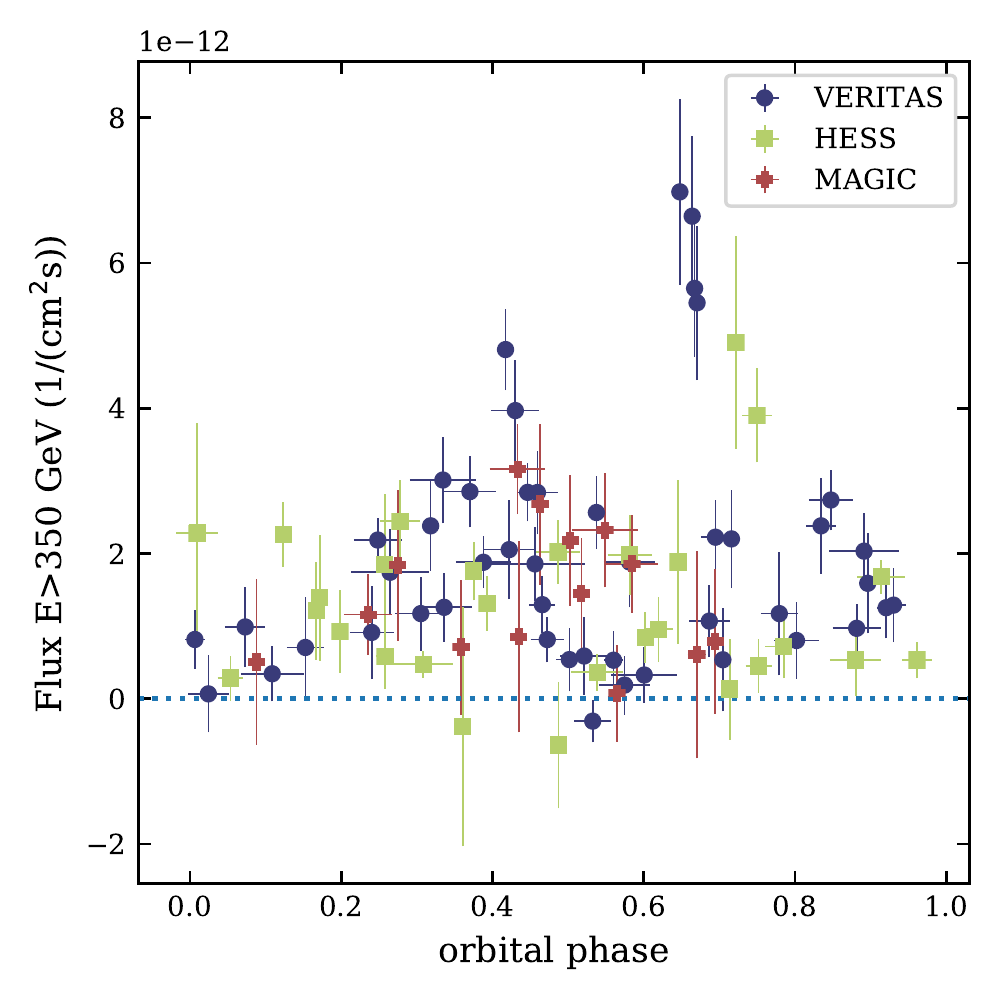}{0.45\textwidth}{(d) Gamma-rays ($P=308 $ days)}
}
\caption{\label{fig:lc-phaseFolded-uncertaintiesG}
Gamma-ray ($>$350\,GeV) light curves as function of orbital phase assuming orbital periods as indicated below the figures.
For further details, see Figure \ref{fig:lc-phaseFolded} and text.}
\end{figure}

Figures \ref{fig:lc-phaseFolded-uncertaintiesX} and \ref{fig:lc-phaseFolded-uncertaintiesG} show the impact of  the variation of the assumed orbital period:
the three most prominent features of a peak around phases 0.3, a minimum around phases 0.4, and a second maximum region around phases 0.6 are clearly visible in all cases.
This shows that the choice of $P_{orbit}=317.3$ does not influence the discussion of the physical properties presented in this work, except for the shortest period of 308 days.
It should be noted that assuming such a short orbital period, as derived from optical H$\alpha$ observations \citep{Moritani-2018}, would change this picture significantly, as most of the discussed features disappear (Figures \ref{fig:lc-phaseFolded-uncertaintiesX} (d) and \ref{fig:lc-phaseFolded-uncertaintiesG} (d).

\section{Light curves per orbital cycle}
\label{appendix:detailedLCs}

\begin{figure}
\plotone{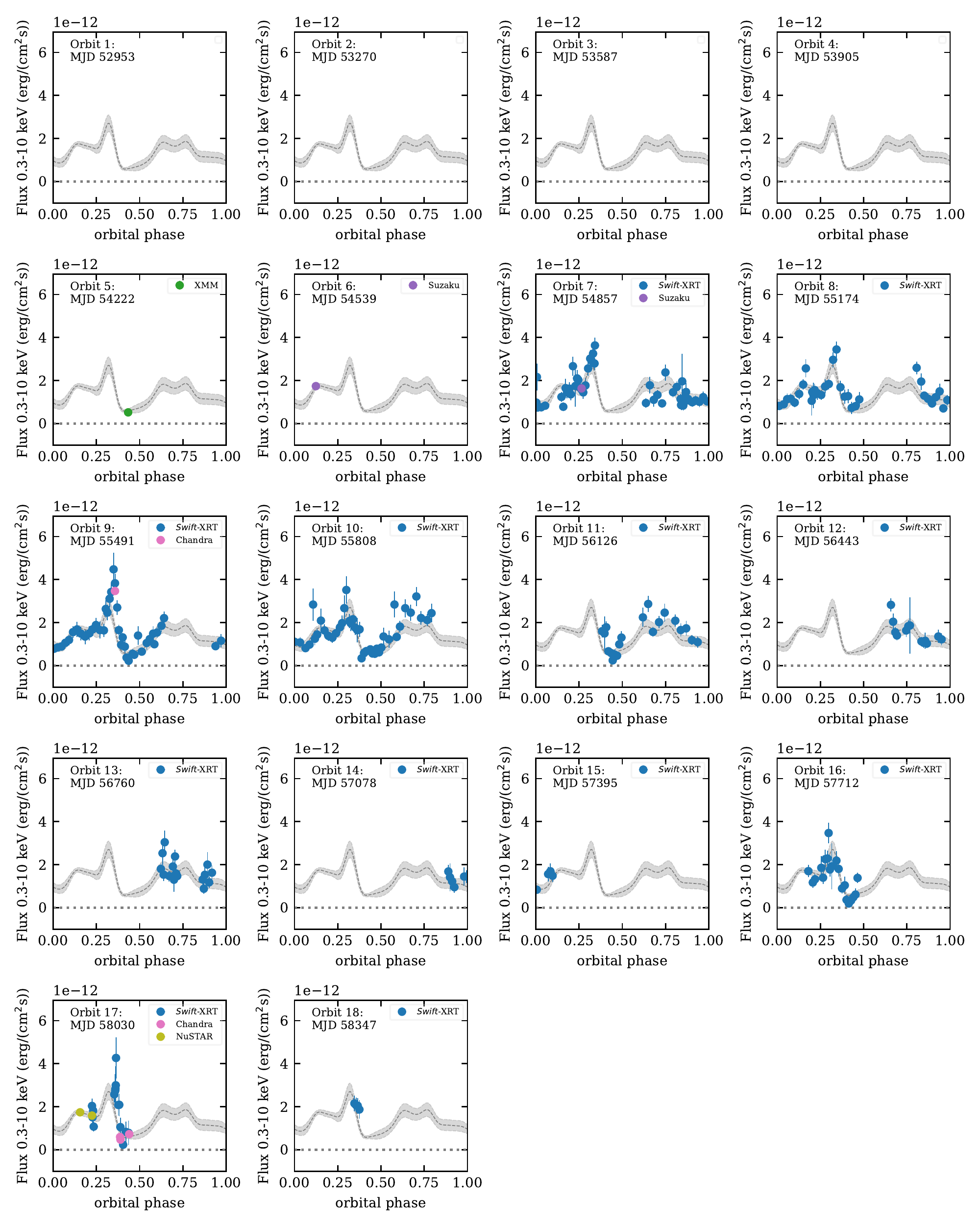}
\caption{\label{fig:lc-phaseFolded-perOrbit-Xray}
X-ray  (0.3--10\,keV) light curves as function of orbital phase for each of the observed orbital cycles.
Orbits are numbered following the start of the gamma-ray observations (no X-ray observations are available for the first four orbits; empty panels are shown for easier comparisons with  Figure \ref{fig:lc-phaseFolded-perOrbit-GammaRay}). 
The MJD given in each panel indicates the start of each orbit.
An orbital period of 317.3\,days is assumed.
Vertical lines show statistical uncertainties; note that these are smaller than the marker size for all instruments but \emph{Swift}-XRT.
The thin blue line and gray-shaded band in each canvas indicates the average \emph{Swift}-XRT light curve and its 68\% containment region calculated from all measurements and smoothed by applying a cubic spline fit.
}
\end{figure}

\begin{figure}
\plotone{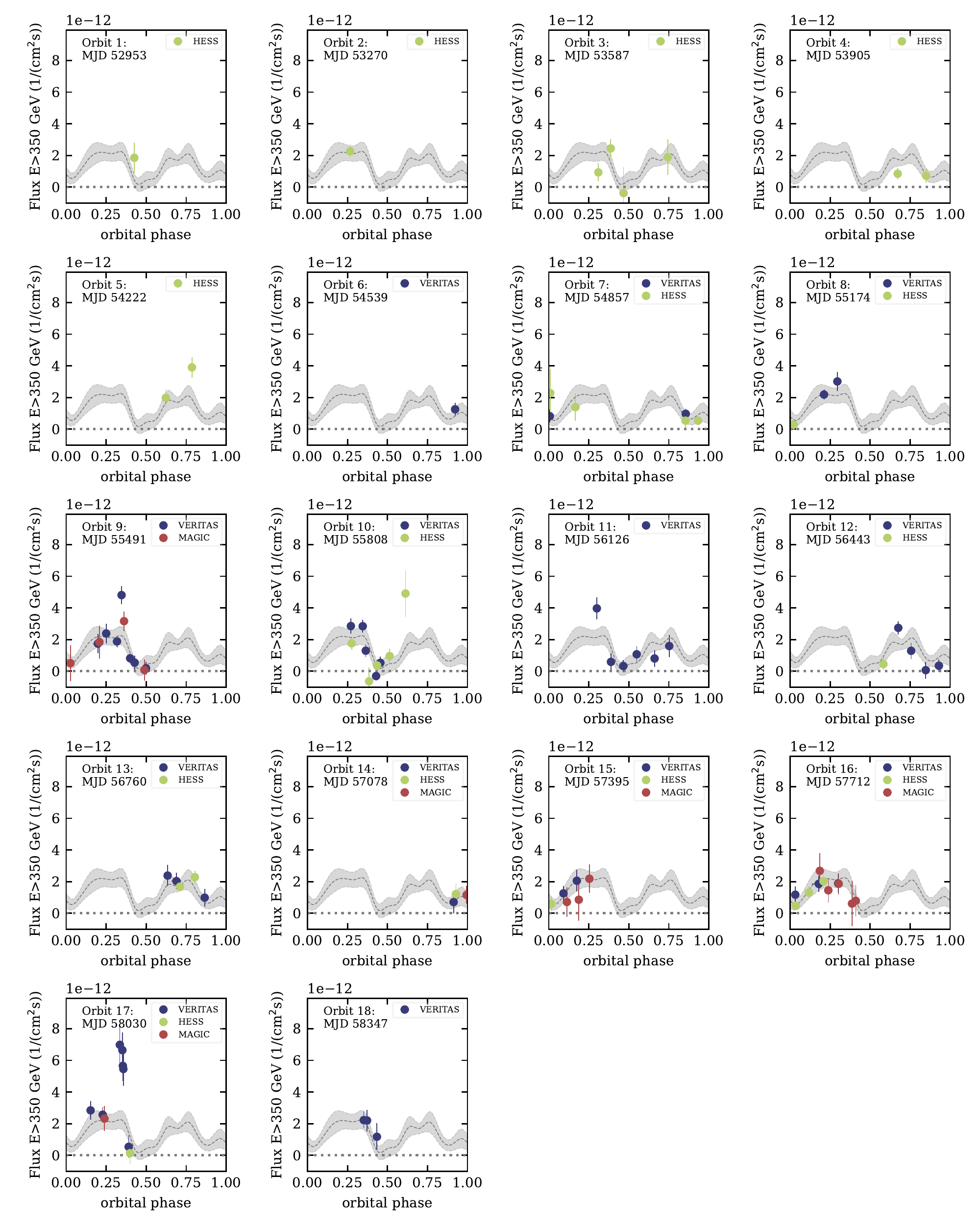}
\caption{\label{fig:lc-phaseFolded-perOrbit-GammaRay}
Gamma-ray ($>$ 350\,GeV) light curves as function of orbital phase for each of the observed orbital cycles (see Figure \ref{fig:lc-phaseFolded-perOrbit-Xray} for further details).
The thin blue line and gray-shaded band in each canvas indicates the average gamma-ray light curve and its 68\% containment region calculated from all measurements and smoothed by applying a cubic spline fit. 
}
\end{figure}

The gamma-ray observations with \hess, MAGIC, and VERITAS sum up to a total observation time of $\approx 450$\,h spanning 18 orbits of \h covering the period of 2004--2019.
The X-ray observations by the Swift\emph{XRT}, Chandra, XMM-Newton, Suzaku, and NuSTAR are obtained during 14 orbits of the binary system.
Despite this large amount of data, there is no good coverage along most of these orbital cycles due to observational constraints and the long binary period of 317\,days.
This is illustrated in figures \ref{fig:lc-phaseFolded-perOrbit-Xray} and \ref{fig:lc-phaseFolded-perOrbit-GammaRay}, which shows the light curve per orbital cycle both at X-ray and gamma-ray energies assuming an orbital period of 317.3\,days.


\section{Contemporaneous spectral energy distributions}

Contemporaneous X-ray and gamma-ray spectral energy distributions for 38\,periods of VERITAS and \emph{Swift}-XRT observations are available through the Zenodo data repository\footnote{\url{https://doi.org/10.5281/zenodo.5157848}}.

\end{appendix}





\end{document}